\input harvmac
 \noblackbox

 \input epsf
 
 \newcount\figno
 \figno=0
 \def\fig#1#2#3{
 \par\begingroup\parindent=0pt\leftskip=1cm\rightskip=1cm\parindent=0pt
 \baselineskip=11pt
 \global\advance\figno by 1
 \midinsert
 \epsfxsize=#3
 \centerline{\epsfbox{#2}}
 \vskip 12pt
 {\bf Fig.\ \the\figno: } #1\par
 \endinsert\endgroup\par
 }
 \def\figlabel#1{\xdef#1{\the\figno}}
 \def\encadremath#1{\vbox{\hrule\hbox{\vrule\kern8pt\vbox{\kern8pt
 \hbox{$\displaystyle #1$}\kern8pt}
 \kern8pt\vrule}\hrule}}
 %
 %


 \font\cmss=cmss10
 \font\cmsss=cmss10 at 7pt
 \def\rlx{\relax\leavevmode}
 \def\inbar{\vrule height1.5ex width.4pt depth0pt}
 \def\IC{\relax\,\hbox{$\inbar\kern-.3em{\rm C}$}}
 \def\IN{\relax{\rm I\kern-.18em N}}
 \def\IP{\relax{\rm I\kern-.18em P}}
 \def\ZZ{\rlx\leavevmode\ifmmode\mathchoice{\hbox{\cmss Z\kern-.4em Z}}
  {\hbox{\cmss Z\kern-.4em Z}}{\lower.9pt\hbox{\cmsss Z\kern-.36em Z}}
  {\lower1.2pt\hbox{\cmsss Z\kern-.36em Z}}\else{\cmss Z\kern-.4em
  Z}\fi}

 \def\IZ{\relax\ifmmode\mathchoice
 {\hbox{\cmss Z\kern-.4em Z}}{\hbox{\cmss Z\kern-.4em Z}}
 {\lower.9pt\hbox{\cmsss Z\kern-.4em Z}}
 {\lower1.2pt\hbox{\cmsss Z\kern-.4em Z}}\else{\cmss Z\kern-.4em
 Z}\fi}
 \def\IZ{\relax\ifmmode\mathchoice
 {\hbox{\cmss Z\kern-.4em Z}}{\hbox{\cmss Z\kern-.4em Z}}

 {\lower.9pt\hbox{\cmsss Z\kern-.4em Z}}
 {\lower1.2pt\hbox{\cmsss Z\kern-.4em Z}}\else{\cmss Z\kern-.4em Z}\fi}

 \def\narrowplus{\kern -.04truein + \kern -.03truein}
 \def\narrowminus{- \kern -.04truein}
 \def\narrowminussub{\kern -.02truein - \kern -.01truein}

 \def\frac#1#2{{#1\over #2}}

\def\IP{{\bf P}}

\def\Tr{\rm Tr}
\def\tr{\rm tr}
\def\SL2z{{\rm SL}(2,\ZZ)}

\def\bfone{\relax{\rm 1\kern-.35em 1}}
\def\inbar{\vrule height1.5ex width.4pt depth0pt}
\def\IC{\relax\,\hbox{$\inbar\kern-.3em{\rm C}$}}
\def\ID{\relax{\rm I\kern-.18em D}}
\def\IF{\relax{\rm I\kern-.18em F}}
\def\IH{\relax{\rm I\kern-.18em H}}
\def\II{\relax{\rm I\kern-.17em I}}
\def\IN{\relax{\rm I\kern-.18em N}}
\def\IQ{\relax\,\hbox{$\inbar\kern-.3em{\rm Q}$}}
\def\us#1{\underline{#1}}
\def\IR{\relax{\rm I\kern-.18em R}}
\font\cmss=cmss10 \font\cmsss=cmss10 at 7pt
\def\ZZ{\relax\ifmmode\mathchoice
{\hbox{\cmss Z\kern-.4em Z}}{\hbox{\cmss Z\kern-.4em Z}}
{\lower.9pt\hbox{\cmsss Z\kern-.4em Z}} {\lower1.2pt\hbox{\cmsss
Z\kern-.4em Z}}\else{\cmss Z\kern-.4em Z}\fi}
\def\nup#1({Nucl.\ Phys.\ $\us {B#1}$\ (}
\def\plt#1({Phys.\ Lett.\ $\us  {B#1}$\ (}
\def\cmp#1({Comm.\ Math.\ Phys.\ $\us  {#1}$\ (}
\def\prp#1({Phys.\ Rep.\ $\us  {#1}$\ (}
\def\prl#1({Phys.\ Rev.\ Lett.\ $\us  {#1}$\ (}
\def\prv#1({Phys.\ Rev.\ $\us  {#1}$\ (}
\def\mpl#1({Mod.\ Phys.\ Let.\ $\us  {A#1}$\ (}
\def\ijmp#1({Int.\ J.\ Mod.\ Phys.\ $\us{A#1}$\ (}
\def\tit#1|{{\it #1},\ }

\def\Coe#1.#2.{{#1\over #2}}

\def\coe#1.#2.{\relax{\textstyle {#1 \over #2}}\displaystyle}

\def\br{\hfill\break}

\def\frac#1#2{{#1\over #2}}

\def\tst{{\theta_t}}
\def\tss{{\theta_s}}

\def\dd{{\rm d}}

 \def\IZ{\relax\ifmmode\mathchoice
 {\hbox{\cmss Z\kern-.4em Z}}{\hbox{\cmss Z\kern-.4em Z}}
 {\lower.9pt\hbox{\cmsss Z\kern-.4em Z}}

 {\lower1.2pt\hbox{\cmsss Z\kern-.4em Z}}\else{\cmss Z\kern-.4em Z}\fi}
 \def\IB{\relax{\rm I\kern-.18em B}}
 \def\IC{{\relax\hbox{$\inbar\kern-.3em{\rm C}$}}}
 \def\Ic{{\relax\hbox{$\inbar\kern-.22em{\rm c}$}}}
 \def\ID{\relax{\rm I\kern-.18em D}}
 \def\IE{\relax{\rm I\kern-.18em E}}
 \def\IF{\relax{\rm I\kern-.18em F}}
 \def\IG{\relax\hbox{$\inbar\kern-.3em{\rm G}$}}
 \def\IGa{\relax\hbox{${\rm I}\kern-.18em\Gamma$}}
 \def\IH{\relax{\rm I\kern-.18em H}}
 \def\II{\relax{\rm I\kern-.18em I}}
 \def\IK{\relax{\rm I\kern-.18em K}}
 \def\IP{\relax{\rm I\kern-.18em P}}

 \font\cmss=cmss10 \font\cmsss=cmss10 at 7pt
 \def\IR{\relax{\rm I\kern-.18em R}}

 %

 %

 %
 \def\eqnn#1{\xdef #1{(\secsym\the\meqno)}\writedef{#1\leftbracket#1}%
 \global\advance\meqno by1\wrlabeL#1}
 \def\eqna#1{\xdef #1##1{\hbox{$(\secsym\the\meqno##1)$}}
 \writedef{#1\numbersign1\leftbracket#1{\numbersign1}}%
 \global\advance\meqno by1\wrlabeL{#1$\{\}$}}
 \def\eqn#1#2{\xdef #1{(\secsym\the\meqno)}\writedef{#1\leftbracket#1}%
 \global\advance\meqno by1$$#2\eqno#1\eqlabeL#1$$}
\Title
 {\vbox{
 \baselineskip12pt
 \hbox{HUTP-01/A023}
 \hbox{HU-EP-01/21}
 \hbox{hep-th/0105045}\hbox{}\hbox{}
}}
{\vbox{
 \centerline{Disk Instantons, Mirror Symmetry}
 \vglue .5cm
 \centerline{and the}
 \vglue .5cm
 \centerline{Duality Web}}}
 \centerline{ Mina Aganagic$^1$, Albrecht Klemm$^2$ and
Cumrun Vafa$^1$}
\bigskip
\centerline{$^1$ Jefferson Physical Laboratory}
\centerline{Harvard University, Cambridge, MA 02138, USA}
\centerline{$^2$ Institut f\"ur Physik, Humboldt Universit\"at zu Berlin}
\centerline{Invaliden Stra\ss e 110, D-10115, Germany}

 \smallskip
 \vskip .5cm \centerline{\bf Abstract}

We apply the methods recently developed
for computation of
type IIA disk instantons using mirror
symmetry to a large class of D-branes wrapped over
Lagrangian cycles of non-compact Calabi-Yau 3-folds.
Along the way we clarify
the notion of ``flat coordinates'' for the boundary theory.
We also discover an integer IR ambiguity needed to define
the quantum theory of D-branes wrapped over non-compact
Lagrangian submanifolds.  In the large $N$ dual Chern-Simons theory,
this ambiguity is mapped to the UV choice of the framing
of the knot.  In a type IIB dual description involving
$(p,q)$ 5-branes, disk instantons of type IIA get mapped to
$(p,q)$ string instantons.
The M-theory
lift of these results lead to computation of superpotential terms
generated by M2 brane instantons wrapped over
3-cycles of certain manifolds of $G_2$ holonomy.

\smallskip \Date{May 2001}

\newsec{Introduction}
D-branes wrapped over non-trivial cycles of a Calabi-Yau threefold
provide an interesting class of theories with 4 supercharges
(such as $N=1$ supersymmetric theories in $d=4$).
As such, they do allow the generation of a superpotential on their
worldvolume.  This superpotential depends holomorphically on the chiral fields
which parameterize normal deformations of the wrapped D-brane.

On the other hand F-terms are captured by topological
string amplitudes
\ref\bcov{M.~Bershadsky, S.~Cecotti, H.~Ooguri and C.~Vafa,
``Kodaira-Spencer theory of gravity and exact results for quantum string amplitudes,''
Commun.\ Math.\ Phys.\  {\bf 165}, 311 (1994), hep-th/9309140.}
and in particular the superpotential is computed
by topological strings at the level of the disk amplitude
\bcov\ref\bdlr{I.~Brunner, M.R.~Douglas, A.~Lawrence and
C.~R\"omelsberger,``D-branes on the quintic,'' JHEP 0008, 015 (2000),
hep-th/9906200.}\ref\kkl{S.~Kachru, S.~Katz,
A.~E.~Lawrence and J.~McGreevy,``Open string instantons and superpotentials,
Phys.\ Rev.\ D {\bf 62}, 026001 (2000), hep-th/9912151; ``Mirror symmetry for open strings,'' Phys.\
Rev.\ D {\bf 62} (2000) 126005, hep-th/0006047.}\ref\ov{H.~Ooguri
 and C.~Vafa, ``Knot invariants and topological strings,''
Nucl.\ Phys.\ B {\bf 577}, 419 (2000), hep-th/9912123.}.  More generally
the topological string amplitude
at genus $g$ with $h$ holes computes superpotential corrections
involving the gaugino superfield $W$
 and the $N=2$
graviphoton multiplet ${\cal W}$
 given by $h\int d^2\theta
(\Tr \, W^2)^{h-1} ({\cal W}^2)^g$ \ref\vaug{C. Vafa,
``Superstrings and Topological
Strings at Large $N$,'' hep-th/0008142.}.
So the issue of computation of topological string amplitudes becomes
very relevant for this class of supersymmetric theories.

In the context of type IIA superstrings such disk amplitudes
are given by non-trivial worldsheet instantons, which are holomorphic
maps from the disk to the CY with the boundary ending on the D-brane.
Such computations are in general rather difficult. The same questions
in the context of type IIB strings involve classical considerations
of the worldsheet theory.
In a recent paper
\ref\av{ M.~Aganagic and C.~Vafa,``Mirror symmetry, D-branes
and counting holomorphic discs,''hep-th/0012041.}\ it was shown how one can use mirror symmetry
in an effective way to transform the type IIA computation
of disk instantons to classical computations in the context of a
mirror brane on a mirror CY for type IIB strings.
The main goal of this paper is to extend this method to more
non-trivial
Calabi-Yau geometries.

One important obstacle to overcome in generalizing \av\ is
a better understanding of ``flat coordinates'' associated
with the boundary theory, which we resolve by identifying it
with BPS tension of associated domain walls.
 We also uncover a generic IR ambiguity
given by an integer in defining a quantum Lagrangian D-brane.
We relate this ambiguity to the choice of the regularizations
of the worldsheet theory associated to the
 boundaries of moduli
space of Riemann surfaces with holes
(the simplest one being two disks connected by an infinite strip).
In the context of the Large $N$ Chern-Simons dual \ref\gopv{R.~Gopakumar and C.~Vafa,
``On the gauge theory/geometry correspondence,''
Adv.\ Theor.\ Math.\ Phys.\  {\bf 3}, 1415 (1999), hep-th/9811131.}\
applied to Wilson Loop observables \ov\
this ambiguity turns out to be related to the UV choice of the framing of
the knot, which is needed for defining the Wilson loop observable
by point splitting \ref\witcso{E.~Witten, ``Quantum Field Theory and
the Jones Polynomial,'' Commun. Math. Phys. 121, (1989) 351-399.}.

Along the way, for gaining further insight, we consider other
equivalent dual
theories, including the lift to M-theory, involving M-theory
in a $G_2$ holonomy background.  In this context we are able
to transform the generation of superpotential by Euclidean
M2 branes (with the topology of $S^3$) to disk instantons of
type IIA\foot{More generally we can map the generation of
superpotential-like terms associated to topological strings
at genus $g$ with $h$ boundaries to Euclidean
M2 brane instantons on a closed 3-manifold with $b_1=2g+h-1$.}
for M-theory on $G_2$ holonomy manifolds
and use mirror symmetry to compute them!  We also relate this theory to
another dual type IIB theory in a web
of $(p,q)$ 5-branes in the presence of ALF-like geometries.

The organization of this paper is as follows:
In section 2 we review the basic setup of \av .  In section 3
we consider the lift of these theories to M-theory in the context
of $G_2$ holonomy manifolds, as well as to type IIB theory
with a web of $(p,q)$ 5-branes in an ALF-like background.
In section 4 we identify the flat coordinates
for boundary fields by computing the BPS tension
of D4 brane domain walls ending on D6 branes wrapping Lagrangian
submanifolds.
In section 5 we discuss the integral ambiguity in the computation of topological
string amplitudes
 and its physical meaning.  This is
discussed both in the context of Large N Chern-Simons/topological
string duality, as well as in the context of the type IIB theory
with a web of $(p,q)$ 5-branes.
 In section 6 we present a large
class of examples, involving non-compact CY 3-folds where
the D6 brane wraps a non-compact Lagrangian submanifold.
In appendix A we perform some of the computations relevant
for the framing dependence for the unknot and verify that in the large
$N$ dual description  this UV choice maps to the integral IR ambiguity we have
discovered for the quantum Lagrangian D-brane.

\newsec{Review of Mirror Symmetry for D-branes}
In this section we briefly recall the mirror symmetry construction
for non-compact toric Calabi-Yau manifolds (specializing to the
case of threefolds), including the mirror of some particular class
of (special) Lagrangian D-branes on them.

Toric Calabi-Yau threefolds arise as symplectic quotient spaces $X
= {\bf C}^{3+k}//G$, for $G= U(1)^k$. The quotient is obtained by
imposing the $k$ $D$-term constraints
\eqn\df{D^a = Q_1^a |X^1|^2 + Q_2^a |X^2|^2 + \ldots Q_{3+k}^a
|X^{3+k}|^2 - r^a=0}
where $a=1,\ldots k$, and dividing by $G$ \eqn\gauge{ X^i
\rightarrow e^{i Q_i^a \epsilon_a} X^i.} The $c_1(X)=0$ condition
is equivalent to $\sum_i Q_i^a=0$. The K\"ahler structure is
encoded in terms of
the  $r^a$ and varying them changes the sizes of various 2 and 4 cycles.
 In the linear sigma model realization \ref\witph{E.~Witten,
``Phases of N = 2 theories in two dimensions,''
Nucl.\ Phys.\ B {\bf 403}, 159 (1993), hep-th/9301042. }\ this
 is realized as a (2,2) supersymmetric
 $U(1)^k$ gauge theory with $3+k$ matter fields $X^i$
 with charges given by $Q_i^a$, and with $k$ FI terms for the $U(1)^k$
gauge group given by $r^a$.

The mirror theory is given in terms of $n+k$ dual $\bf{C}^*$
fields $Y^i$ \ref\hv{K.~Hori and C.~Vafa,
``Mirror symmetry,'' hep-th/0002222.},
 where
 \eqn\dual{{\rm Re}(Y^i) = -|X^i|^2}
with the periodicity $Y^i\sim Y^i+2\pi i$. The D-term equation \df\ is
mirrored by
\eqn\mdf{Q_1^a Y^1 + Q_2^a Y^2 + \ldots Q_{3+k}^a Y^{3+k} = -t^a}
where $t^a = r^a+i \theta ^a$ and $\theta^a$ denotes the
$\theta$-angles of the $U(1)^a$ gauge group. Note that \mdf\ has a
three-dimensional family of solutions. One parameter is trivial
and is given by $Y^i\rightarrow Y^i+c$. Let us pick a parameterization
of the two non-trivial solutions by $u,v$.

The mirror theory can be represented as a theory of variations of
complex structures of a hypersurface $Y$
\eqn\mirr{xz = e^{Y^1(u,v)} + \ldots +e^{Y^{k+3}(u,v)}\equiv
P(u,v),} where \eqn\sol{Y^i(u,v) = a^i u + b^i v +t^i(t)}
is a solution to \mdf\ (in obtaining this form, roughly speaking
the trivial solution of shifting of all the $Y^i$ has been
replaced by $x,z$ whose product is given by the above equation).
We choose the solutions so that the periodicity condition of the
$Y^i \sim Y^i+2\pi i$ are consistent with those of $u,v$ and that
it forms a fundamental domain for the solution.
 Note that this in particular requires $a^i, b^i$ to be
integers.
 Even after taking these constrains into account
 there still is an $\SL2z$ group action on the
space of solutions via
$$ u \rightarrow a u +b v$$
$$ v \rightarrow cu + d v.$$
Note that the holomorphic 3-form for CY is given by
$$\Omega = \frac{ \dd x\, \dd u\, \dd v}{x},$$
and is invariant under the $\SL2z$ action.

\subsec{Special Lagrangian Submanifolds and Mirror Branes}

In \av\ a family of special Lagrangian submanifolds of the A-model
geometry was studied, characterized by two charges $q_i^\alpha$
with $i=1,...,k+3$ and $\alpha =1,2$, subject to
$$\sum_i q_i^\alpha =0$$
and in terms of which the Lagrangian submanifold is given by three
constrains. Two of them given by
\eqn\gna{\sum q_i^{\alpha} |X^i|^2=c^{\alpha}}
and the third is $\sum \theta^i=0$ where $\theta^i$ denotes the
phase of $X^i$.  The worldsheet boundary theory for this
class of theories has been further studied in \ref\khor{K. Hori,
``Linear Models of  Supersymmetric D-Branes,'' hep-th/0012179.}.

 The submanifolds in question project to one
dimensional subspaces of the toric base (taking into account the
constrains \df ,\gna ),
 \eqn\Abrane{ |X^i|^2 = r +b ^i }
for some fixed $b^i$ (depending on $c^{\alpha},r^a$) and $r \in
\bf{R^+}$. In order to get a smooth Lagrangian submanifold one has
to double this space (by including the $\sum \theta^i=\pi$). The
topology of the Lagrangian submanifold is $R\times S^1 \times
S^1$. There is however a special choice of $c^{\alpha}$ which
makes the Lagrangian submanifold pass through the intersection
line of two faces of the toric base. The topology of the
Lagrangian submanifold will be different in this limit.  It
corresponds to having one of the $S^1$ cycles pinched at a point
of $R$ in the Lagrangian submanifold.  This is topologically the
same as two copies of ${\bf C}\times S^1$ touching at the origin
of $\bf C$. In this limit we view the Lagrangian submanifold as
being made of {\it two} distinct ones intersecting over an $S^1$.
We can now have a deformation, which  moves the two
Lagrangian submanifolds {\it independently}, where the end point
of each one should be a point (not necessarily the same) on the
base of the toric geometry (see the example below).

Under mirror symmetry, the A-brane maps to a holomorphic
submanifold of the $Y$ given by \eqn\Bbrane{ x \;\; = \;0 \; =
P(u,v)=e^{Y^1(u,v)}+\ldots + e^{Y^{3+k}(u,v)\ . }} The mirror brane is
one-complex dimensional, and is parameterized by $z$.  Its moduli
space is one complex dimensional parameterized by a point on a
Riemann surface $P(u,v)=0$.  The choice of the point depends on
$c^{\alpha}$ and the Wilson line around $S^1$ and it is possible
to read it off in the weak coupling limit of large volume of
Calabi-Yau and large parameters $c^{\alpha}$ as discussed in \av .

\subsec{Example}

For illustration consider $X =O(-1)\oplus O(-1)\rightarrow P^1$,
which is also called small resolution of conifold.  This sigma
model is realized by $U(1)$ gauge theory with $4$ chiral fields,
with charges $Q=(1,1,-1,-1)$. The D-term potential vanishes on
$|X^1|^2+|X^2|^2-|X^3|^2-|X^4|^2 = r$, and $X$ is a quotient of
this by $U(1)$. The D-term equations can be regarded as linear
equations by projecting $X^i \rightarrow |X^i|^2$, and solved
graphically in the positive octant of $R^3$ (see Fig. 1).
\bigskip
\centerline{\epsfxsize 4.truein\epsfbox{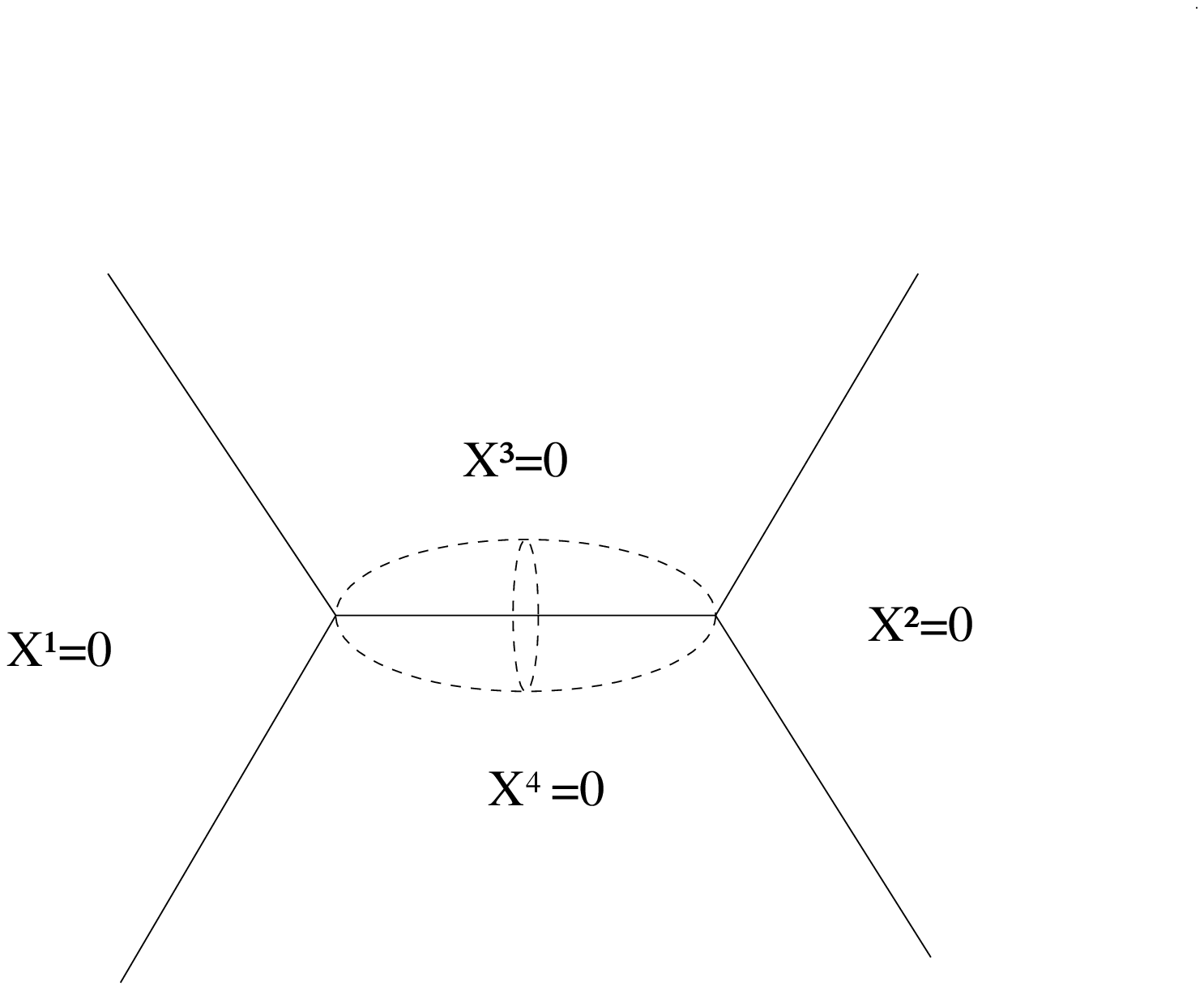}}
\rightskip 2pc
\noindent{\ninepoint\sl \baselineskip=8pt {\bf Fig.1}{
$X=O(-1)\oplus O(-1)\rightarrow {\bf P^1}$ viewed as a toric fibration.
The base is $(|X^1|^2,|X^3|^2,|X^4|^2)$ as generic solution to
the vanishing of D-term potential, but is bounded by $|X^2|^2 \geq 0$
hyperplane. Over the faces of the bounding hyperplanes some cycles of the fiber
shrink. For example, there is a minimal ${\bf
P^1}$ in $X$ which lies over the finite edge.}
}
\bigskip
$X$ is fibered over this base with fiber which is torus of phases
of $X^i$'s modulo $U(1)$, $T^3 =T^4/U(1)$. Note that $r$ is the size of a minimal
$\bf{P^1}$ at $X^3=0=X^4$.

Consider a special Lagrangian D-brane in this background with
$q_1=(1,0,0,-1)$, $q_2=(0,0,1,-1)$. This gives two constrains
$|X^1|^2-|X^4|^2 = c_1$ and $|X^3|^2-|X^4|^2 = c_2$ in the base
which determine a two dimensional family of Lagrangians, but
D-branes of topology $\bf{C} \times S^1$ are further constrained
to live on the one dimensional faces of the base. For this we need
for example $c_2 = 0$, and $c_1$ arbitrary but in $(0,r)$
interval. As discussed above this can be viewed as coming from the
deformation of a Lagrangian submanifold which splits to two when
it intersects the edges of the toric geometry and move them
independently on the edge.  See Fig. 2.  Typically we would be
interested in varying the position of one brane, keeping the other
brane fixed (or taken to infinity along an edge).

\bigskip
\centerline{\epsfxsize 7truein\epsfbox{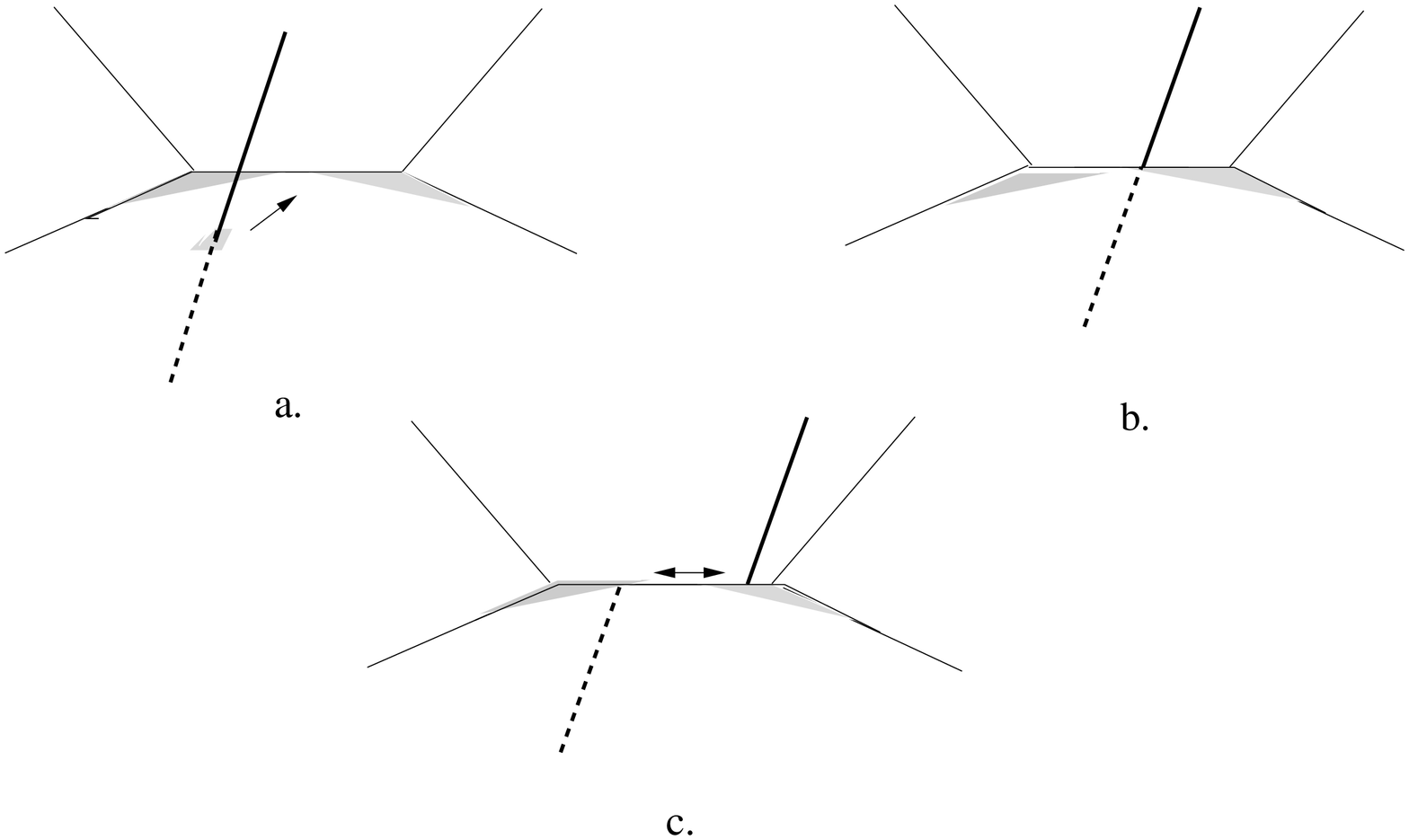}}
\rightskip 2pc \noindent{\ninepoint\sl \baselineskip=8pt {\bf
Fig.2}{ The special Lagrangian submanifold which has topology
${\br R} \times T^2$ for generic values of $c_i$ (case a) can
degenerate (case b) and split (case c) into two Lagrangian
submanifolds, when it approaches a one-dimensional edge of the
toric base. The two resulting components have topology ${\bf
C\times S^1}$, and can move independently, but only along
one-dimensional edges.}}
\bigskip

The mirror of $X$ is $$xz = e^u+e^v+e^{-t-u+v}+1$$ obtained
by solving
$Y^1+Y^2-Y^3-Y^4=-t$ for $Y^2$, fixing the trivial solution by
setting $Y^4=0$, and putting $Y^1=u$ and $Y^3=v$.

The mirror B-brane propagates on the Riemann surface
$0=P(u,v)=e^u+e^v+e^{-t-u+v}+1$ shown in Figure 3.
\bigskip
\centerline{\epsfxsize 4truein\epsfbox{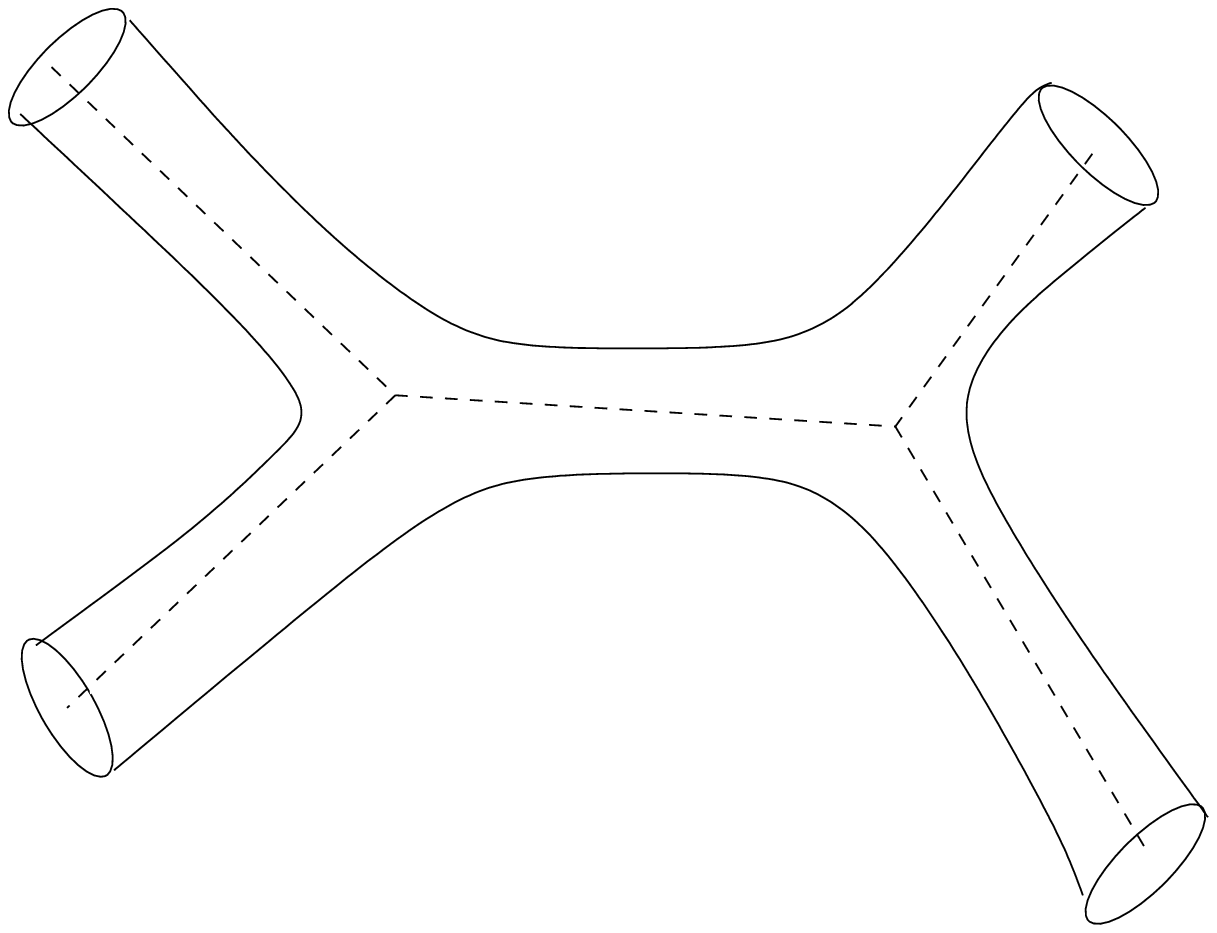}}
\rightskip 2pc \noindent{\ninepoint\sl \baselineskip=8pt {\bf
Fig.3} {Riemann surface $\Sigma \;: \; P(u,v)=0$ corresponding to
the mirror of $X=O(-1)\oplus O(-1)\rightarrow {\bf P^1}$.
$\Sigma$ is related to the toric diagram of $X$ by thickening out
the one-dimensional edges of the base in Fig.1.} }
\bigskip

Note that  mirror map \dual \ gives the B-brane at ${{\rm Re}}(u) = -
c_1 $ and ${{\rm Re}}(v) = 0$ which is on the Riemann surface in the
large radius limit, $r \gg 0$ and $ r/2 > c_1 \gg 0$. In other
words, in the large radius limit, classical geometry of the
D-brane moduli space is a good approximation to the quantum
geometry given by $\Sigma$.

We can also construct, as a limit, Lagrangian submanifolds
of ${\bf C}^3$ by considering the limit $r+i \theta =t\rightarrow \infty$
holding $c_1$ fixed, as shown in figure 4.  In this limit the
mirror geometry become $xz=e^{u}+e^{v}+1$.
\bigskip
\centerline{\epsfxsize 4truein\epsfbox{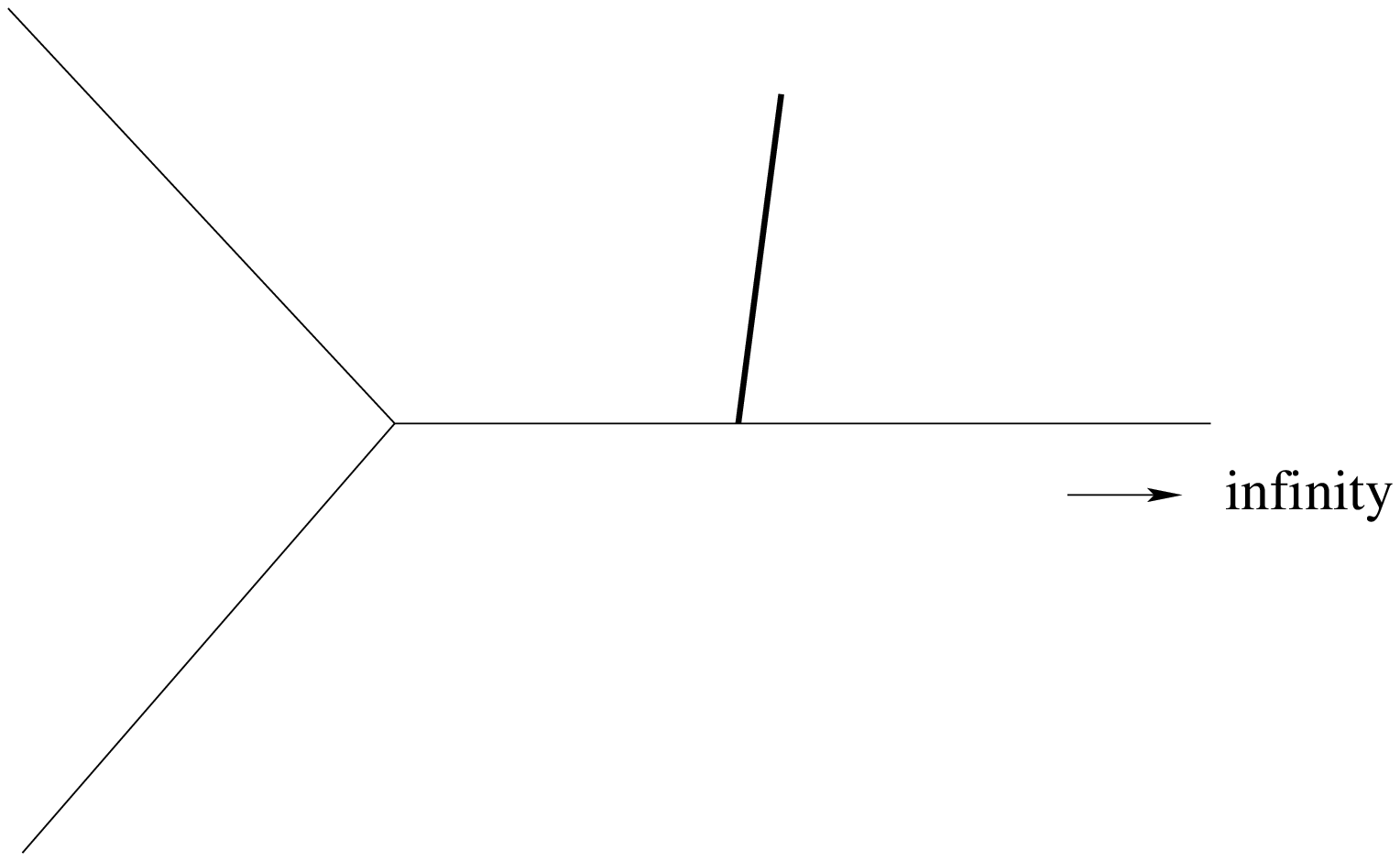}}
\rightskip 2pc
\noindent{\ninepoint\sl \baselineskip=8pt {\bf Fig.4}{ In the limit
in which the size $t$ of the ${\bf P^1}$ in $X=O(-1)\oplus O(-1)
\rightarrow {\bf P^1}$ goes to infinity, the
manifold looks locally like $\bf{C^3}$, together with
a Lagrangian D-brane.}}
\bigskip
This case was studied in detail in \ref\joyce{D. Joyce, ``On counting
special Lagrangian homology 3-spheres,'' hep-th/9907013.}.

\subsec{Disk Amplitude}
The disk amplitudes of the topological A-model give rise to an
$N=1$ superpotential in the corresponding type IIA
superstring theory \bcov\bdlr\kkl\ov\ where we view the D6 brane as
wrapping the Lagrangian submanifold and filling the spacetime.
  The corresponding superpotential for the mirror
of the Lagrangian submanifolds we have discussed above was
computed in \av , and is given in terms of the Abel-Jacobi map
\eqn\suppo{W(u) = \int_{u_*}^{u} v(u) du.}
where $u_*$ is some fixed point on the Riemann surface $P(u,v)=0$
and the line integral is done on this surface.
  This defines the
superpotential up to an addition of a constant.
  More physically
if we construct the `splitting' of the Lagrangian brane
over the toric edges, we can view $u_*$ as the location
of one of the Lagrangian halves, which we consider fixed.

Note that if we move the point $u$ on the Riemann surface over a
closed cycle and come back to the same point, the superpotential
\suppo\ may change by an overall shift, which depends on the
choice of the cycle as well as the moduli of the Riemann surface
(given by $t$'s).  It is natural to ask what is the interpretation
of this shift.  This shift in superpotential can be explained both
from the viewpoint of type IIA and type IIB.  In the context of
type IIA this corresponds to taking the Lagrangian D6 brane over a
path, whose internal volume traces a 4-dimensional cycle $C_4$ of
CY (fixing the boundary conditions at infinity). By doing so we
have come back to the same Brane configuration, but in the process
we have {\it shifted} the RR 2-form flux.  The 4-cycle $C_4$ is
dual to a 2-form which we identify with the shift in the RR 2-form
flux.  In the type IIA setup this process changes the
superpotential by (the quantum corrected) $\int_{C_4} k\wedge k $,
as discussed  in \ref\tv{T.~R.~Taylor and C.~Vafa, ``RR flux on
Calabi-Yau and partial supersymmetry breaking,'' Phys.\ Lett.\ B
{\bf 474}, 130 (2000), hep-th/9912152 .}\ref\ma{P.~Mayr, ``On
supersymmetry breaking in string theory and its realization in
 brane  worlds,''
Nucl.\ Phys.\ B {\bf 593}, 99 (2001), 
hep-th/0003198. }\ref\cklt{G.~Curio,
 A.~Klemm, D.~L\"ust and S.~Theisen, ``On the
vacuum structure of type II string compactifications on Calabi-Yau
spaces with H-fluxes,'' hep-th/0012213.}.   The Type IIB version
of this involves varying D5 brane wrapped over a 2-cycle over a
path and bringing it back to the original place.  During this
process the brane traces a 3-cycle in the internal Calabi-Yau which
contributes integral of the holomorphic 3-form $\Omega$ over the
3-cycle to the superpotential.  This is interpreted as shifting
the RR flux of $H$ along the dual 3-cycle. Note that we can use
this idea to generate fluxes by bringing in branes not
intersecting the toric edge, to the edges, splitting them on the
edge and bringing it back together and then moving it off the
toric edge.  The process leads to the same CY but with some RR
flux shifted.

The superpotential \suppo\ is not invariant under different
choices of parameterization of the fundamental domain for $u,v$
given by an $\SL2z$ transformation, but transforms as
$$ W (u) \rightarrow W(u)+ \int d[ac u^2 /2 + bd v^2/2 - bc uv]
=$$
$$W(u)+ac u^2 /2 + bd v^2/2 - bc uv,$$
where $v$ is defined implicitly in terms of $u$ by $P(u,v)=0$.
Note that if we added a boundary term it could have canceled this
change in superpotential, which can be viewed as a
choice of boundary condition at infinity on the non-compact brane
\av .  Thus this  IR choice is needed for the definition of the
brane, and as we see it affects the physics by modifying the
superpotential.  As discussed in \av\ the choice of the splitting
to $u,v$ depends on the boundary conditions at infinity on the
fields normal to the brane. Each $\SL2z$ action picks a
particular choice of boundary conditions on the D-brane.  Using
the mirror symmetry and what A-model is computing, below we will
be able to fix a canonical choice, up to an integer, which we will
interpret physically.

As noted above, in terms of the topological A-model,
superpotential $W$ is generated by the disk amplitudes. The
general structure of these amplitudes has been determined in \ov\
where it was found that
\eqn\intg{W=\sum_{k,n,{\vec m}}{1\over n^2} N_{k,{\vec m}} {\rm
exp}(n[ku-{\vec m}\cdot{\vec t}])}
Here $u$ parameterizes the size of a non-trivial holomorphic disk
and where $N_{k,{\vec m}}$ are integers capturing the number of
domain wall $D4$ branes ending on the $D6$ brane, which wrap the
CY geometry in the 2-cycle class captured by ${\vec m}$, and $k$
denotes the wrapping number around the boundary.

In the large volume limit (where the area of 2-cycles ending or
not ending on the D-brane are large) the A-model picture is
accurate enough.  In this case we do not expect a classical
superpotential as there is a family of special Lagrangian
submanifolds. Since $dW/du=v$ and $W$ should be zero for any
moduli of the brane, we learn that $v=0$ on the brane.  This in
particular chooses a natural choice of parameterization of the
curve adapted to where the brane is. In particular the D-brane is
attached to the line which is classically specified by $v=0$
(which can always be done). $u$ should be chosen to correspond to
the area of a basic disk instanton. However this can be done in
many ways.  In particular suppose we have one choice of such $u$.
Then
$$u\rightarrow u+nv$$
is an equally good choice, because $v$ vanishes on the Lagrangian
submanifold in the classical limit.  So even though this
ambiguity by an integer is irrelevant in the classical limit, in
the quantum theory since $v$ is non-vanishing due to worldsheet
instanton corrections this dramatically changes the quantum
answer. Thus we have been able to fix the $u,v$ coordinates up to
an integer choice $n$  for each particular geometry of brane.
   We
will discuss further the meaning of the choice of $n$ in section 3
and 4.

Later we will see that there is a further correction to what $u,v$
are quantum mechanically. In particular as we will discuss in
section 3 this arises because the quantum area of the disk differs
from the classical computation which gives $u$.  This is similar
to what happens for the closed string theory where the parameter
$t$ which measures the area of the basic sphere is replaced by the
quantum corrected area $T$.  This is usually referred to
as the choice of the ``flat coordinates'' for the Calabi-Yau moduli.

\newsec{$G_2$ holonomy and type IIB 5-brane Duals}
Consider type IIA superstrings on a non-compact Calabi-Yau
threefold $X$ with a special Lagrangian submanifold $L\subset X$. Consider
wrapping a D6 brane around $L$ and filling $R^4$. This theory has
$N=1$ supersymmetry on $R^4$ and we have discussed the
superpotential generated for this theory. In this section we would
like to relate this to other dual geometries.

\subsec{M-theory Perspective} D6 branes are interpreted as KK
monopoles of M-theory.  This means that in the context of M-theory
the theories under consideration should become purely geometric.
This in fact was studied in
\ref\ach{B.~S.~Acharya, ``On Realising $N=1$ Super Yang-Mills
 in M theory,''
hep-th/0011089. }\ref\amv{ M.~Atiyah, J.~Maldacena and C.~Vafa,
``An M-theory flop as a large $N$ duality,'' hep-th/0011256.
}\ref\Go{J.~Gomis, ``D-branes, holonomy and M-theory,''
hep-th/0103115. }\ref\Nu{J.~D.~Edelstein and C.~Nunez, ``D6 branes
and M-theory geometrical transitions from gauged  supergravity,''
hep-th/0103167.}\ref\kg{S.~Kachru and J.~McGreevy, ``M-theory on
manifolds of G(2) holonomy and type IIA orientifolds,''
hep-th/0103223.}\ where it was seen that the M-theory geometry
corresponds to a 7 dimensional manifold with $G_2$ holonomy. In
other words we consider a 7-fold which is roughly $Y\sim X\times
S^1$ where $S^1$ is fibered over the CY manifold $X$ and vanishes
over the location of the Lagrangian submanifold $L \subset X$. In
this context the superpotentials that we have computed must be
generated by M2 brane instantons wrapping around non-trivial
3-cycles. Some examples of Euclidean M2 brane instantons for $G_2$
holonomy manifolds has been studied in
 \ref\hamo{J.~A.~Harvey and G.~Moore,
``Superpotentials and membrane instantons,''
hep-th/9907026.}.
 In fact
there is a direct map from the disks ending on $L$ to a closed
3-cycle with the topology of $S^3$. In order to explain this we
first discuss some topological facts about $S^3$.

We can  view $S^3$ as
$$|z_1|^2+|z_2|^2=1$$
with $z_i$ complex numbers.  Let $x=|z_1|^2$. The range for $x$
varies from $0$ to $1$. There is an $S^1\times S^1$ of $S^3$ which
project to any fixed $x$ with $0<x<1$, given by the phases of
$z_1$ and $z_2$.  At $x=0$ the circle corresponding to the phase
of $z_1$ shrinks and at $x=1$ the circle corresponding to the
phase of $z_2$ shrinks. So we can view the $S^3$ as the product of
an interval with two $S^1$'s where one $S^1$ shrinks at one end and
the other $S^1$ shrinks at the other end. See Fig.5.
\bigskip
\centerline{\epsfxsize 3truein\epsfbox{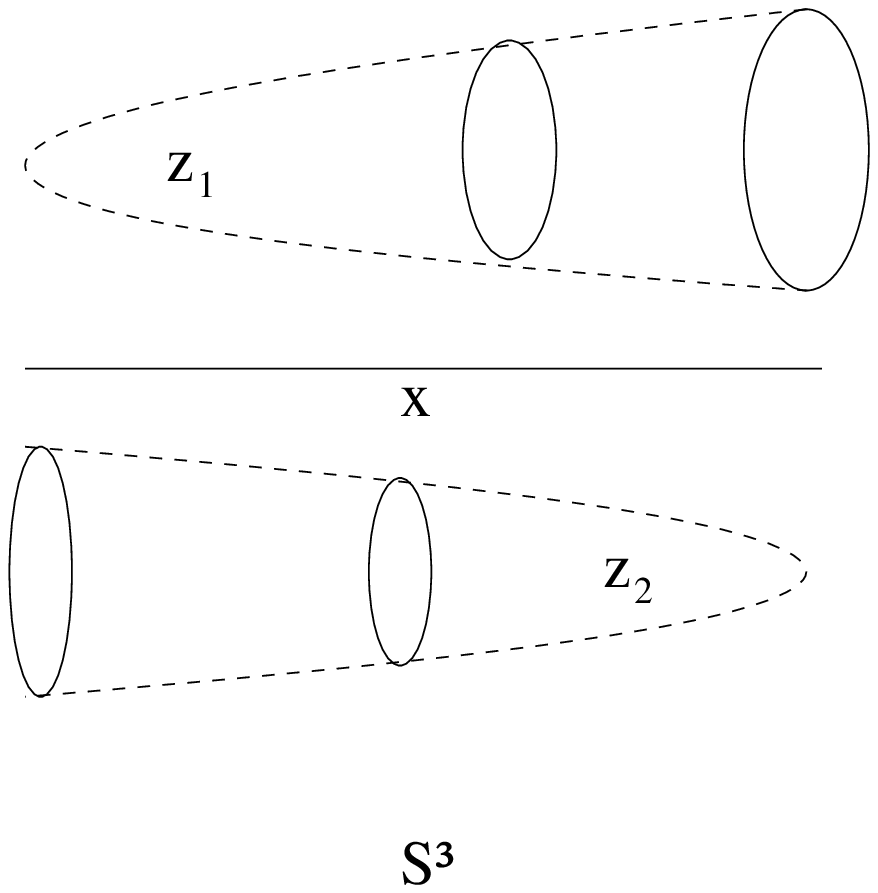}}
\rightskip 2pc \noindent{\ninepoint\sl \baselineskip=8pt {\bf
Fig.5}{ We can view $S^3$ as an $S^1\times S^1$ fibration over an
interval.  Near the ends of the interval it can be viewed as a
complex plane ${\bf C} \times S^1$, where the complex plane is
$z_1$ at one end and $z_2$ at the other. This gives two
inequivalent descriptions of $S^3$ in terms of a circle fibered
over a disk.}}
\bigskip

 We can
also view $S^3$ as a disk times a circle where the circle vanishes
on one boundary--this can be done in two different ways, as shown
in Fig.5.

Now we are ready to return to our case. Consider a disk of type
IIA.  The M2 brane Euclidean instanton can be viewed as the disk
times an $S^1$, where the $S^1$ is the `11-th' circle. Note that
on the boundary of the disk, which corresponds to the Lagrangian
submanifold, the 11-th circle shrinks.  Therefore, from our
discussion above, this three dimensional space has the topology of
$S^3$.

We have thus seen that using mirror symmetry, by mapping the type
IIA geometry with a brane to an equivalent type IIB with a brane,
and computing the superpotential, in effect we have succeeded in
transforming the question of computation of superpotentials
generated by M2 brane instantons in the context of $G_2$ holonomy
manifolds, to an application of
mirror symmetry in the context of D-branes!
More generally, one can in principle compute, using mirror symmetry
\ref\cvext{C. Vafa, ``Extending Mirror Conjecture to Calabi-Yau with Bundles,''
hep-th/9804131.},
partition function for higher genus Riemann surfaces with boundaries.
This computes
 F-term corrections to the
 spacetime theory \vaug\
 given by $  h\int d^4 x d^2 \theta F_{g,h}({\cal W}^2)^{g}{ \Tr(W^2)}^{h-1}$
where ${\cal W}$ is the gravi-photon multiplet, and ${ W}$
the $N=1$ gaugino superfield containing the $U(M)$ field strength
on the worldvolume of M coincident $KK$ monopoles (if we wish
to get infinitely many such contributions we need $M\rightarrow \infty$).
It is easy to see that the topology of the corresponding
M2 brane instantons is a closed 3-manifold with $b_1=2g+h-1$. The case
of the ordinary superpotential is a special case of this with
$g=0,h=1$.

\subsec{Dual Type IIB perspective} We have already given one dual
type IIB theory related to our type IIA geometry, and that is
given by the mirror symmetry we have been considering. However
there is another type IIB dual description which is also rather
useful.

Consider M-theory on a non-compact Calabi-Yau $X$ compactified to
5 dimensions, which admits a $T^2$ action, possibly with fixed
points.  We can use the duality of M-theory on $T^2$ with type IIB
on $S^1$ \ref\schw{
J.~H.~Schwarz,``The power of M theory,''
Phys.\ Lett.\ B {\bf 367}, 97 (1996),
hep-th/9510086.}\ to give a dual type IIB description
for this class of Calabi-Yau manifolds.  Note that the complex
structure of the $T^2$ gets mapped to the coupling constant of
type IIB.
 The non-compact Calabi-Yau manifolds we have been
considering do admit $T^2$ actions and in this way they can be
mapped to an equivalent type IIB theory.  This in fact has been
done in \ref\lev{N.~C.~Leung and C.~Vafa,
 ``Branes and toric geometry,''
Adv.\ Theor.\ Math.\ Phys.\  {\bf 2}, 91 (1998), hep-th/9711013.}\ where it was shown that this class of
CY gets mapped to type IIB propagating on a web of $(p,q)$
5-branes considered in \ref\ahhan{O.~Aharony, A.~Hanany and B.~Kol,
``Webs of (p,q) 5-branes, five dimensional field theories and grid  diagrams,''
JHEP {\bf 9801}, 002 (1998)
hep-th/9710116.}. In this
picture the 5-branes fill the 5 dimensional space time and extend
along one direction in the internal space, identified with various
edges of the toric diagram.  The choice of $(p,q)$ 5-branes
encodes the $(p,q)$ cycle of $T^2$ shrinking over the
corresponding edge. The 5-branes are stretched along straight
lines ending on one another and making very specific angles
dictated by the supersymmetry requirement (balancing of the
tensions) depending on the value of the type IIB coupling constant
$\tau$.  In particular each $(p,q)$ fivebrane is stretched
along 1 dimensional line segments
 on a 2-plane
which is parallel to the complex vector given by $p+q \tau$.
An example of a configuration involving a D5 brane, NS 5 brane
and a (1,1) 5-brane is depicted in Fig. 6.

\bigskip
\centerline{\epsfxsize 3truein\epsfbox{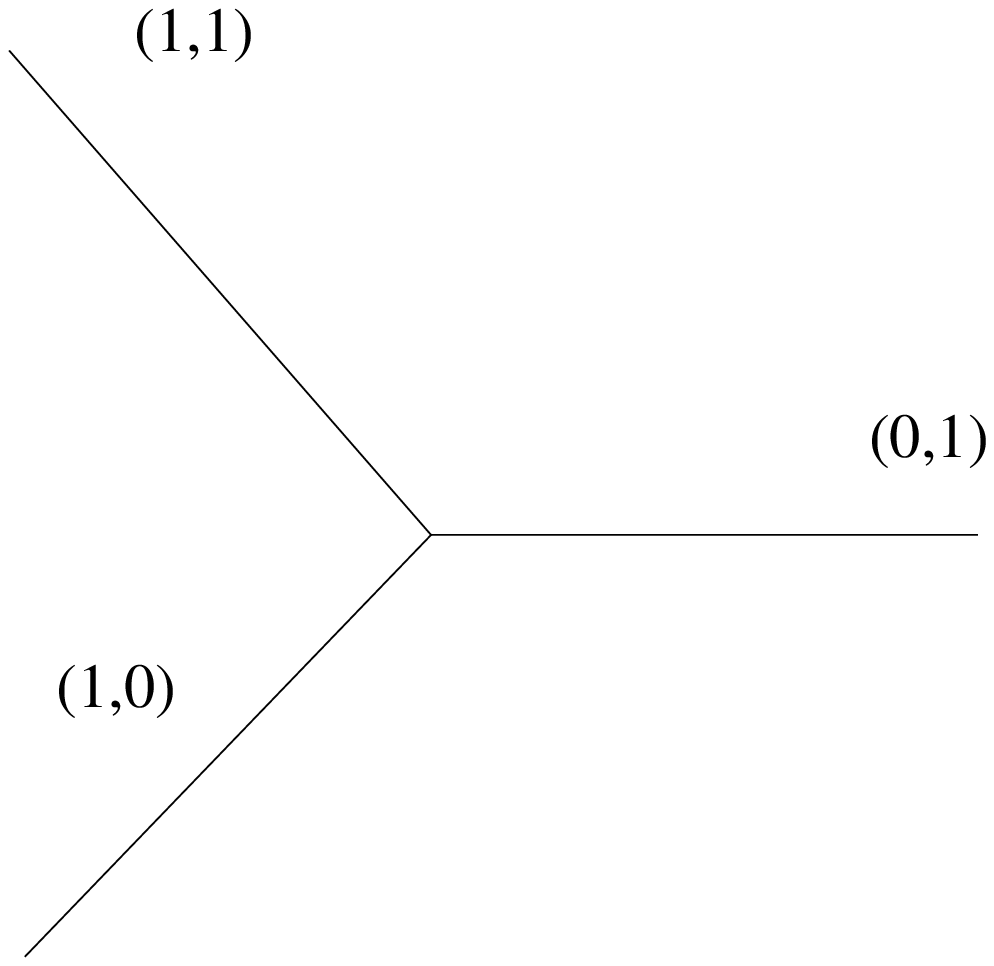}}
\rightskip 2pc
\noindent{\ninepoint\sl \baselineskip=8pt {\bf Fig.6}{ A $(p,q)$ web
of 5-branes which
is dual to M-theory on ${\bf C^3}$.
This web is a junction involving a D5 brane which has $(p,q)=(1,0)$,
an NS5 brane which is $(0,1)$ and a $(1,1)$ brane.}}
\bigskip

Now we recall from the previous discussion that to get the $G_2$
holonomy manifold we need to consider an extra $S^1$ which is
fibered over the corresponding CY.  In other words we are now
exchanging the `5-th' circle with the `11-th' circle.  So we
consider going down to 4 dimensions on a circle which is varying
in size depending on the point in Calabi-Yau. In particular the
circle (i.e. the one corresponding to the 5-th dimension) vanishes
over a 2-dimensional subspace of 5-dimensional geometry of type
IIB (it vanishes along the radial direction of the Lagrangian
submanifold on the base of the toric geometry as well as on the
$S^1$ which is dual to the $T^2$ of M-theory).  Indeed it
corresponds to putting the IIB 5-brane web in a background of ALF
geometry dictated by the location of the Lagrangian submanifold in
the base times $S^1$, and varying the geometry and splitting the
ALF geometry to two halves, as shown in Fig.2. In this picture the
worldsheet disk instantons of type IIA get mapped to $(p,q)$
Euclidean worldsheet instantons, wrapping the 5-th circle and
ending on the 5-branes. In particular if we follow the map of the
Euclidean instanton to this geometry it is the other disk
realization of $S^3$ (see Fig.7) .\foot{ Note that a D6 branes
wrapped around $S^3$ is realized in type IIB as an $NS$ 5-brane in
the $x$ direction, a $D5$ brane in the $y$ direction separated in
the $z$ direction, and where the $5$-th circle vanishes along the
interval in the $z$ direction joining the two branes.}

\bigskip
\centerline{\epsfxsize 5truein\epsfbox{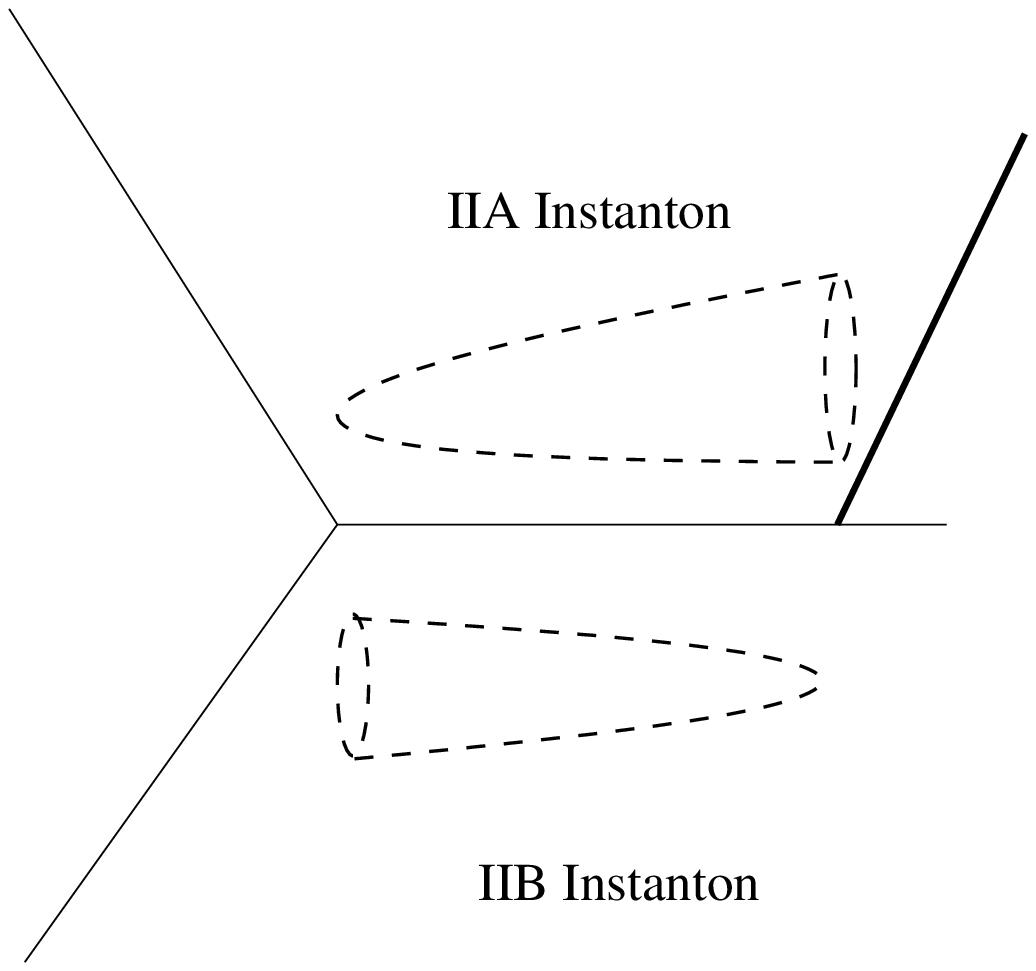}}
\rightskip 2pc \noindent{\ninepoint\sl \baselineskip=8pt {\bf
Fig.7}{ The M2 brane instanton with a topology of $S^3$ wrapping
over a 3-cycle of a local $G_2$ manifold gets mapped to two
alternative disk projections of $S^3$, depending on which duality
we use. In one case we get the description involving a type IIA
string theory on a CY with D-brane wrapped over Lagrangian
submanifold and in the other we get a web of IIB 5-branes in the
presence of ALF like geometries.}}
\bigskip

\newsec{Choice of Flat Coordinates}

In this section we consider the map between the moduli of the
brane between the A- and B-model.

As discussed in section 2, the moduli of the D-branes in the
A-model, are labeled by $c$ which measures the size of the disk
instanton ending on the Lagrangian submanifold.  In the quantum
theory $c$ gets complexified by the choice of the Wilson line on
the brane, and get mapped to the choice of a complex point on the
mirror type IIB geometry. The choice is characterized by the
choice of a point on a Riemann surface $F(u,v)=0$, which we choose
to be our `$u$' variable.  However it could be that the `size' of
the disk instanton, receives quantum corrections, and this, as we
will now discuss is relevant for finding the natural (``flat'')
coordinates parameterizing the moduli space of Lagrangian
D-branes.

First we have to discuss what we mean by the ``natural'' choice of
coordinate for the A-model. This is motivated by the integrality
structure of the A-model expansion parameter. There is a special
choice of  coordinates \ov \ on the moduli space of D-branes in
terms of which $A$-model disk partition function has integer
expansion \intg , and this is the coordinate which measures the
tension of the $D4$ brane domain walls. There is no reason to
expect this to agree with the classical size of the disk that the
B-model coordinate measures, and in general the two are not the
same, as we will discuss below. This is what we take as the
natural coordinates on the A-model side.

The B-model and the A-model are equivalent theories,
and this dictates the corresponding flat coordinate on the
$B$ model moduli space.  This is the tension of
the domain-wall D-brane which is mirror brane to the D4 brane
of the A-model.

The D4 brane wrapping a minimal disk $D$ is magnetically
 charged under the gauge field
on the D6 brane. Consider the domain wall which in the $R^{3,1}$ is at
a point in $x_3$ and fills the rest
of the spacetime. The Bianchi identity
for the gauge field-strength $F$ on $L$, modified by the presence of the
D4 brane, says that if the $B$ is the cycle Poincare dual to
the boundary of the disk $\partial D \subset L$.  Recall that
our brane $L$ has the topology of ${\bf C}\times S^1$, so $B$
can be identified with ${\bf C}$.  Then
 the charge $n$ of the domain wall is measured by
$$2 \pi n =\int_{B}F(x_3=\infty)-F(x_3=-\infty)
= \int_{\partial B}A(x_3=\infty)-A(x_3=-\infty)$$
Recall that ${\rm Im}u$ and ${\rm Im}v$ map to the one forms
related to the $S^1\times S^1$ cycles of the Lagrangian geometry,
viewed as a cone over $T^2$. Thus ${\rm Im}(u)$ is the mirror of
the Wilson-line $\int_{\partial D} A$, and the Wilson-line
around the dual $S^1=\partial B$ is identified with ${\rm Im}(v)$.
Thus we find that the $v$ jumps over the mirror domain wall by
$$ v \rightarrow v + 2\pi i n.$$
The case of $n=1$ is
depicted in Fig. 8.
%
\bigskip
\centerline{\epsfxsize 7truein\epsfbox{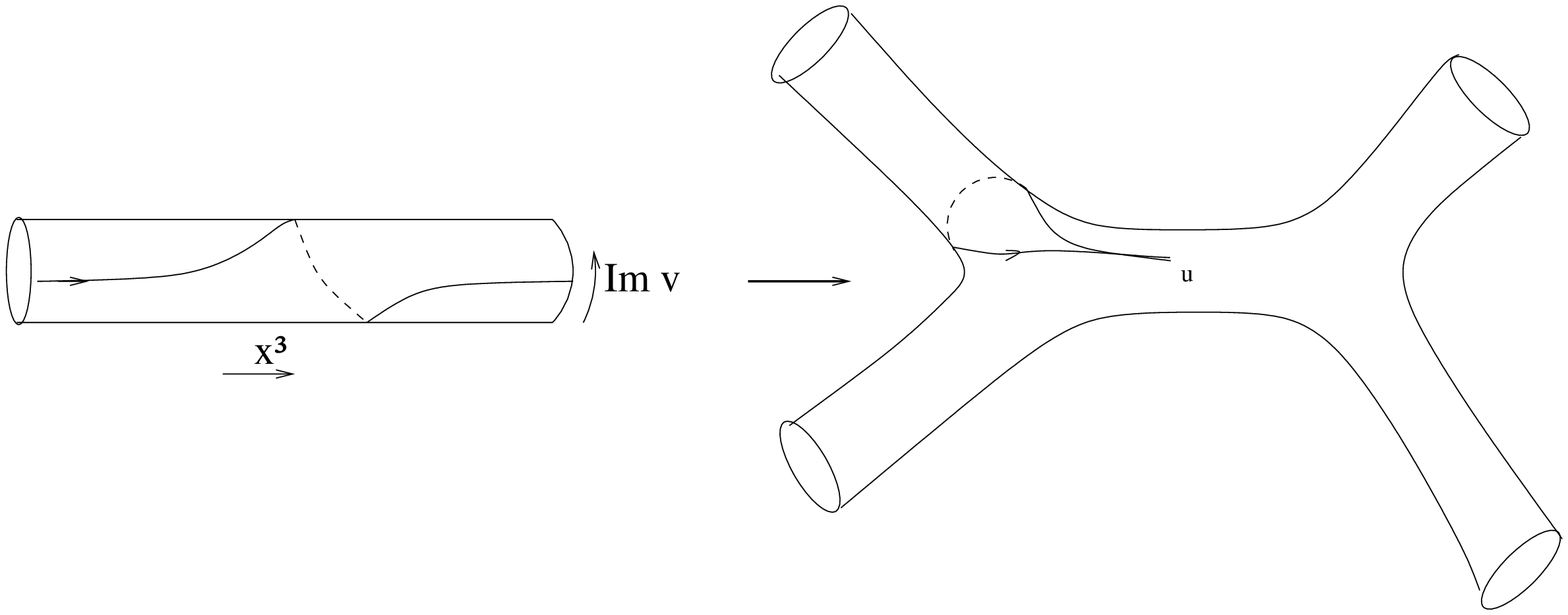}}
\rightskip 2pc \noindent{\ninepoint\sl \baselineskip=8pt {\bf
Fig.8}{ The D4 brane ending on the D6 brane is mirrored to a
domain wall in type IIB where $v$ shifts by $2\pi i$ across it.
This projects to a closed cycle on the Riemann surface
$P(u,v)=0$.}}
\bigskip
This allows us to find the tension of the
mirror domain wall as discussed in \av .
The BPS tension of the domain wall is given
by $\Delta W$ the difference of the superpotentials
on the two sides of the domain wall. Since $W =  \frac{1}{2\pi
i}\int^{u}_{u_*}v du$, the tension of the BPS domain wall is simply
the integral $\frac{1}{2\pi i}\int_{{C}_u} vdu$ where $C_u$ denotes the
appropriate cycle shifting $v\rightarrow v+2\pi i$, beginning
and ending on a given $u$.  We thus define the flat coordinate
\eqn\flate{{\hat u}(u,{\vec t})=\frac{1}{2\pi i}\int_{C_u} vdu}

To summarize, we predict that the disk partition function
\suppo , expanded in terms of $\hat u = \hat u (u,{\vec t})$ and the
corresponding closed string counterpart $\hat t(t)$--
has integral expansion \intg \ the coefficients of which count the
``net number'' of $D4$ brane domain walls ending on the Lagrangian
submanifold $L$ (for a more precise definition see \ov \ref\lmv{
J.~M.~Labastida, M.~Marino and C.~Vafa,
 ``Knots, links and branes at large N,'' JHEP {\bf 0011},
 007 (2000), hep-th/0010102.}).

It is not hard to see that $u$ and $\hat{u}$ as defined above
agree at the classical level and differ by instanton generated
corrections. In the large radius limit, the local A-model geometry
in the neighborhood of the disk $D$ is just $\bf{C^3}$ -- all
toric vertices other than the one supporting $D$ go away to
infinity. In this limit, the equation of the mirror simply becomes
$\;P(u,v)\;\; \rightarrow \;\;e ^u +e^v +1\;$ so
$$\hat u = \frac{1}{2\pi i}\int_{C_u}v(u) du \rightarrow \frac{1}{2\pi i}
\int_{C_u} \;
[\log( 1 + e^u ) + i\pi ]\;\;du$$ This has
a branch point in the $u-$plane around which $v$ has monodromy $v
\rightarrow v +2\pi i$. The contour $C_u$
receives contribution only from difference of values of $v$ on the
two sides of the cut, and thus for a single domain-wall $\hat u = u+ i
\pi$.

Away from the classical limit we can have subleading corrections
to the above relation that can in principle be subleading in
$e^{-t}$ and in $e^{-u}$.  But we will now argue that it is of the form
$$ \hat u = u + {\rm const.}+O(e^{-t}).$$
In other words, we show that
 $\delta \hat u /\delta u = 1$ is exact even away from the classical
 limit. This in fact is obvious from the definition of \flate\ because
 as we change $u$, the change in \flate\ can be computed
 from the beginning and the end of the 
 path.  But the integrands are the same except
 for the shift of $v$ by $2\pi i$, and therefore the difference is given
 by $\delta {\hat u} =\delta \int du =\delta u$.  We have thus shown that
$\hat u$ differs from $u$ by closed string instanton corrections only.
Another way to see this is to note that
$${\hat u}={1\over 2\pi i}\int_{C_u} vdu=
{1\over 2\pi i}\Delta (uv)-{1\over 2\pi i}\int_{C_u}
udv$$
Noting that $\Delta (uv)=2\pi i u$ due to the shift in $v$, and using
the fact that $u$ is not shifting and that $dv$ is well defined,
we deduce that the $-{1\over 2\pi i}\int_{C_u} udv=\Delta$ is independent
of $u$, by the deformation of the contour $C_u$, and only depends
on the class of the contour $C$.  It thus depends only on the
bulk moduli.

A cautionary remark is in order.  We have talked about cycle $C_u$
on the B-model side as a cycle on the Riemann surface $F(u,v)=0$.
In general the cycles on the Riemann surface $F(u,v)=0$ can be divided
to those that lift to cycles where $u$ and $v$ come back to the original
values, or those that shift by an integer multiples of $2\pi i$.
The cycles that come back to themselves without any shifts in
$u$ and $v$
correspond to closed 3-cycles
in the underlying
CY. Integration of $vdu$ over those cycles
correspond to computation of electric and magnetic BPS masses
for the underlying $N=2$ theory in 4 dimensions
(and are relevant for the computation of the
``flat'' coordinate for the bulk field $\hat t(t)$).  However, the cycles
whose $u$ or $v$ values
shift by an integer multiple of $2\pi i$ do not give rise to closed
3-cycles in the CY (as the CY in question does have non-trivial cycles
corresponding to shifting $u$ or $v$ by integer multiples of $2\pi i$).
Nevertheless as we discussed above such cycles are important
for finding the natural coordinates in the context of D-branes.

Note that closed string periods which determine $\hat t$ can also be expressed
in terms of linear combinations of periods where $u$'s and $v$'s
shift.  Thus computing periods where $u$
and $v$ shift are the fundamental quantities to compute.
We will discuss these in the context of examples in section 6.

Just as we have defined ${\hat u}$ as the quantum corrected
tension of domain wall, we can define ${\hat v}$ at the quantum
corrected tension of the domain wall associated with shifting
$u\rightarrow u+2\pi i$.  Note that in the derivation of the superpotential
\av\
$u,v$ are conjugate fields
of the holomorphic Chern-Simons field.  Thus replacing $u\rightarrow {\hat u}$
will require\foot{To see this from the target space viewpoint
it is natural to consider
the 1+1 realization of this theory as D4 brane wrapped over
the Lagrangian submanifold.  Then, as discussed in \ov\
the disk amplitude computes $S=\int d^2xd^2\theta (dW/d{\Sigma}) \Sigma$
for the $U(1)$ gauge theory in $1+1$ dimension where
$\Sigma $ is the twisted chiral gauge field strength multiplet, whose
bottom component is $\hat u$.
In this formulation the domain wall associated with shifting
of $\Sigma \rightarrow \Sigma +2\pi i$  is realized by $u\rightarrow u+2\pi i$,
whose BPS mass we have denoted by $\hat v$. From $S$, the change
in the value of the superpotential under shifting $\Sigma $ is given
by $dW/d{\hat u}$ which leads to the statement
that $dW/d{\hat u}={\hat v}$. This provides
an alternative, and more physical derivation of the main formula
we use for computation of $W$.}
 replacing $v$ by the quantum corrected conjugate field
${\hat v}$ and so the equation satisfied by the superpotential changes
to
$${\partial W\over \partial u}=v\rightarrow {\partial W\over \partial
{\hat u}}=
{\hat v}$$
which is the equation we will use in section 6 to compute $W$.

\newsec{Quantum Ambiguity for Lagrangian Submanifold}

We have seen that the choice of flat coordinates naturally adapted
to the A-model Lagrangian D-branes are fixed up to an integer
choice.  In particular we found that if $u,v$ are complex
coordinates satisfying $P(u,v)=0$ and if the brane is denoted in
the classical limit by $v=0$ and $u$ classically measures the size
of the disk instanton, then we can consider a new $u$ given by
$$u\rightarrow u+nv$$
for any $n$, which classically still corresponds to the disk
instanton action. In this section we explain why fixing the arbitrary
choice is indeed needed for a {\it quantum} definition of the
A-model Lagrangian D-brane.  In particular specifying the A-model
Lagrangian D-brane just by specifying it as a classical subspace
of the CY {\it does not uniquely fix the quantum theory}, given by
string perturbation theory. The choice of $n$ reflects choices to
be made in the quantum theory, which has no classical counterpart.
In this section we show how this works in two different ways:
First we map this ambiguity to an UV Chern-Simons ambiguity related to
framing of the Wilson Loop observables.  Secondly we relate it to
the choice of the Calabi-Yau geometry at infinity, and for this we
use the type IIB 5-brane web dual, discussed in section 3.

\subsec{Framing Choices for the Knot}

To see how this works it is simplest to consider the case where
the D-brane topological amplitudes were computed using the
observables of Chern-Simons theory \ov \ref\labm{J.~M.~Labastida
and M.~Marino, ``Polynomial invariants for torus knots and
topological strings,'' Commun.\ Math.\ Phys.\  {\bf 217}, 423
(2001), hep-th/0004196.}\lmv .
  These were obtained
by considering expectation values for Wilson loop observables in
the large $N$ Chern-Simons theory, in the context of the large $N$
duality of Chern-Simons/closed topological strings proposed in
\ref\gopv{R.~Gopakumar and C.~Vafa,
``M-theory and topological strings I and II,''
hep-th/9809187 and hep-th/9812127.}. Let us briefly recall this setup.

Consider $SU(N)$ Chern-Simons theory on $S^3$.  As was shown in
\ref\witcss{E. Witten, ``Chern-Simons gauge theory as a string theory,''
hep-th/9207094.}\ if we consider
topological A-model on the conifold, which has the same symplectic
structure as $T^*S^3$, and consider wrapping $N$ D3 branes on
$S^3$, the open string field theory living on the D3 brane is
$SU(N)$ Chern-Simons theory where the level of Chern-Simons theory
(up to a shift by $N$, i.e., $g_s=2\pi i/(k+N)$)
 is identified with the inverse of string coupling constant.  The large
$N$
 duality proposed in \gopv\ states that this topological string theory
is equivalent to topological strings propagating on the
non-compact CY 3-fold $O(-1)\oplus O(-1)\rightarrow {\bf P}^1$, which is
the resolution of the conifold, where the complexified K\"ahler
class on ${\bf P}^1$ has size $t=N g_s$.  In \ov\ it was shown how
to use this duality to compute Wilson loop observables. The idea
is that for every knot $\gamma \subset S^3$ one considers a
non-compact Lagrangian submanifold $ L_\gamma \subset T^* S^3$
such that $L_\gamma \cap S^3=\gamma$.  We wrap $M$ D3 branes over
$L_\gamma$ which gives rise to an $SU(M)$ Chern-Simons gauge
theory on $L_\gamma$.  In addition bi-fundamental fields on
$\gamma$ transforming as $(N,{\overline M})$, arise from open
strings with one end on the
 D-branes wrapped over $S^3 $ and with the other end on
D-branes wrapped over $L_{\gamma}$.
  Integrating out these fields give rise to the insertion of
$$\langle {\rm exp}(\sum_n {\tr U^n \tr V^n\over n}) \rangle$$
where $U$ and $V$ denote the holonomies of the $SU(N)$ and $SU(M)$
gauge groups around $\gamma$ respectively. Considering the $SU(M)$
gauge theory as a spectator we can compute the correlations of the
$SU(N)$ Chern-Simons theory and obtain
\eqn\larn{\langle {\rm exp}(\sum_n {\tr U^n \tr V^n\over n} )
\rangle= {\rm exp}(-F(V,t,g_s))}
It was shown in \ov\ that the right-hand side can be interpreted
as the topological string amplitude
in the large $N$ gravitational dual, where the $N$ D-branes have
disappeared and replaced by $S^2$.  In this dual
geometry the $M$ non-compact
D-branes are left-over and wrapped over some
Lagrangian submanifold in $O(-1)\oplus O(-1)\rightarrow {\bf P}^1$. This
Lagrangian submanifold was constructed for the case of the unknot
explicitly in \ov\ and extended to algebraic knots in \lmv\
 (this
latter construction has been recently generalized to all knots
\ref\ctau{C. Taubes, To appear.}).  Moreover
$F(V,t,g_s)$ denotes the topological string amplitude in the
presence of the $M$ D-branes wrapped over some particular
Lagrangian submanifold in $O(-1)\oplus O(-1)\rightarrow {\bf P}^1$ with
a non-trivial $S^1$ cycle. Note that a term in $F(V,t,g_s)$ of the form
$\prod_{i=1}^{b} \tr V^{k_i}$ comes from a worldsheet with $b$
boundaries where the $i$-th boundary of the worldsheet wraps the
$S^1$ of the Lagrangian submanifold $k_i$ times.

The left-hand side of \larn\ is computable by the
methods initiated in \witcso\
and in this way gives us a way to compute the open string
topological amplitudes for this class of D-branes.  Note in
particular that the disk amplitude corresponds to the $1/g_s$ term
in $F(V,t,g_s)$.  A particular case of the brane we have
considered in $O(-1)\oplus O(-1)\rightarrow {\bf P}^1$ corresponds
to the unknot.  This is depicted in the toric Fig. 2.

The match between the computation in this case, using mirror
symmetry and the result expected from Chern-Simons theory was
demonstrated in \av .    However as we have discussed here the
disk amplitudes have an integer ambiguity, when we use mirror
symmetry for their computation. Thus apparently the right hand
side of \larn\ is defined once we pick an integer, related to the
boundary conditions at infinity on the B-brane in
the type IIB mirror.  If the right hand side of \larn\ is
ambiguous, then so should the left hand. In fact
the computation of Wilson loop observables also has an
ambiguity given by an integer! In particular we have to choose a
framing on the knot $\gamma$ to make the computation well defined
in the quantum theory \witcso . A framing, is the choice of a normal
vector field on the knot $\gamma$, which is non-vanishing everywhere on
the knot.
 Note that if we are given a framing of the knot,
any other topologically distinct framing is parameterized by an integer,
given by the class of the map $S^1 \rightarrow  S^1$, where the
domain $S^1$ parameterizes $\gamma$ and the range denotes the relative
choice of the framing which is classified by the direction
of the vector field on the normal plane to the direction along
the knot.  The
framing of the knot enters the gauge theory computation by
resolving UV divergencies of the Chern-Simons theory in the
presence of Wilson loops.  It arises when we take the Greens
function for the gauge field coming from the same point on the
knot. The framing of the knot allows a point splitting definition
of the Greens function.

We have thus seen that both sides of \larn\ have a quantum
ambiguity that can be resolved by a choice of an integer.  On the
left hand side the ambiguity arises from the UV.  On the right
hand side the ambiguity arises from the IR (i.e. boundary
conditions on the brane at infinity). We have
checked that the two ambiguities match for the case of the unknot,
by comparing the disk amplitudes on both sides (using CS
computation of the framing dependence of the knot on the left and
comparing it with the mirror symmetry computation of the knot on
the right). Some aspects of this computation is presented in the
appendix A.  The computation of the disk amplitude for this case
using mirror symmetry is presented in section 6.

Here let us discuss further how this match arises. Consider the
disk amplitude at large $N$ corresponding to a given knot.  In the
gauge theory side the computation arises from open string diagrams
of a planar diagram with the outer hole on the Lagrangian
submanifold $L_{\gamma}$, and rest of the holes ending on D-branes
wrapping the $S^3$, as shown in Fig.9.  In the large $N$ limit,
the interior holes get ``filled'' and we get the topology of the
disk.

\bigskip
\centerline{\epsfxsize 5truein\epsfbox{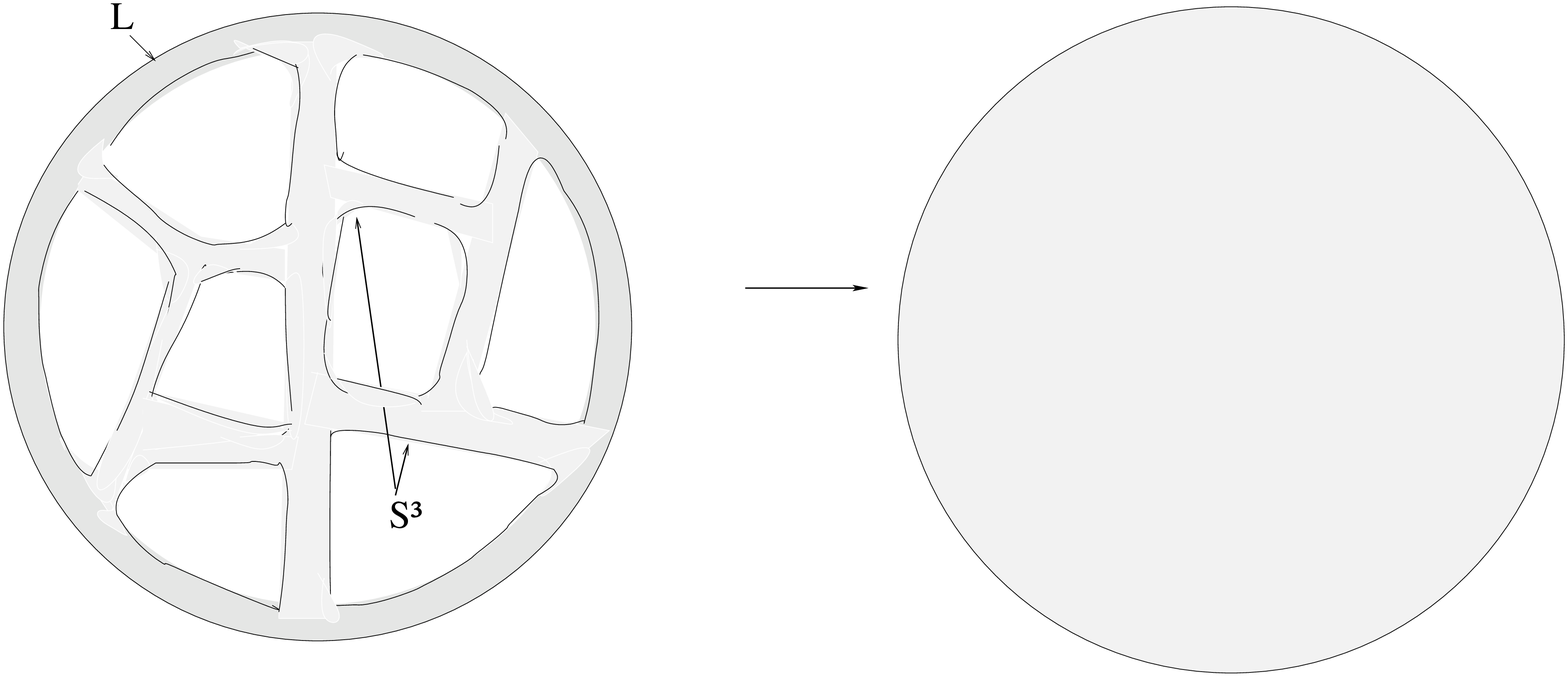}}
\rightskip 2pc
\noindent{\ninepoint\sl \baselineskip=8pt {\bf Fig.9}{
The Wilson loop observables arise from worldsheet diagrams where
some  boundaries end on $L_{\gamma}$ and $N$ branes wrapping $S^3$.
In the large $N$ limit the holes ending on $S^3$ get ``filled''
and we end up with a Riemann surface which has only
the boundaries associated with $L_\gamma$. In the above
figure the outer hole is the only one ending on $L_\gamma$.   All
of the interior holes end on $S^3$ and disappear in the large $N$
limit, leaving us with a disk.}}
\bigskip

Now consider where the UV divergencies of the gauge theory would
arise. They would arise from Feynman diagrams where the Schwinger
parameter for the gauge field goes to zero--an example of this is
depicted in Fig.10a.  Note that the two end points of the short
propagator will be mapped to the same point on the knot $\gamma$
in the limit of zero length propagator.  In the large $N$ limit,
where the disk gets filled these get mapped to configurations such
as that shown in Fig.10b.    In
this dual description by a conformal transformation the worldsheet
can be viewed as that depicted in Fig.10c.  In
other words we have in the dual channel a long schwinger time, of
an open string ending on $L_{\gamma}$.  This means that the issue of
ambiguity is mapped to an IR behaviour of fields living on $L_{\gamma}$,
and that is exactly where we found the ambiguity in the
computation of the superpotential in the context of mirror
symmetry.

\bigskip
\centerline{\epsfxsize 5truein\epsfbox{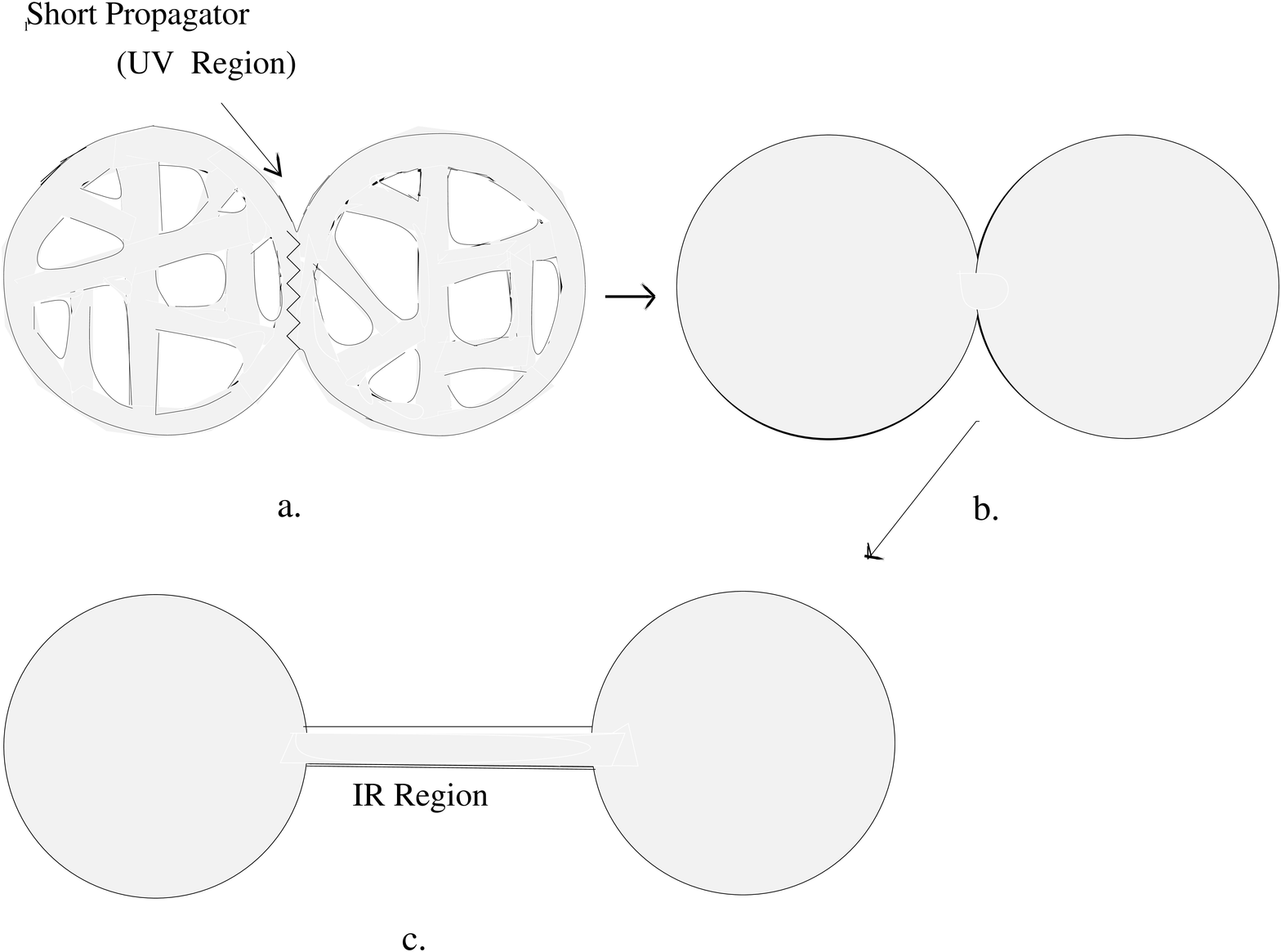}}
\rightskip 2pc \noindent{\ninepoint\sl \baselineskip=8pt {\bf
Fig.10}{ The Wilson loop observable has UV divergencies which need
to be regulated, coming from points along the knot where the gauge
boson propagator is of zero size (a).  In the
large $N$ limit these map to disks touching at a point (b), which can
be viewed via a conformal transformation as a long propagator (c).
Note that the boundaries of the long propagator are on the
Lagrangian submanifold $L_{\gamma}$ (or more precisely its large
$N$ dual) and correspond to open string propagating on it. Thus
 the UV framing choice of CS
gets mapped to the choice of large distance (IR) physics of modes
on the brane.}
}
\bigskip

Other examples of the large $N$ limit
of UV regions for the computation of the Wilson
loop observable along the knot get mapped to disks shown in Fig.11,
and look like branched trees of disks.

\bigskip
\centerline{\epsfxsize 5truein\epsfbox{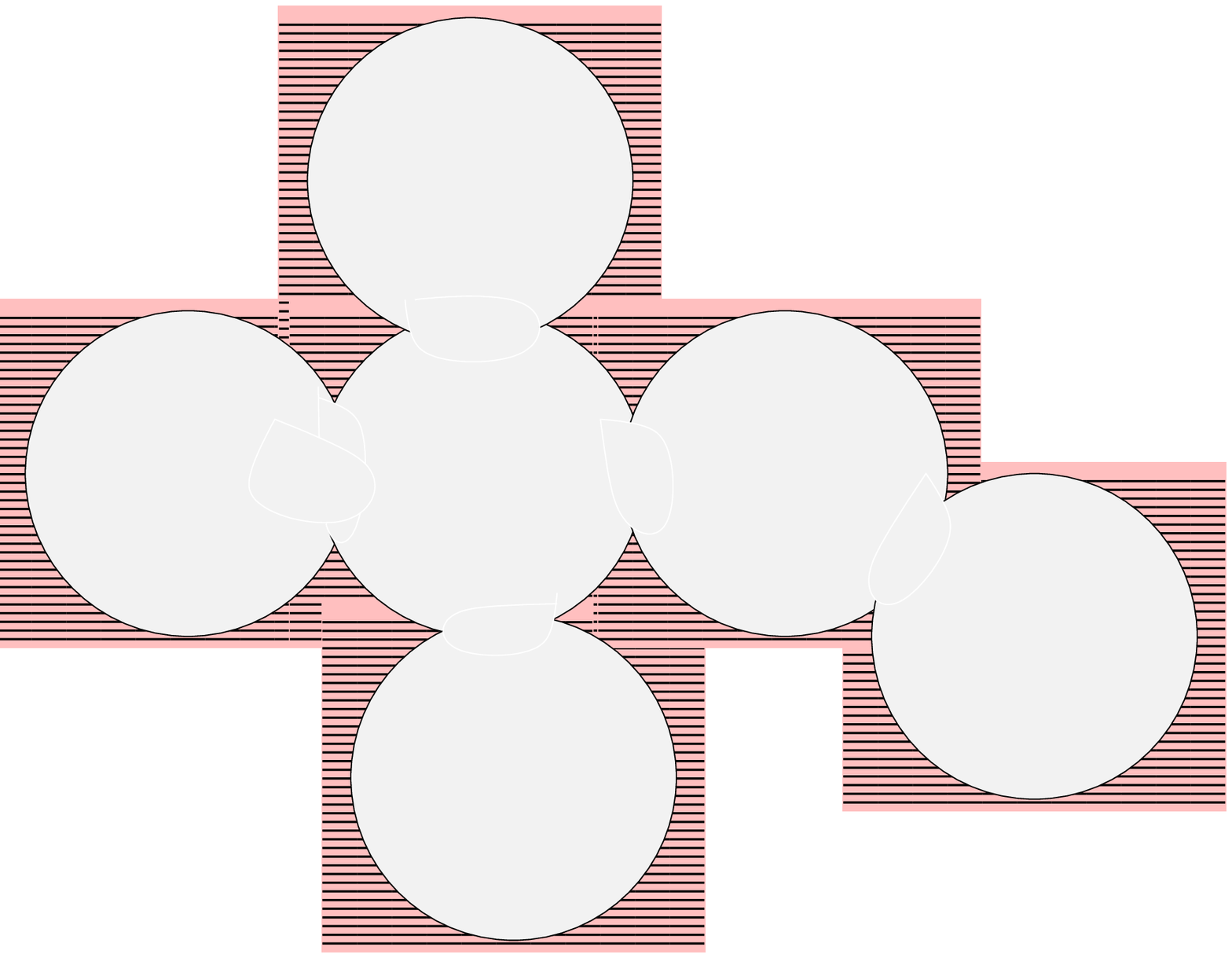}}
\rightskip 2pc
\noindent{\ninepoint\sl \baselineskip=8pt {\bf Fig.11}{
Other examples of large $N$ limit of UV divergencies
of Wilson loop observables.}}
\bigskip

\subsec{Calabi-Yau Geometry, 5-brane Perspective and the Integral Ambiguity}

As discussed in section 3 we have dual descriptions of type IIA
geometry with D6 branes wrapped over Lagrangian submanifolds in
terms of M-theory on $G_2$ holonomy manifolds (viewed as circle
fibration over CY manifolds) or in terms of type IIB web of 5
branes in an ALF-like background in $R^6$. To be precise, the
M-theory on $T^2$/type IIB on $S^1$ duality relates the coupling
constant of type IIB to the complex structure of the $T^2$.
Fibering this duality gives rise to the duality just mentioned.
However, in the type IIB picture we typically fix the type IIB
coupling at infinity. This means that, by this chain of duality,
the $G_2$ holonomy manifold is a circle fibration over the CY
where the complex structure of the $T^2$ fibration of CY is fixed
at infinity.  Turning this around,
 this duality predicts the existence of particular class of CY
 and $G_2$ holonomy metrics with particular behaviour of the metric
at infinity (we can also fix the area of $T^2$ at infinity as that
gets mapped to the inverse of the radius of type IIB $S^1$).  We
now argue that the choice of the complex structure of $T^2$, or
equivalently type IIB coupling constant at infinity {\it changes}
the number of Euclidean M2 branes, exactly as is expected by the
ambiguity.  Instead of being general we simply illustrate this
idea in the context of a simple example, namely the ${\bf C}^3$
geometry discussed before and represented by Fig.6, consisting of
NS 5-brane $(0,1)$ and D5-brane $(1,0)$ and the $(1,1)$ 5-brane.
We already discussed that M2 brane instantons get mapped to
worldsheet disk instantons consisting of the $(p,q)$ string shown
in Fig.7 times an extra circle (the `5-th' circle) which vanishes
at the position of the ALF space. However now consider changing
the coupling constant.  Then the angles between the 5-branes will
be changed. Beyond some critical angles we {\it can} get new
worldsheet instantons.  Basically we can consider the web of
strings and the only data that we need to take into account is
that a BPS $(p,q)$ string can end only on a $(p,q)$ 5-brane (and
perpendicularly).  This web of strings however, can end on the
projection of the ALF space on the web at any angle since at that
point now the `5-th' circle is shrinking (again in a perpendicular
fashion). To be concrete let us consider the limiting choices of
the type IIB coupling constant given by $\tau \rightarrow -1/n$.
This can be obtained from $\tau \rightarrow i\infty$ by the
modular transformation ${-1\over \tau}\rightarrow {-1\over
\tau}+n$.  Said differently we can take the weak coupling limit
$\tau \rightarrow i\infty$ by considering the inverse modular
transformation where we consider the 5-brane web given by NS
fivebrane $(0,1)$ and the fivebranes $(1,n)$ and $(1,n+1)$.  It is
easy to see that this geometry of webs has an $n\rightarrow
-(n+1)$ symmetry, when we reverse the direction of the handedness
on the D5 brane.  Thus, the counting of the worldsheet instantons
here will exhibit this symmetry. This is exactly the symmetry we
will find for the ambiguity of the mirror to ${\bf C}^3$, as
discussed in section 6.  Moreover it is easy to see that except
for $n=0,-1$ where there is exactly one choice of disk instanton,
for all other $n$'s we get a large number of disk instantons,
allowed by the geometry of the 5-branes, as shown in Fig.12. This
is also in line with the result we find from mirror symmetry
in section 6.
Thus as the geometry of Calabi-Yau changes, we get jumps in the
number of instantons as predicted by this picture.

\bigskip
\centerline{\epsfxsize 6truein\epsfbox{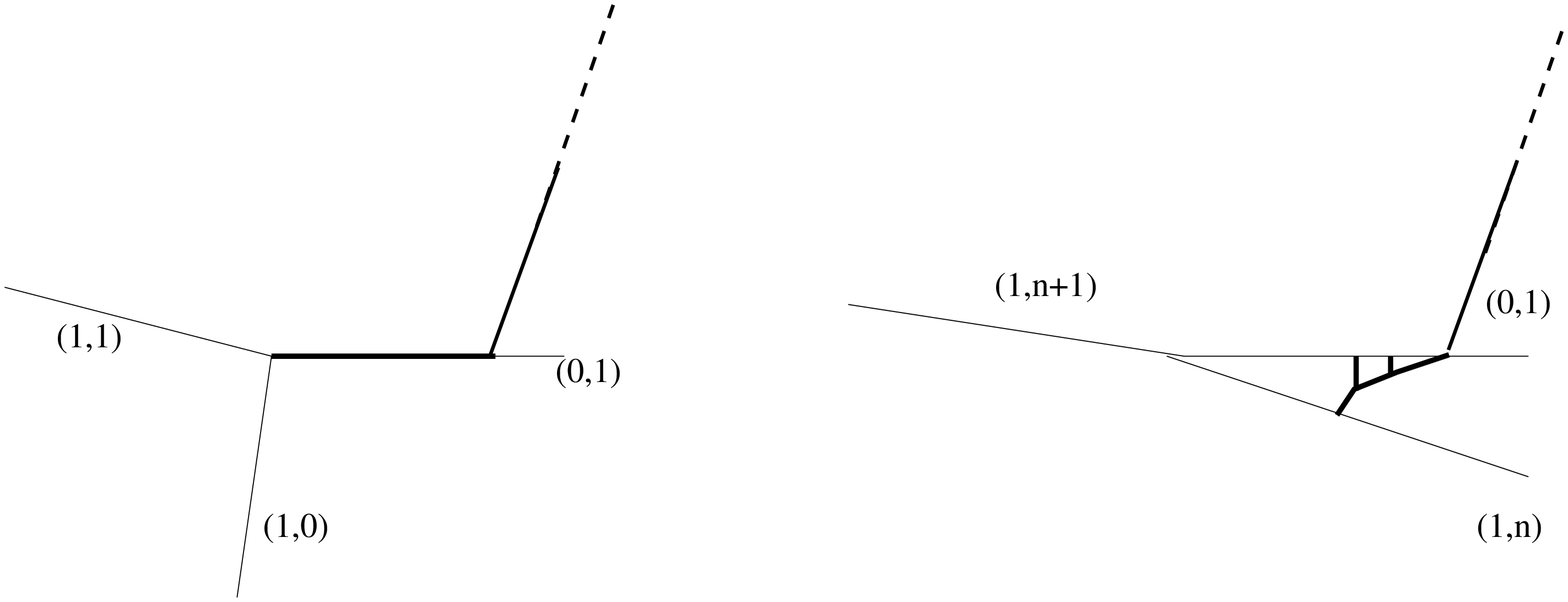}}
\rightskip 2pc
\noindent{\ninepoint\sl \baselineskip=8pt {\bf Fig. 12}{ In the
weak coupling limit, we consider an NS $(0,1)$ 5-brane with
an ALF like space ending on it.  If this is connected to
$(1,0)$ and $(1,1)$ 5-branes in a junction, the only
instantons allowed involve the worldsheet instanton stretched
along the interval shown on the left-hand side of the figure above,
times the 5-th circle which shrinks at the attachment point of the
ALF space.  There is no other instanton allowed in this geometry.
However if NS $(0,1)$ 5-brane is attached to $(1,n)$ and $(1,n+1)$
5-branes the geometry of the intersection dramatically changes (as long
as $n\not=0,-1$) and now we can have many more allowed configurations for
the worldsheet $(p,q)$ instantons. This is depicted on the right hand
side of the above figure.}}
\bigskip

It is natural to ask what IR aspect of the type IIA CY geometry
with the brane, this choice of $\tau$ is reflected in.  One's
natural guess is that the normalizability of the 1-form on the
brane represented by the Wilson line vev, which represents the
mode corresponding to moving the brane, is the relevant issue.  If
this is the case the possible choices of metrics for which a
particular 1-form is normalizable will have to be classified by
the integer ambiguity $n$ which in turn is related to the choice
of the CY metric with fixed $\tau$ at infinity, labeled by $n$.
Similarly in the M-theory lift this would be related to the
possible choices of normalizable deformations of the $G_2$
holonomy metric.  It would be interesting to study these issues
further.

\newsec{Examples}

In this section we will consider a number of examples which
illustrate the general discussion we have presented for the
computation of disk amplitudes for D-branes wrapping Lagrangian
submanifold of non-compact CY 3-folds. The examples include
considerations of Lagrangian branes in ${\bf C}^3$, ${\bf
P}^1\times {\bf P}^1$, ${\bf P}^2$ and its blow up at a point
${\bf F}_1$.  In these examples we consider inequivalent
configurations of Lagrangian branes.  For the case of ${\bf C}^3 $
we exhibit the dependence of the topological string computations
on the integer ambiguity we discussed earlier. For all of these
cases we find the predicted integrality properties of the disk
amplitudes.\foot{The integrality requirement is highly non-trivial.
Originally, we experimentally found the highly non-trivial
flat coordinates only
by requiring the integrality of the domain wall degeneracies.
Later we showed that they agree with
the BPS tension of domain walls, which is the definition
used in this section.}
  We also consider different Lagrangian
branes (for example ending on the ``internal'' or ``external''
edges of the toric diagrams).

As discussed in section 4, a non-trivial part of the story
involves finding the flat coordinates for the open string variables.
As discussed there we find the flat coordinates to be given by
$${\hat u}=u+ \Delta(t_i)$$
where $\Delta (t_i)$ (up to an addition of a constant) is  an exponentially
suppressed function of K\"ahler moduli of the Calabi-Yau
($O(e^{-t_i})$), and $\Delta$ depends on the choice of the brane.
We compute $\Delta$ using the methods discussed in section 4 for all
the examples.

Note that using $\partial_{\hat u} W={\hat v}(\hat u)$ and \intg\
we have
$$
\partial_{\hat u} W={\hat v}(\hat u) =-\sum_k k N_{k,{\vec m}}
{\rm log}(1-{\rm exp}[{k{\hat u}-{\vec m}\cdot {\vec {\hat t}}}])$$
and so by solving for $\hat v$ in terms of $\hat u$
(from $P(u,v)=0$ and the correction to $u, v \rightarrow \hat u, \hat v$ 
due to mirror map)
we find a prediction of an expansion
in terms of integers $N_{k,{\vec m}}$.  We verify
these  highly non-trivial
integrality predictions in the examples below which we now
turn to.

\subsec{Almost $\bf{C}^3$}

We consider the simplest Calabi-Yau $X=\bf{C^3}$.
It is described by just three chiral fields $X^i$ and no gauge group.
The special Lagrangian D-brane $L$ we are interested in is given by
$q_1=(1,0,-1)$ and $q_2=(0,1,-1)$:
$$|X^1|^2-|X^3|^2 =c_1 \quad |X^2|^2 -|X^3|^2 = 0$$
This geometry can be regarded as a local approximation, in the
limit all radii are large, to a more involved geometry it is
embedded in, for example the one discussed in section 2. The
value of $c_1$ on the brane and the Wilson-line around the one
finite circle on $L$ form
one complex modulus $u$ of the A-brane, which measures the
classical BPS tension of the D4 brane domain wall.

The mirror manifold can be written as  \eqn\mirrC{xz = P(u,v)
=e^u+e^v+1} where we have set $Y^3$ to zero,
and $Y^1 =u$, $Y^2=v$.

Equations \dual \ fix the classical limit of the brane
$$ c^1 ={{\rm Re}}(u)\;\;$$
so the transverse coordinate to the B-brane is
$u$. In this limit, $v$ is zero on the brane corresponding to
the vanishing of the classical superpotential.

The $\SL2z$ group of reparametrizations acts on \mirrC\ by
linear transformations $\left(\matrix{\hat u'\cr \hat
v'}\right)=\left(\matrix{a&b\cr c& d}\right) \left(\matrix{\hat
u\cr \hat v}\right)$ that leave the holomorphic $(3,0)$ form
$\Omega={\dd z \over z} \dd u \dd v$ invariant. The subgroup of
$\SL2z$  that preserves the classical limit above
consists of transformations  $\left(\matrix{a&b\cr c&
d}\right)=\left( \matrix{1&p\cr 0 &  1}\right)$.
These transformations result in a family of parameterizations
of the mirror geometry
$$xz= P_p(u,v)=e^{u+pv} +e^v+1,$$
and a family of superpotentials obtained by solving the order
$p$ polynomial in $e^v$,
$$W_p (u) = \int v_p(u) du.$$
To compute the numbers of disk domain walls we need to know the flat
coordinates.

For $p=0$ we have computed in section 4 the BPS tension of the
D-brane in $u$-phase and also the  $v-$phase, by trivial relabeling, to be
$\hat u_{p=0} = u+i\pi$, $\hat v_{p=0}= v+i\pi.$ The knowledge of these
suffices to find the BPS tensions for a different choice of framing $p$.

The $\SL2z$ transformations $u\rightarrow u+p v$, $v \rightarrow v$
act on the basis of one-cycles on the
Riemann surface that are associated to periodicity of $Im(u)$ and $Im(v)$.
Since $\Delta_u$ and $\Delta_v$
which provide the quantum corrections to $u$ and $v$,
$$\; \hat u = u + \Delta_u ,\quad \hat v = v + \Delta_v$$
are obtained by evaluating periods of the one-form $\lambda = vdu$
over these, changing the basis of one-cycles by an $\SL2z$ transformation to
$u' = au+bv$ and $v'=cu+dv$ acts in an analogous way on the
periods i.e.
$$\Delta_u' = a \Delta_u + b\Delta_v .$$
Thus, we find that
$$\hat u_p= u+ (p+1) i\pi , \quad \hat v_p = v+i\pi.$$
%
The equation \eqn\facIII{P_p(\hat u, \hat v) =0 =1-e^{\hat u + p
\hat v} -e^{\hat v}} is solved iteratively using the ansatz
$e^{\hat v}=\sum_{k=0}^\infty a_k e^{k \hat u}$. Using
$(\sum_{k=0}^{\infty} a_k x^k)^p=\sum_{m=0}^{\infty} c_m x^m,$
where $a_0=c_0=1$ and $c_m={1\over m} \sum_{k=1}^m (k p - m +k)a_k
c_{m-k}$ we get immediately a recursive formula $a_m={1\over m-1}
\sum_{m=1}^{m-1} (k p-m+k+1) a_k a_{m-k}$ for the coefficients.
This can be summed using Stirlings coefficients of the first kind
(see e.g. \ref\grad{I.~S.~Gradshteyn and
 I.~M.~Ryzhik, ``Table of Integrals, Series
and Products,'' Academic Press, Boston (1994).}), which are defined by
$x (x-1) (x-2)\cdot \ldots \cdot (x-n+1)=\sum_{m=0}^n S^{(m)}_n x^m$.
Using the
relation $\left(\matrix{m\cr r}\right) S^{(m)}_n=\sum_{k=m-r}^{n-r}
 \left(\matrix{n\cr k}\right)
S^{(r)}_{n-k} S_k^{(m-r)}$ the result of the summation is
\eqn\summation{a_m={(-1)^m\over m!}\prod_{j=0}^{m-2}(mp-j)\ . }
Solving now for $\hat v=\partial_{\hat u} W$ and integrating
we get, up to trivial integration constants, the superpotential
\eqn\supocIII{W=\sum_{m=1}^\infty {1\over m m!} \prod_{j=1}^{m-1} (m p -j)
e^{m u}\ . }
We can write this in the general form \intg .  In the case at hand
that is given by $W=\sum_{m,k}
\frac{N_m}{k^2}e^{k m \hat u}$  which yields the following
integers for $N_m$:
$$\eqalign{
N_1&=(-1)^p\cr N_2&=-{p\over 2}+{1-(-1)^p\over 4}=-\left[p\over
2\right]\cr N_3&={(-1)^p\over 2}p (p-1)\cr N_4&=-{1\over 3} (2
p-1)(p-1)p \cr N_5&={5\over 24} ( -1)^p p (p-1) (5 p^2-5 p+2)\cr
N_6&={1\over 20} (p-1) p (36 p^3-54 p^2+31 p -9)\cr
N_7&={7\over 720} ( -1)^p p (p-1) (343p^4-686 p^3 +539 p^2-1
96p+36)\cr \vdots &}$$
To show integrality for a  given
expression $N_m(p)$ for all $p\in \ZZ$, one may factorize the
denominator into $\prod_i p_i^{n_i}$ with $p_i$ prime and check
that one can factor $p_i^{n_i}$ from the numerator for all
$p=p_i^{n_i} n-k_i$ with $k_i=1,\ldots, p_i^{n_i}$. That has been
checked for $m\le 50$.

%

Note that $N_m$ is a polynomial in $p$ of degree $m-1$.
As discussed in section 5, the $p$ dependence of the superpotential
for this particular Lagrangian brane (viewed
as a large $t$ limit of the $O(-1)\oplus O(-1)\rightarrow {\bf P}^1$ geometry)
can be mapped to the framing ambiguity of the unknot for the
Chern-Simons theory.  In appendix A we discuss this computation
and verify that the large $N$ Chern-Simons computation
leads to a polynomial of degree $m-1$ in $p$ for $N_m$.  Moreover
we have verified that for $N_i$ for $i=1,2,3$ the $p$ dependence
of the framing choice agrees
with the above result.  We also have shown from the
computation of the framing dependence
of the CS theory that for all $m$ the
coefficient of the leading term $p^{m-1}$ is given by
$m^{m-2}/m!$ in agreement with the above result.

Results of the prediction of \ov\ for the topological
string amplitudes for the case of the unknot have recently
been verified mathematically using
localization methods in \ref\kliu{S. Katz and C.-C. M.\ Liu,
``Enumerative Geometry of
Stable Maps with Lagrangian Boundary Conditions and Multiple Covers
of the Disc,'' Math.AG/0103074.}\ref\wwr{J. Li and Y. S.
Song,``Open String Instantons
and Relative Stable Morphisms,'' hep-th/0103100.}.
In verifying the predictions of \ov\
there were some choices made for the toric action in these works.
The authors of \kliu\ have recently checked
and verified that the integer
choice of ambiguity changes the topological string
amplitudes exactly as predicted by the above results\foot{We are grateful
to S. Katz and C.-C. Liu for performing these computations upon our request.
The form we have presented our answer \supocIII\ was chosen
to simplify the comparison with their formula.}.
Moreover this ambiguity arises from the boundary of moduli
space of Riemann surfaces with holes, as we already
discussed.

\subsec{${\cal O}(K)\rightarrow \bf{P^1 \times P^1}$}

This manifold involves a compact ${\bf P}^1 \times {\bf P}^1$
geometry inside a CY 3-fold with a non-compact
direction given by the canonical line bundle.  It
 can be described by a linear sigma model
with $G=U(1)^2$ and five chiral fields $X^i$, for $i=0,\ldots 4$ with charges
$Q^1=(-2,1,1,0,0)$ and $Q^2=(-2,0,0,1,1)$
which leads to the $D$-terms
$$|X^1|^2+|X^2|^2-2
|X^0|^2=r_t \quad  \quad |X^3|^2+|X^4|^2-2 |X^0|^2=r_s$$
The solutions to the D-term equation, projected to the toric
base are given in Fig.13. The
${\bf F_0} = {\bf{P^1}}\times {\bf P^1}$ corresponds to the divisor
class represented by $X^0=0$ and is visible in Fig.13 as the minimal
parallelogram in the toric base. The fiber $O(K)$ corresponds to
the normal direction.
The special Lagrangian D-branes of the A-model for
this Calabi-Yau have inequivalent phases, depending on where one puts
the brane, together with a $\bf{Z}$-family of choices of framing in each.
The D-brane charge can be taken to be $q_1=(-1,0,1,0,0),
q_2=(-1,0,0,1,0)$ and so
$$|X^2|^2-|X^0|^2=c^1, \quad |X^3|^2-|X^0|^2=c^2$$
The constants $c_i$ are required to be chosen
so that the Lagrangian D-brane lies on one dimensional toric edges,
and different toric legs give rise to generally different phases of
the theory.

The mirror variables $Y^i$ satisfy
 $$Y^1+Y^2-2Y^0 =-t, \quad Y^3+Y^4 - 2 Y^0 =
- s $$ 
where the real parts of complex structure parameters $t,s$
measure the classical sizes of the two minimal $P^1$'s in the toric base,
${\rm Re}(t)=r_t$, ${\rm Re}(s)=r_s$.
The mirror manifold  is given by
$$xz = e^u + e^{-t-u} +e^v + e^{-s-v}+1$$
where $Y^2 = u$, $Y^3=v$, $Y^0 = 0$ and the two holomorphic
constrains are solved in terms of these. The Riemann surface $\Sigma$
which is the configuration space of the mirror B-brane is given
by $\;0=P(u,v)=e^u + e^{-t-u} +e^v + e^{-s-v}+1\;$.
When viewed in terms of
the single valued variables $e^u$ and $e^v$, $\Sigma$ is related to the
$A-$model geometry by ``thickening'' of the one-dimensional edges
of the toric diagram (or more precisely, their projection onto the
$X^0=0$ plane). In the large radius limit $0\ll t,s$ the A-model
geometry becomes classical and the legs of the
Riemann surface of the B-model become long and thin, so the A- and
the B-model can be related already at the classical level.
\bigskip
\centerline{\epsfxsize 7.truein\epsfbox{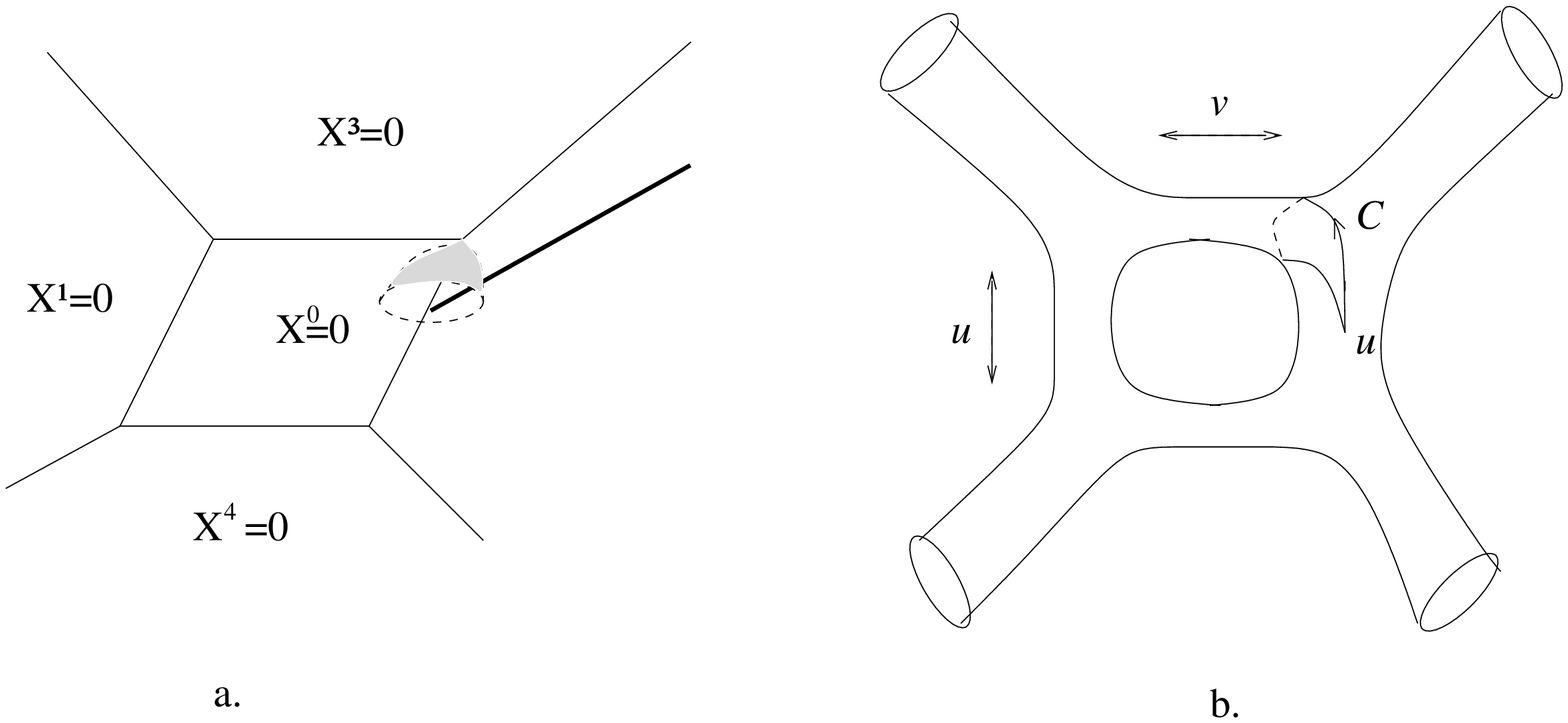}}
\rightskip 2pc \noindent{\ninepoint\sl \baselineskip=8pt {\bf
Fig.13}{ The Figure a. depicts the base of the toric fibration for
$X = O(K) \rightarrow {\bf P^1} \times {\bf P^1}$ together with a
special Lagrangian brane in $X$. The minimal disk which ends on
the A-brane in the figure wraps partly the ${\bf P^1}$ over which
the brane is. Figure b. depicts the Riemann surface $\Sigma$ of
the mirror geometry, and the mirror B-brane as a choice of a point
on $\Sigma$. D4 brane wrapping the disk in figure a. is mirror to
a domain-wall which projects to path ${\cal{C}}$ in Figure
b.} }
\bigskip
For example, the A-brane on Fig.13 is on the internal leg
which is given by $c_2 = 0$, and $c_1$ in the $[0,r_t]$ interval. The large
radius limit is, in addition to $r_t$ and $r_s$ being large, the
limit of large disk size. For example in the regime $c_1 < r_t/2$ the basic
disk is the one on the Fig.13a and the classical limit corresponds
to $c_1$ being of order of $r_t/2$. The size parameter $c_1$
together with the Wilson line $u = c_1+i\int A$ is the
classical tension of the D4 brane domain wall wrapping this disk
and ending on the D6 brane.

In the limit the A-model 
geometry is classical, it follows from the
mirror map \dual \ that the mirror B-brane is located 
in the region on $\Sigma$ where $v$ is constant, $v \sim 0 $, 
and $u$ a parameter which is large, $u \sim t/2$. Away
from the large radius limit, this deforms to
a root of the equation $P(u,v)=0$. This has two solutions
for $v$ at every value of $u$,
$$v=v_{1,2}(u) = \log[\frac{1+e^u+e^{-t-u}}{2} \pm
\frac{\sqrt{(1+e^u+e^{-t-u})^2 - 4e^{-s}}}{2}]+i \pi$$ and the
mirror B-brane propagates along a region of the root $v=v_1(u)$.
As we discussed above, the superpotential on this leg is given by
$$W(u) = \int v_1(u) du.$$
This vanishes in the classical limit since 
$\partial_u W =v_1(u)\rightarrow 0$.

The flat coordinate computes the tension $\hat u$ of the D4 brane
domain wall wrapped on the basic disk as in Fig.13a, and we
will compute this by computing the tension of the mirror B-brane
domain wall. As explained in section 4. the mirror domain wall
projects to the loop ${\cal C}$ on the Riemann surface which
starts at the location of the B-brane at a fixed $u$, winds around
$v \rightarrow v + 2\pi i$ before ``attaching'' again (see
 Fig.13b). In this configuration, the tension is given by a classical
integral on the Riemann surface
$$\hat u = \frac{1}{2\pi i}\int_{\cal C} v_1(u) du.$$

%

The period integral $\frac{1}{2\pi i}\int_{\cal C} v_1(u) du$
is, as explained in Section 4, is the sum of two pieces. 
There is the classical
contribution to the domain wall tension which is equal to
$\frac{1}{2\pi i}\Delta(v_1 u) = u$ as the 
the initial and final end-point of $\cal{C}$
differ by $v\rightarrow v+2\pi i$.
The quantum correction to
the BPS mass 
comes from the ``small period'', the contour integral around
$\beta_1$ cycle (see Fig.14) of $\frac{1}{2\pi i}\int_{\beta_1}udv$.
\bigskip
\centerline{\epsfxsize 5truein\epsfbox{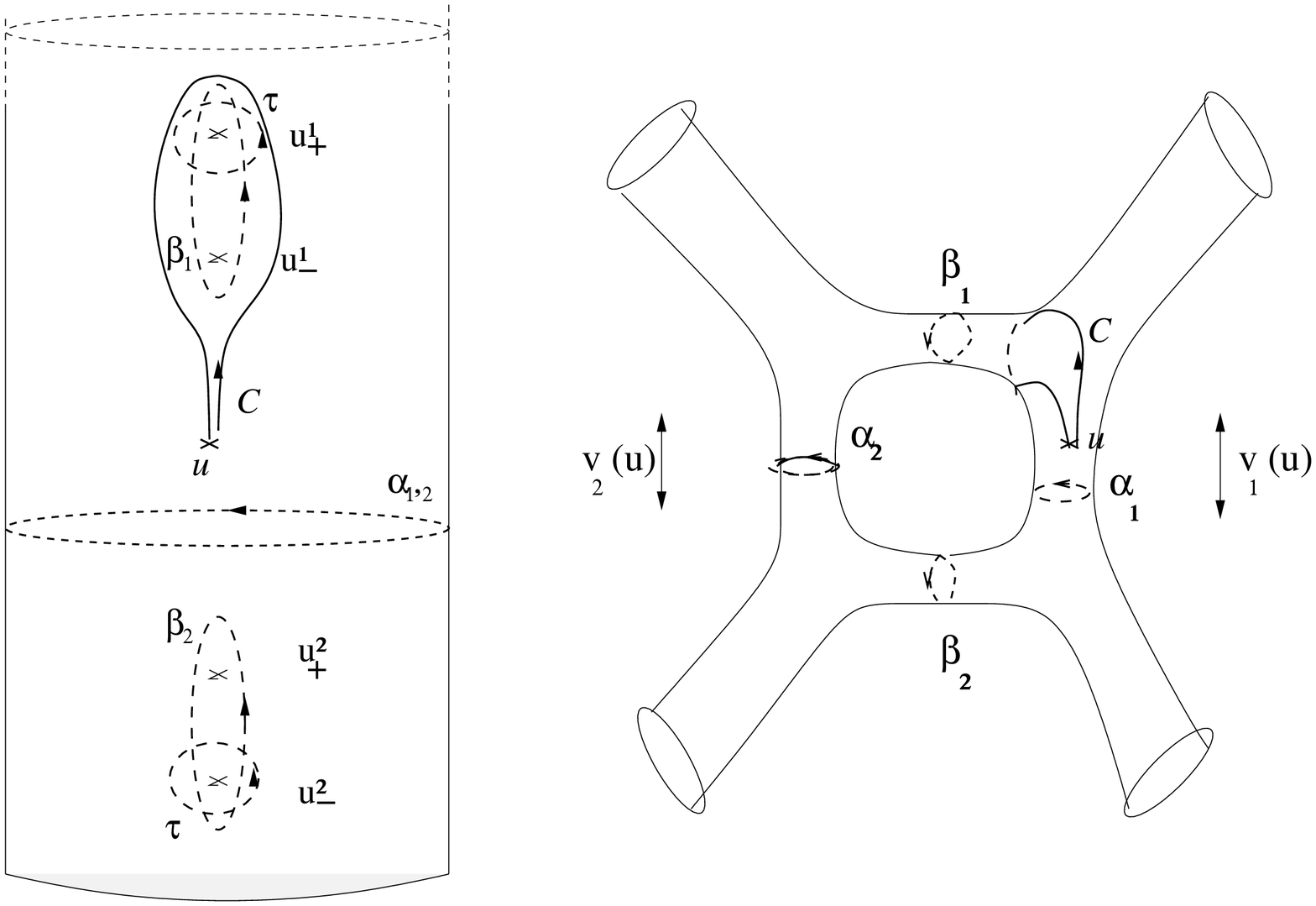}}
\rightskip 2pc \noindent{\ninepoint\sl \baselineskip=8pt {\bf
Fig.14} { Riemann surface $\Sigma:  P(u,v)=0$ can be viewed as a
branched cover of the $u$-cylinder by two solutions $v_{1,2}(u)$.
The solutions are branched over $4$ points $u^{\pm}_{1,2}$, with
monodromies $\tau_{1,2}$ which exchanges $v_1 \leftrightarrow
v_2$, and $\beta_{1,2}$ which take $v_{1,2}\rightarrow
v_{1,2}+2\pi i$ corresponding to the $v$-cylinders which glue the
two roots on the figure to the right. Monodromies $\alpha_{1,2}$
arise from periodicity of $u$ itself.}}
\bigskip

The small periods can be found as follows.
Notice that since  $v_{1} +  v_{2} = -s +2\pi i$, the sum of
two periods around $u\rightarrow u+2\pi i$ is 
$$\frac{1}{2\pi i}\int_{\alpha_1} v_{1} du +
\frac{1}{2\pi i} \int_{\alpha_2} v_{2}du =s + 2\pi i.$$
On the other hand, the closed string period $\hat s$,
that measures the mass of the D4 brane wrapping a
$\bf{P^1}$ of size $r_s$
can also be expressed in terms of small periods as
it is computed along the
contour $\alpha_{s} = \alpha_1 - \alpha_{2}$, (orientations are fixed
up to an over-all sign by the requiring that $both$ $u$ and $v$ have
trivial monodromy around $\beta_{s}$),
$$\frac{1}{2\pi i}\int_{\alpha_1} v_{1} du -\frac{1}{2\pi i}
\int_{\alpha_2} v_{2}du =-\hat{s}.$$

{}From this we can find for example the small period along 
$\alpha_1$ as\foot{This method of calculating the domain-wall tension
gives the result up to factors of $i\pi$. The direct evaluation
of the period around $\cal C$ can be done, and it determines the answer
to be the one we presented above.}
$\frac{s - \hat s }{2}+ i \pi = \frac{1}{2\pi i}\int_{\alpha_1} v_{1} du$.
%
\bigskip
\centerline{\epsfxsize 3.5truein\epsfbox{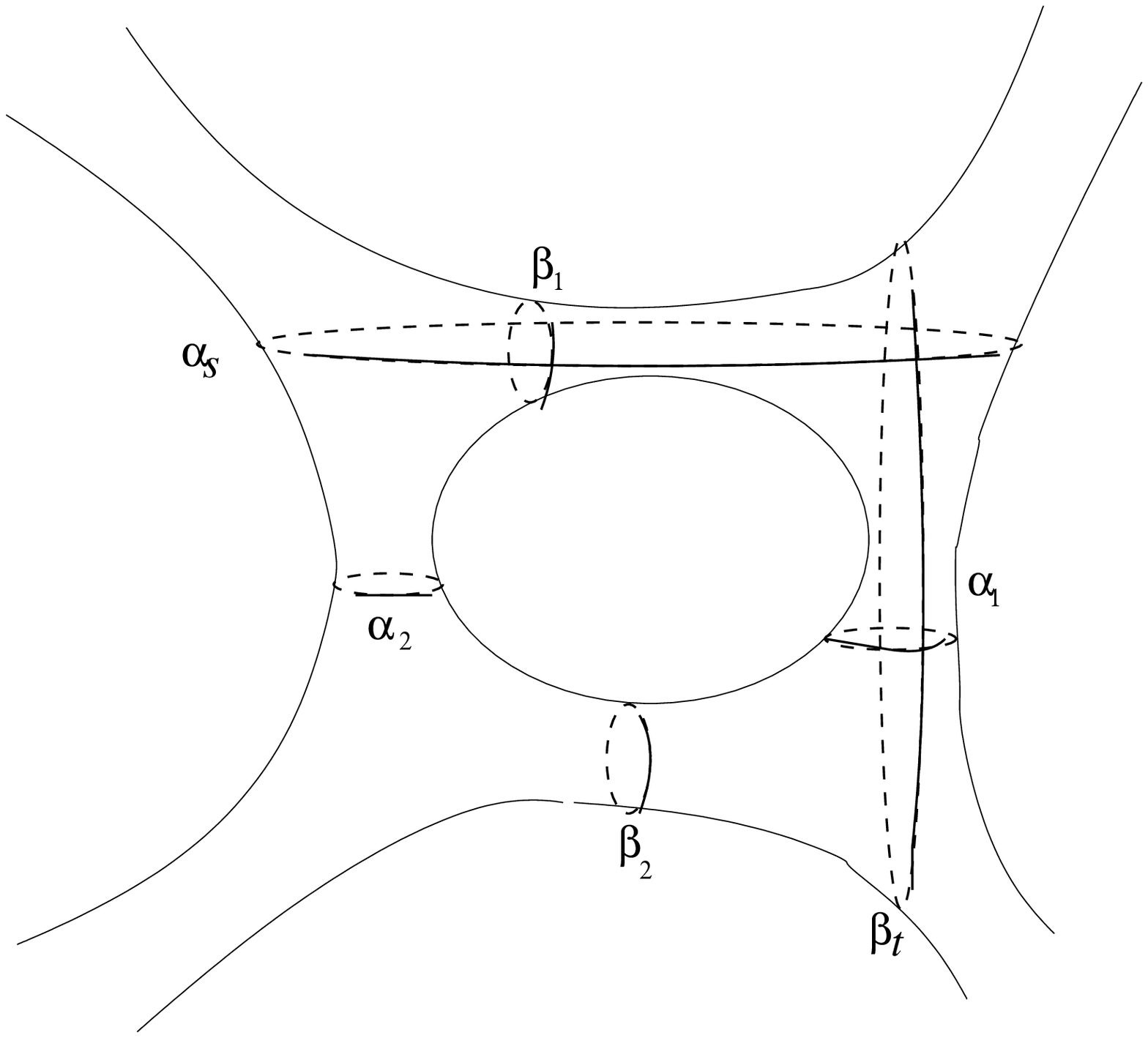}}
\rightskip 2pc
\noindent{\ninepoint\sl \baselineskip=8pt {\bf Fig.15}{ Closed string
periods $\beta_t$, $\beta_s$ are linear combinations of small periods
$\alpha_{i}$, $\beta_{i}$ for $i=1,2$ which come from periodicity of
$u$ and $v$ variables.}}
\bigskip

For the $B-$brane at hand we need the small period around
$\beta_1$ cycle on the $v-$leg, and by computation
analogous to the one we just did,
$$\hat u = u + \frac{ t-\hat t}{2} + \pi i.$$
In fact the small period of $\alpha_1$ is the correction for the B-brane on
the other leg of the Riemann surface -- the leg parameterized by
$v$ which corresponds to the phase $II$ (see
Fig.16). In this phase we have therefore
$$\hat v = v + \frac{s - \hat s }{2} + i \pi.$$
\bigskip
\centerline{\epsfxsize 3.5truein\epsfbox{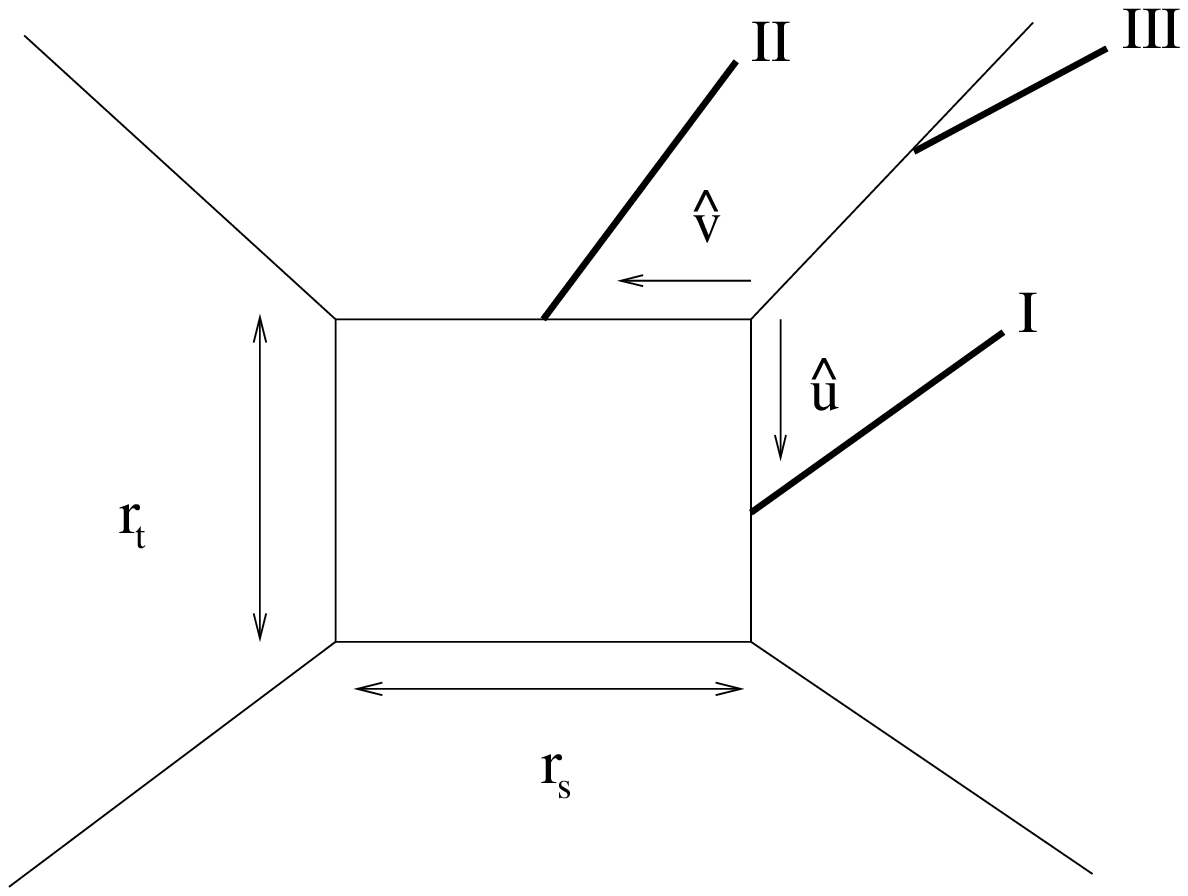}}
\rightskip 2pc
\noindent{\ninepoint\sl \baselineskip=8pt {\bf Fig.16}{ Three phases
of the $A-$brane on $O(K)\rightarrow {\bf P^1}\times {\bf
P^1}$. Phases
$I,II$ are related by the exchange of the two $\bf{P^1}$'s in the
base, and this is reflected in the disc domain wall numbers.}}
\bigskip

According to the discussion in the $\bf{C^3}$ case and which carries
over here as well, the correction terms we computed
$$\Delta_u = \frac{t -\hat t}{2} + i\pi \quad \Delta_v = \frac{s -\hat
s}{2} +i\pi$$
form a doublet under the $\SL2z$ transformations, just as $u,v$ do.
Consider now an $\SL2z$ group element which takes:
$$ u \rightarrow u' =u - v,  \quad v \rightarrow v' =v$$
The new equation of the curve becomes
$$ e^{u' - v'} + e^{-t - u' + v'} + e^{v'} + e^{-s-v'} +1=0.$$
This change of coordinates does not change the classical value of
the superpotential in phase $I$ (the phase originally parameterized by
$u$) since $v'=v \sim 0$ 
but corresponds to a different choice of framing with $n = - 1$
which changes the fluctuating field on the brane from $u$ to $u'=u-v$.
The quantum corrected domain wall tension is now\foot{As we will see
when we discuss the closed string flat coordinates in more detail
below, $\Delta_{u'}$  turns out to be zero.}
$$\Delta_{u'} = \Delta_{u} - \Delta_{v}$$
For the brane in phase $II$ it is no longer true that the classical
superpotential is zero, since $u' = u-v$ is not zero after the
$\SL2z$ transformation.

With this choice of parameterization, consider the superpotential 
$$W = \int u'(v') d v',$$
with $u'(v)$ obtained by solving $P(u',v)=0$.
This superpotential has no classical piece, as $v'\rightarrow
\infty$, $u'\sim 0$ is on the curve. In fact, 
$u'\sim 0$ is the equation of the outer leg of $\Sigma$ (see Fig. 16),
so this computes the superpotential of the brane in phase III.
The domain wall tension in the phase III (the outer leg) is given by
$$\Delta_{v'} = \Delta_{v} = \frac{s - \hat s }{2} +i \pi .$$
To compute disk numbers in these various phases, we need to
write the superpotential in each case in terms of the open and
closed string flat coordinates.

The closed string flat coordinates have a complicated dependence
on the classical, linear-sigma model, coordinates $t$.
Quite analogously to what we found in the open string,
the $A$-model closed string amplitudes have integrality
properties when expanded in terms of flat coordinates which measure
the BPS mass of D2-branes wrapped on rational curves in $X$.
The corrected quantities are related transcendentally by the mirror
map  $\hat{t_i} = \hat{t_i}(t_j)$
to the complex structure variable $z_i=e^{-t_i}$.
In fact, as discussed above, $\hat t$ and $\hat s$ are among the
periods of the Riemann surface $\Sigma$. Alternatively, they can
be obtained from the solutions to Picard-Fuch's equations ${\cal
L}_i f =0$, \eqn\periodintegrals{{\cal L}_i =\prod_{Q_i>0}
\left(\partial_i \right)^{Q_i}- \prod_{Q_i<0}
\left(\partial_i\right)^{-Q_i}} where $\partial_i =
\partial/\partial z^i$. For the $O(K)\rightarrow P^1\times P^1$ the linear
differential operators ${\cal L}_{t,s}$  are:
\eqn\pfsystem{\eqalign{ {\cal L}_t&=\tst^2-2
z_t(\tst+\tss)(2\tss+2\tst+1)\cr {\cal L}_s&=\tss^2-2
z_s(\tst+\tss)(2\tss+2\tst+1)\ , }} where $\theta_{t,s} =
z_{t,s}\partial/\partial {z_{t,s}}$. There is a constant
solution $f_0=1$ and near $z_t=z_s=0$ there are two logarithmic solutions
$f_1,f_2$. The flat coordinates  $\hat t, \hat s$ are  given by
(linear combinations) of ratios of the periods $f_i/f_0, i=1,2$
picked by correct classical behavior: \eqn\logsolutions{\eqalign{
\hat {t}&= t-(2 z_t + 2 z_s+3 z_t^2 + 12z_t z_s + 3
z_s^2+O(z^3))\cr \hat{s}&=s -( 2 z_t + 2 z_s+3 z_t^2 + 12z_t z_s +
3 z_s^2+O(z^3))\ .}} Note that the solutions are symmetric in
$z_{t,s}$ except for the logarithmic term.
%
%
To expand the disk amplitude in terms of the
flat coordinates we need the inverse relations $z_{t,s}(q_{t,s})$ for $q_t =
e^{-\hat{t}}$ and $q_s = e^{-\hat{s}}$. The first few
 terms of the expansion are
\eqn\mm{\eqalign{ z_s=&q_s-2q_s^2 - 2q_sq_2+ 3q_s^3 + 3q_sq_t^2
      -4q_s^4 - 4q_s^3q_t - 4q_s^2q_t^2 - 4q_sq_t^3 +O(q^5)\cr
z_t=&q_t-2q_sq_t - 2q_t^2+
      3q_s^2q_t + 3q_t^3
      -4q_s^3q_t - 4q_s^2q_t^2 - 4q_sq_t^3 - 4q_t^4 +O(q^5)\ .}}
where $q_t = e^{-\hat{t}}$ and $q_s = e^{-\hat{s}}$.

To get the disk numbers the only remaining task is to expand
the $W(u)$ in terms of open and closed string flat coordinates.
The integrality of the disk
amplitude \intg \ implies that
we write $ \partial_{\hat u} W(\hat u) =\hat v$ as
$$\partial_{\hat u} W(\hat u) =
-\sum_{k_s=0,k_t=0\atop m=-k_t }^\infty
m N_{k_s,k_t,m} \log\left(1-q_s^{k_s} q_t^{k_t} e^{m \hat u}\right)$$
In phase $I$ we get the following values for numbers of 
primitive disks $N_{k_s,k_t,m}$:

\bigskip
{\vbox{\ninepoint{
$$
\vbox{\offinterlineskip\tabskip=0pt \halign{\strut
\vrule#&
&  \hfil ~$#$ \hfil & \vrule#&\hfil ~$#$ &\hfil ~$#$ &\hfil ~$#$ &\hfil ~$#$ &\hfil ~$#$ &\hfil ~$#$
&\vrule#\cr \noalign{\hrule}
&m=1&& & & & & & & \cr
\noalign{\hrule}
&k_s &&k_t=0 &1 &2 &3 &4 &5
& \cr \noalign{\hrule} &0&&1& 0& 0& 0& 0& 0&  \cr
&1&&1& 2& 3& 4& 5& 6& \cr
&2&&1& 10& 45& 140& 350& 756&\cr
&3&&1& 30& 300& 1776&7650&26532  &\cr
&4&&1& 70& 1332& 13400& 91070 & 472368& \cr
&5&&1&140& 4590& 72856 & 736270 & 5437530 & \cr
\noalign{\hrule}}\hrule}$$}}}

{\vbox{\ninepoint{
$$
\vbox{\offinterlineskip\tabskip=0pt \halign{\strut
\vrule#&
&  \hfil ~$#$ \hfil & \vrule#&\hfil ~$#$ &\hfil ~$#$ &\hfil ~$#$ &\hfil ~$#$ &\hfil ~$#$ &\hfil ~$#$
&\vrule#\cr \noalign{\hrule}
&m=2& & & & & & & & \cr
\noalign{\hrule}
&k_s &&k_t=0 &1 &2 &3 &4 &5
& \cr \noalign{\hrule}
&0&& 0& 0& 0& 0& 0& 0& \cr
&1&& 1& 2& 3& 4&5& 6& \cr
&2&& 2& 16& 62& 180& 428& 896&  \cr
&3&& 4& 70& 552& 2856&11280&36828 &  \cr
&4&& 6& 224& 3130& 26336& 159078& 759200 &  \cr
&5&& 9& 588& 13420& 171720& 1503135 & 10016490&  \cr
\noalign{\hrule}}\hrule}$$}}}

{\vbox{\ninepoint{
$$
\vbox{\offinterlineskip\tabskip=0pt \halign{\strut
\vrule#&
&  \hfil ~$#$ \hfil & \vrule#&\hfil ~$#$ &\hfil ~$#$ &\hfil ~$#$ &\hfil ~$#$ &\hfil ~$#$ &\hfil ~$#$
&\vrule#\cr \noalign{\hrule}
&m=3&& & & & & & & \cr
\noalign{\hrule}
&k_s&&k_t=0 &1 &2 &3 &4 &5
& \cr \noalign{\hrule}
&0&&0& 0& 0& 0& 0& 0&  \cr
&1&&    1& 2& 3&4& 5& 6&  \cr
&2&&     4& 24& 85& 230& 525& 1064&  \cr
&3&&    11&146& 977& 4542& 16644& 51420 &  \cr
&4&&    25& 618& 6975& 50912&278134& 1231230&  \cr
&5&&    49& 2070& 36637&395818  & 3068331 &
18655290 & \cr
\noalign{\hrule}}\hrule}$$}}}

{\vbox{\ninepoint{
$$
\vbox{\offinterlineskip\tabskip=0pt \halign{\strut \vrule#&
&  \hfil ~$#$ \hfil & \vrule#&\hfil ~$#$ &\hfil ~$#$ &\hfil ~$#$ &\hfil ~$#$ &\hfil ~$#$ &\hfil ~$#$
&\vrule#\cr \noalign{\hrule}
&m=4&& & & & & & & \cr
\noalign{\hrule}
&k_s&&k_t=0 &1 &2 &3 &4 &5
& \cr \noalign{\hrule}
&0&& 0& 0& 0& 0& 0& 0& \cr
&1&& 1& 2& 3& 4&5& 6& \cr
&2&& 6& 34& 112& 290& 638& 1260&  \cr
&3&& 25& 276& 1645&7040& 24246& 71400 & \cr
&4&& 76& 1498& 14496& 94830& 476900&1979098&  \cr
&5&& 196& 6248& 91935& 870220 &6103867& 34309080&\cr
\noalign{\hrule}}\hrule}$$}}}

{\vbox{\ninepoint{
$$
\vbox{\offinterlineskip\tabskip=0pt \halign{\strut \vrule#&
&  \hfil ~$#$ \hfil & \vrule#&\hfil ~$#$ &\hfil ~$#$ &\hfil ~$#$ &\hfil ~$#$ &\hfil ~$#$ &\hfil ~$#$
&\vrule#\cr \noalign{\hrule}
&m=5&& & & & & & & \cr
\noalign{\hrule}
&k_s&&k_t=0 &1 &2 &3 &4 &5
& \cr \noalign{\hrule}
&0&& 0& 0& 0& 0& 0& 0&  \cr
&1&&1& 2& 3& 4&5& 6&\cr
&2&&9& 46& 145& 360& 770& 1484&\cr
&3&&49& 482& 2640&10592& 34674& 98028& \cr
&4&&196& 3270& 28240& 169402&795998 &3126928& \cr
&5&&635& 16642& 213083&1816038&11729677& 61675880&\cr
\noalign{\hrule}}\hrule}$$}}}

{\vbox{\ninepoint{
$$
\vbox{\offinterlineskip\tabskip=0pt \halign{\strut \vrule#&
&  \hfil ~$#$ \hfil & \vrule#&\hfil ~$#$ &\hfil ~$#$ &\hfil ~$#$ &\hfil ~$#$ &\hfil ~$#$ &\hfil ~$#$
&\vrule#\cr \noalign{\hrule}
&m=6&& & & & & & & \cr
\noalign{\hrule}
&k_s&&k_t=0 &1 &2 &3 &4 &5
& \cr \noalign{\hrule}
&0&&  0& 0& 0& 0& 0& 0&  \cr
&1&&    1& 2& 3&4& 5& 6&  \cr
&2&&    12& 60& 182& 440& 918& 1736&  \cr
&3&&    87&790& 4060& 15478& 48600&  132732&  \cr
&4&&    440& 6560& 51906&290600& 1290870& 4840248&  \cr
&5&&     1764& 40050&460355&3604656& 21737688& 107979508& \cr
\noalign{\hrule}}\hrule}$$}}}
\noindent{\ninepoint\sl \baselineskip=2pt {\bf Tab.1\ }{The
disk domain wall degeneracies for brane $I$ in the $O(K)\rightarrow
{\bf P^1}\times {\bf P^1}$ geometry for $m>0$. Exchanging $s$ with $t$ yields the result for brane $II$. }}
\bigskip

There is a symmetry which relates the numbers of
disk instantons with negative $m$ to those above. We have
\eqn\symmetry{N_{k_s,k_t,-m}=\cases{ 0 & if $k_t-m< 0$\cr
-N_{k_s,k_t-m,m}& if $k_t-m\geq 0$ }}
so for e.g

{\vbox{\ninepoint{
$$
\vbox{\offinterlineskip\tabskip=0pt \halign{\strut \vrule#&
&  \hfil ~$#$ \hfil & \vrule#&\hfil ~$#$ &\hfil ~$#$ &\hfil ~$#$ &\hfil ~$#$ &\hfil ~$#$ &\hfil ~$#$
&\vrule#\cr \noalign{\hrule}
&m=-1&& & & & & & &\cr
\noalign{\hrule}
&k_s&&k_t=0 &1 &2 &3 &4 &5
& \cr \noalign{\hrule}
&0&& 0& -1& 0& 0& 0& 0&  \cr
&1&&0& -1& -2&-3& -4& -5&  \cr
&2&&0& -1& -10& -45& -140& -350&  \cr
&3&&0& -1&-30& -300& -1776& -7650&\cr
&4&&0& -1& -70& -1332& -13400& -91070 &\cr
&5&&0& -1& -140& -4590& -72856& 736270&  \cr
\noalign{\hrule}}\hrule}$$}}}
\noindent{\ninepoint\sl \baselineskip=2pt {\bf Tab.2\ }{Disk degeneracies
for brane $I$ in the $O(K)\rightarrow {\bf P^1}\times {\bf P^1}$
geometry for $m<0$.}}
\bigskip

The numbers of primitive disks for the brane III are

\bigskip
{\vbox{\ninepoint{
$$
\vbox{\offinterlineskip\tabskip=0pt \halign{\strut \vrule#&
&  \hfil ~$#$ \hfil & \vrule#&\hfil ~$#$ &\hfil ~$#$ &\hfil ~$#$ &\hfil ~$#$ &\hfil ~$#$ &\hfil ~$#$
&\vrule#\cr \noalign{\hrule}
&m=1&& & & & & & & \cr
\noalign{\hrule}
&k_s&&k_t=0 &1 &2 &3 &4 &5
& \cr \noalign{\hrule}
&0&&2& 2& 2&  2& 2& 2&  \cr
&1&&0& 2& 12& 40& 100&210& \cr
&2&&0& 2& 40& 310& 1520& 5628&\cr
&3&&0& 2&100& 1520&12908& 76488  &\cr
&4&&0& 2&210& 5628& 91070 & 680940 & \cr
&5&&0& 2&392& 17184 & 353316 & 4515558 & \cr
\noalign{\hrule}}\hrule}$$}}}

{\vbox{\ninepoint{
$$
\vbox{\offinterlineskip\tabskip=0pt \halign{\strut \vrule#&
&  \hfil ~$#$ \hfil & \vrule#&\hfil ~$#$ &\hfil ~$#$ &\hfil ~$#$ &\hfil ~$#$ &\hfil ~$#$ &\hfil ~$#$
&\vrule#\cr \noalign{\hrule}
&m=2&& & & & & & & \cr
\noalign{\hrule}
&k_s&&k_t=0 &1 &2 &3 &4 &5
& \cr \noalign{\hrule}
&0&& 0& 0& -2& -4& -8& -12& \cr
&1&& 0& 0& -4& -32& -140& -448& \cr
&2&& 0& 0& -8& -140& -1188& -6580&  \cr
&3&& 0& 0& -12&-448&  -6580& -58240 &  \cr
&4&& 0& 0& -28&-1176& -27840&  -370428 &  \cr
&5&& 0& 0& -24&-2688& -97020 & -1859648&  \cr
\noalign{\hrule}}\hrule}$$}}}

{\vbox{\ninepoint{
$$
\vbox{\offinterlineskip\tabskip=0pt \halign{\strut \vrule#&
&  \hfil ~$#$ \hfil & \vrule#&\hfil ~$#$ &\hfil ~$#$ &\hfil ~$#$ &\hfil ~$#$ &\hfil ~$#$ &\hfil ~$#$
&\vrule#\cr \noalign{\hrule}
&m=3&& & & & & & & \cr
\noalign{\hrule}
&k_s&&k_t=0 &1 &2 &3 &4 &5
& \cr \noalign{\hrule}
&0&&0& 0& 0& 2& 10     & 28      &  \cr
&1&&0& 0& 0&10& 100    & 540     &  \cr
&2&&0& 0& 0&28& 540    & 5012    &  \cr
&3&&0& 0& 0&62& 2100   & 317072  &  \cr
&4&&0& 0& 0&120&6600   & 147420  &  \cr
&5&&0& 0& 0&210&17820  & 576212  & \cr
\noalign{\hrule}}\hrule}$$}}}

{\vbox{\ninepoint{
$$
\vbox{\offinterlineskip\tabskip=0pt \halign{\strut \vrule#&
&  \hfil ~$#$ \hfil & \vrule#&\hfil ~$#$ &\hfil ~$#$ &\hfil ~$#$ &\hfil ~$#$ &\hfil ~$#$ &\hfil ~$#$
&\vrule#\cr \noalign{\hrule}
&m=4&& & & & & & & \cr
\noalign{\hrule}
&k_s &&k_t=0 &\ldots &4 &5 &6 &7
& \cr \noalign{\hrule}
&0&& 0& \ldots& 0& -4& -28& -104& \cr
&1&& 0& \ldots& 0& -28& -336&-2156&  \cr
&2&& 0&\ldots & 0& -104& -2156& -21888& \cr
&3&& 0& \ldots& 0& -300& -9856 & -149940&\cr
&4&& 0& \ldots& 0& -720&  -36036& -787640& \cr
&5&& 0&\ldots & 0& -1540& -112112& - 3406480&\cr
\noalign{\hrule}}\hrule}$$}}}

{\vbox{\ninepoint{
$$
\vbox{\offinterlineskip\tabskip=0pt \halign{\strut \vrule#&
&  \hfil ~$#$ \hfil & \vrule#&\hfil ~$#$ &\hfil ~$#$ &\hfil ~$#$ &\hfil ~$#$ &\hfil ~$#$ &\hfil ~$#$
&\vrule#\cr \noalign{\hrule}
& m=5&& & & & & & & \cr
\noalign{\hrule}
&k_s &&k_t=0 &\ldots &5 &6 &7 &8
& \cr \noalign{\hrule}
&0&& 0& \ldots  & 0& 10&   84    & 396&  \cr
&1&& 0& \ldots  & 0& 84& 1176    & 8736&\cr
&2&& 0&\ldots   & 0& 396&8736    & 96660&\cr
&3&& 0&\ldots   & 0& 1386&45864  & 724800& \cr
&4&& 0&\ldots   & 0& 4004&191100 & 4273840& \cr
&5&& 0&\ldots   & 0& 10090&672672&    - &\cr
\noalign{\hrule}}\hrule}$$}}}

{\vbox{\ninepoint{
$$
\vbox{\offinterlineskip\tabskip=0pt \halign{\strut \vrule#&
&  \hfil ~$#$ \hfil & \vrule#&\hfil ~$#$ &\hfil ~$#$ &\hfil ~$#$ &\hfil ~$#$ &\hfil ~$#$ &\hfil ~$#$
&\vrule#\cr \noalign{\hrule} &m=6 && & & & & & & \cr
\noalign{\hrule}
&k_s &&k_t=0 &- &6 &7 &8 &9
& \cr \noalign{\hrule}
&0&& 0& \ldots  & 0& -26&   -264    & -1504&  \cr
&1&& 0& \ldots  & 0& -264&  -4224   & -35640&\cr
&2&& 0&\ldots   & 0& -1504& -35640  & -427540&\cr
&3&& 0&\ldots   & 0& -6228& -211200 & -3484800& \cr
&4&& 0&\ldots   & 0& -21028&-987360 &    -    & \cr
&5&& 0&\ldots   & 0& -61152&  -     &    -    &\cr
\noalign{\hrule}}\hrule}$$}}}
\noindent{\ninepoint\sl \baselineskip=2pt {\bf Tab.3\ }{The
disk degeneracies for brane $III$ in the $O(K)\rightarrow
{\bf P^1}\times {\bf P^1}$ geometry for $m>0$. }}
\bigskip

Up to the invariant $N_{0,0,1}=2$ we have $N_{k_s,k_t,m}=0$ for $k_t<m$ and
for $k_t\geq m$ there is a symmetry
\eqn\symmetryIII{N_{k_s,k_t,m}=N_{k_t-m,k_s+m,m}\ , }
which reflects the exchange symmetry of $s$ and $t$ for the brane in
phase $III$. The instantons with negative $m$ are absent for
$k_s \leq |m|$ and their numbers are related to those with positive $m$ by
\eqn\symmetryIII{N_{k_s,k_t,-m}=-N_{k_s+m,k_t-m,m}\ , }
e.g.

\bigskip
{\vbox{\ninepoint{
$$
\vbox{\offinterlineskip\tabskip=0pt \halign{\strut \vrule#&
&  \hfil ~$#$ \hfil & \vrule#&\hfil ~$#$ &\hfil ~$#$ &\hfil ~$#$ &\hfil ~$#$ &\hfil ~$#$ &\hfil ~$#$
&\vrule#\cr \noalign{\hrule}
&m=-1&& & & & & & &\cr
\noalign{\hrule}
&k_s &&k_t=0 &1 &2 &3 &4 &5
& \cr \noalign{\hrule}
&0&& 0&   0&       0&      0& 0& 0&  \cr
&1&& -2& -2&      -2&     -2& -2&          -2&  \cr
&2&& -2& -12&    -40&   -100& -210&      -392&  \cr
&3&& -2& -40&   -310&  -1520& -5628 &    -17184&\cr
&4&& -2& -100& -1520& -12908& -76488 & -353316 &\cr
&5&& -2& -210& -5628& -76488& -680940& -4515558&  \cr
\noalign{\hrule}}\hrule}$$}}}
\noindent{\ninepoint\sl \baselineskip=2pt {\bf Tab.4\ }{The
disk degeneracies for brane $III$ in the $O(K)\rightarrow
{\bf P^1}\times {\bf P^1}$ geometry for $m<0$. }}
\bigskip

\subsec{${\cal O}(-3) \rightarrow {\bf P}^2$}

The linear sigma model for this
geometry has  $G=U(1)$ and four matter fields with charges $Q=(-3,
1,1,1)$
\eqn\dtermpII{|X^1|^2+|X^2|^2+|X^3|^2- 3 |X^0|^2=r \  }
The base of the toric fibration is shown in the Fig.17.

The D-brane charge can be taken to be $q_1=(1,0,-1,0),
q_2=(0,1,-0,-1)$ (the choice is unique in all local models) and so
$$|X^1|^2-|X^0|^2=c^1 ,\quad
|X^2|^2-|X^0|^2=c^2.$$
Consider three phases of the A-brane that are visible classically (see Fig.
17). \eqn\pIIbranes{\eqalign{ {\rm Phase\ I:}\quad\quad& r_t>c^1>
0,
 \quad c^2 = 0\
\cr {\rm Phase\ II:}\quad\quad & c^1=0, \quad r_t> c^2>0 \
\cr {\rm Phase\ III:}\quad\quad & c^1=c^2  , \quad 0 <c^1 \cr }}
\bigskip
\centerline{\epsfxsize 3truein\epsfbox{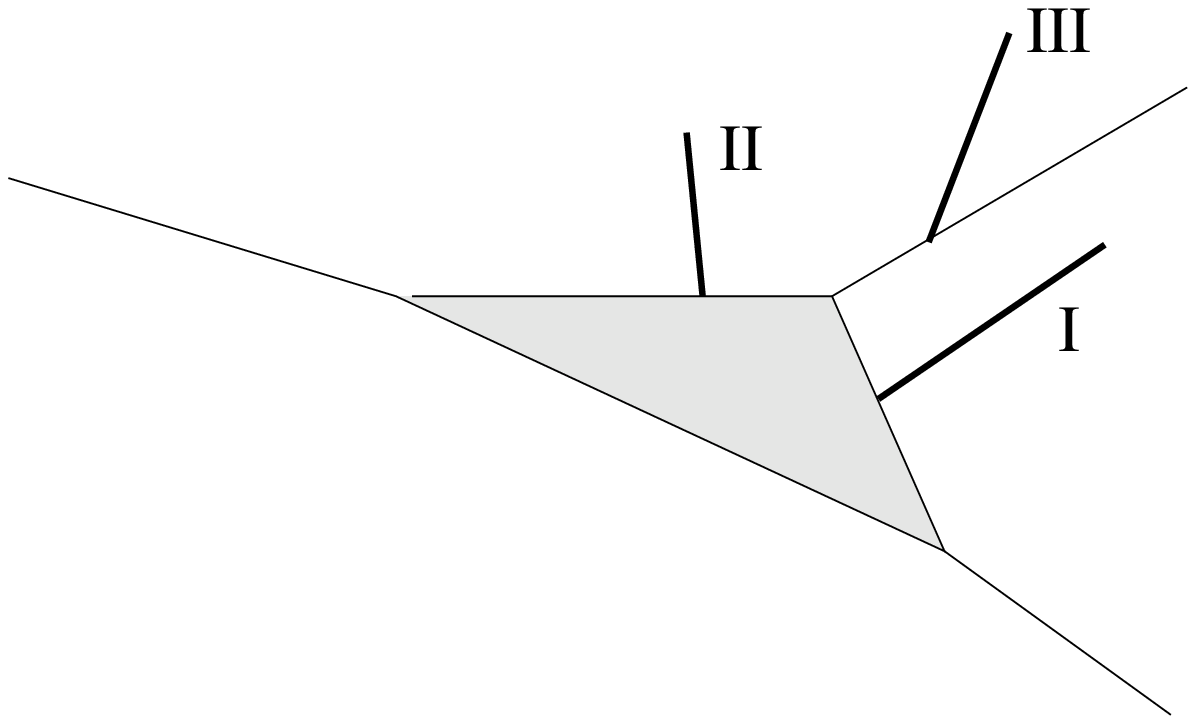}}
\rightskip 2pc
\noindent{\ninepoint\sl \baselineskip=8pt {\bf Fig. 17}{ Three phases
of the $A-$brane on $O(-3)\rightarrow {\bf P^2}$. Phases $I$ and $II$
are related by ${\bf Z_3}$ symmetry of the $\bf{P^2}$}.}
\bigskip
As the branes in phases $I$ and $II$ are related
by the ${\bf Z}_3$ symmetry of ${\bf P}^2$, 
the special Lagrangian D-branes of the A-model for
this Calabi-Yau have two inequivalent phases, together
with a $\bf{Z}$-family of choices of framing in each.

The mirror $B$-model geometry (see e.g.
\ref\ckyz{T.~M.~Chiang, A.~Klemm, S.~T.~Yau and E.~Zaslow, ``Local
mirror symmetry: Calculations and interpretations,'' Adv.\ Theor.\
Math.\ Phys.\ {\bf 3} (1999) 495, hep-th/9903053.}\ref\alet{M.
Aganagic, A. Karch, D. Lust and A. Miemiec, ``Mirror Symmetries
 for Brane Configurations and Branes at Singularities,''
hep-th/9903093.}\ref\hiqbv{K. Hori,
A. Iqbal and C. Vafa, ``D-Branes And Mirror Symmetry,''
hep-th/0005247.}) can be written as
\eqn\mirrorpII{x z=e^u+e^v+e^{-t-u-v}+1}
where we have ``solved''
$Y^1+Y^2+Y^3 = -t +3
Y^0$
by $Y^0=0$, $Y^1=u$, $Y^2=v$ and $Y_3=-t-u-v$.

In terms of these coordinates,
the brane in phase $I$ propagates on the internal leg of
the Riemann surface where $v\sim 0$ and $u$ large of order $-t/2$, the 
brane in phase $II$ is on
$u\sim 0$, and $v \sim t/2$, and the brane on
external leg has $u \sim v$, and both are large.

In the variables of \mirrorpII\ the classical superpotential vanishes in
phase $I$ and
$W(u) = \int v(u) du$
computes the disk instanton generated superpotential.

To compute disk numbers we need the flat coordinate.
According to discussion in section 5, this is given by the
difference of superpotentials on the two sides of the domain wall
as computed along the contour $C$ on Fig.18. The non-trivial
contribution to $\hat u$ is the exponentially suppressed shift
which comes from the small period on the Fig.18.
$$\hat u = \int_C v(u) du = u + \Delta_u$$
We find, using the same kind of methods discussed
for the ${\bf P}^1\times {\bf P}^1$ example that,
$$\Delta_u = {{t-\hat t} \over 3}+ i \pi ,$$ 
where $\hat t$
is the closed string period on contour $\alpha_1$ in Fig.18.

\bigskip
\centerline{\epsfxsize 3.5truein\epsfbox{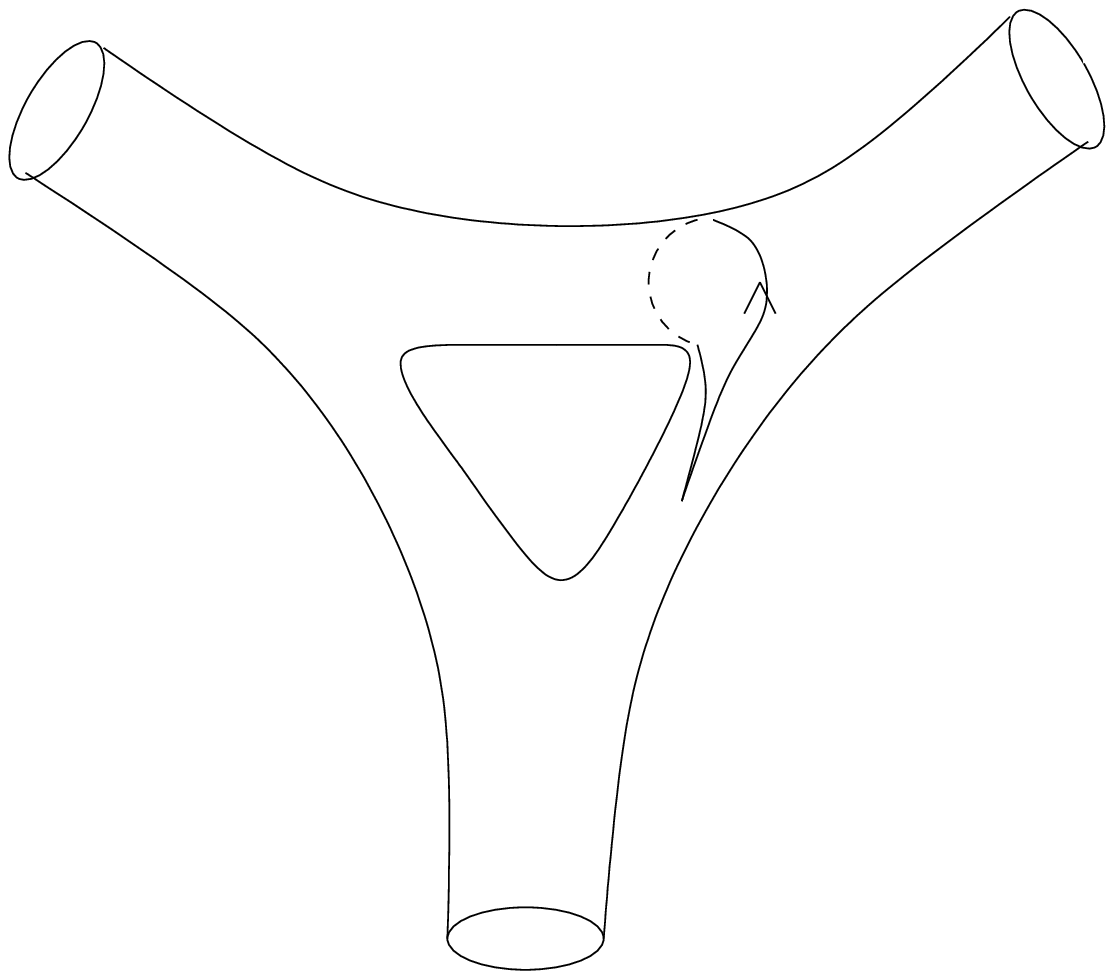}}
\rightskip 2pc
\noindent{\ninepoint \sl \baselineskip=8pt {\bf Fig.18}{ The curve
$P(u,v)=0$
associated to the mirror of $O(-3)\rightarrow {\bf P^2}$.}}
\bigskip

The closed string period can be computed either directly
by integration on $\Sigma$, or from the Picard-Fuchs
equation with ${\cal{L}}=\theta^3+  3 z\theta (3 \theta +1)(3
\theta +2)$ where $\theta = z \partial_z$, and
$z=e^{-t}$.
The solutions to the Picard-Fuchs equation can
be expressed e.g. as Mejer G-functions \ref\kz{A.~Klemm and
E.~Zaslow,``Local mirror symmetry at higher genus,''
hep-th/9906046.} and
$$
\hat t = t -\sum_{n=1}^{\infty}{{(-1)^n}\over n} {{(3
n)!}\over {(n!)^3}} z^n
$$
The inverse $z({\hat t})$
relation is: \eqn\inversemirrormap{z= q+ 6 q^2 + 9 q^3 + 56 q^4 -
300 q^5 + 3942 q ^6 + \ldots} where $q = e^{-\hat t}$.

After rewriting $W(u,t)$ in terms of the flat coordinates $\hat u,
{\hat t}$, and expanding as in \intg \ we obtain the following
integer invariants

\vskip 5 mm
{\vbox{\ninepoint{
$$
\vbox{\offinterlineskip\tabskip=0pt \halign{\strut \vrule#&
&\hfil ~$#$ \hfil &\vrule#& \hfil ~$#$ &\hfil ~$#$ &\hfil ~$#$ &\hfil ~$#$ &\hfil ~$#$
&\hfil ~$#$ &\hfil ~$#$ &\hfil ~$#$ &\hfil ~$#$
&\vrule#\cr \noalign{\hrule} &m &&k=0 &1 &2 &3 &4 &5 &6 &7 &8
& \cr \noalign{\hrule}
&-5&& 0& 0& 0& 0& 0& 5& -84& 1200& -16854&
\cr &-4&& 0& 0& 0& 0& -2& 28& -344& 4360& -57760&
\cr &-3&& 0& 0& 0& 1& -10& 102& -1160& 14274& -185988&
\cr &-2&& 0& 0& -1& 4& -32& 326& -3708& 45722& -598088&
\cr &-1&& 0& 1& -2& 12& -104& 1085& -12660& 159208& -2112456&
\cr &1&& 1& -1& 5& -40& 399& -4524& 55771& -729256& 9961800&
\cr &2&& 0& -1& 7& -61& 648& -7661& 97024& -1293185& 17921632&
\cr &3&& 0& -1& 9& -93& 1070& -13257& 173601& -2371088& 33470172&
\cr &4&& 0& -1& 12& -140& 1750& -22955& 312704& -4396779& 63460184&
\cr &5&& 0& -1& 15& -206& 2821& -39315& 559787& -8136830&
120497011&
\cr
\noalign{\hrule}}\hrule}$$}}}
\noindent{\ninepoint\sl \baselineskip=2pt {\bf Tab.5\ }{Disk
degeneracies for brane $I$ or $II$
in the $O(K)\rightarrow {\bf P}^2$ geometry. }}
\bigskip

For the phase $III$, to compute the integer invariants we need to
change the parameterization of the curve.
Consider the $\SL2z$ transformation
$$u \rightarrow u' = u-v$$
$$v \rightarrow v' = v$$
which will
allow us to compute the superpotential in phase III since in these
variables, the equation of the phase III leg is $u'=0$.
The equation for $\Sigma$ written in the new variables becomes
\eqn\ptt{e^{u'}+e^{v'}+1 +e^{-{u'} + 3{v'} -t} =0,} upon trivial
multiplication by $e^{v'}$. The good
flat coordinate for this phase is $\hat v' = v' +\Delta_v =v' +
\frac{t - \hat t}{3}$. 
For the integer invariants in phase III we get that $d_{k,m}=0$
for $m<0$ and, for positive $m$, we obtain

\vskip 5 mm
{\vbox{\ninepoint{
$$
\vbox{\offinterlineskip\tabskip=0pt \halign{\strut \vrule#& &\hfil
~$#$ \hfil &\vrule#&\hfil ~$#$ &\hfil ~$#$ &\hfil ~$#$ &\hfil ~$#$ &\hfil ~$#$
&\hfil ~$#$ &\hfil ~$#$ &\hfil ~$#$ &\hfil ~$#$ &\vrule#\cr
\noalign{\hrule} &m &&k=0 &1 &2 &3 &4 &5 &6 &7 &8 & \cr
\noalign{\hrule} &1&& -1& 2& -5& 32& -286& 3038& -35870& 454880&
-6073311&\cr &2&& 0& 1& -4& 21& -180& 1885& -21952& 275481&
-3650196&\cr &3&& 0& 1& -3& 18& -153& 1560& -17910& 222588&
-2926959&\cr &4&& 0& 1& -4& 20& -160& 1595& -17976& 220371&
-2869120&\cr &5&& 0& 1& -5& 26& -196& 1875& -20644& 249120&
-3205528&\cr &6&& 0& 1& -7& 36& -260& 2403& -25812& 306095&
-3889116&\cr &7&& 0& 1& -9& 52& -365& 3254& -34089& 397194&
-4981102&\cr &8&& 0& 1& -12& 76& -528& 4578& -46812& 535639&
-6627840&\cr \noalign{\hrule}}\hrule}$$}}}
\noindent{\ninepoint\sl \baselineskip=2pt {\bf Tab.6\ }{Disk  
degeneracies for brane $III$
in the $O(K)\rightarrow {\bf P}^2$ geometry. }}
\bigskip

{}From the equation
$\ptt$ we obtain another description of
phase $I$, which is at the classical level equivalent to the one
already given,
but differes in the quantum theory by relative framing $n=-1$,
The flat coordinate in the phase $I_{n=-1}$ is
$$\hat u' = u',$$
since under the $\SL2z$ transformation
$\Delta_u$ and $\Delta_v$ cancel off.
We have also considered other values of $n$, i.e
the D-branes with the $u \rightarrow u' =u +nv$ as the
dynamical field on the brane.
First, note that $I_n$ and $I_{-(n+1)}$ (where $I_n$ denotes the
brane in phase $I$ and with framing $n$) are related by
$v \rightarrow - v +t$, with $u$ fixed.
Thus, we expect $$N^n_{k,m}=\pm N^{-(n+1)}_{k+m,-m}$$
where $m$ denotes the is boundary class of the disk.
The following are the integer invariants for $n=-1,1,2$,
which clearly respect this.

\vskip 5 mm {\vbox{\ninepoint{
$$\vbox{\offinterlineskip\tabskip=0pt \halign{\strut \vrule#& &\hfil
~$#$ \hfil & \vrule#& \hfil ~$#$ &\hfil ~$#$ &\hfil ~$#$ &\hfil ~$#$ &\hfil ~$#$
&\hfil ~$#$ &\hfil ~$#$ &\hfil ~$#$ &\hfil ~$#$ &\vrule#\cr
\noalign{\hrule} &n=-1 &&    &  &  &  & & & & & &\cr
\noalign{\hrule} &m    &&k=0 &1 &2 &3 &4 &5 &6 &7 &8 & \cr
\noalign{\hrule}
&-5&& 0& 0& 0& 0& 0& 40& -1274& 27885& -528934&\cr
&-4&& 0& 0& 0& 0&10& -253& 4604& -76068& 1214324&\cr
&-3&& 0& 0& 0& 3& -54& 783&-11058& 157347& -2274642&\cr
&-2&& 0& 0& 1& -13& 142& -1657& 20785&-274473& 3769424&\cr
&-1&& 0& 1& -4& 29& -274& 3002& -36144&464522& -6262370&\cr
&1&& 1& -1& 4& -29& 274& -3002& 36144&-464522& 6262370&\cr
&2&& 0& 0& -1& 13& -142& 1657& -20785& 274473&-3769424&\cr
&3&& 0& 0& 0& -3& 54& -783& 11058& -157347&2274642&\cr
&4&& 0& 0& 0& 0& -10& 253& -4604& 76068& -1214324&\cr
&5&& 0& 0& 0& 0& 0& -40& 1274& -27885& 528934&\cr
\noalign{\hrule}}\hrule}$$}}}

\vskip 5 mm {\vbox{\ninepoint{
$$\vbox{\offinterlineskip\tabskip=0pt
\halign{\strut
\vrule#& &\hfil ~$#$ \hfil& \vrule#&~$#$ &\hfil ~$#$ &\hfil ~$#$ \hfil &  \hfil ~$#$ &\hfil ~$#$ &\vrule#\cr \noalign{\hrule}
&n=1 && & & & & & \cr \noalign{\hrule}
&m &&k=0 &1 &2 &3 &4 & \cr
\noalign{\hrule}
&-4&& 0& 0& 0& 0&  0&\cr
&-3&& 0& 0& 0& 0& -1&\cr
&-2&& 0& 0& 0& -1& 7&\cr
&-1&& 0& 1& -1& 5& -40&\cr
&1&&  1& -2& 12&-104& 1085&\cr
&2&& -1& 4& -32& 326& -3708&\cr
&3&&  1& -10& 102&-1160& 14274&\cr
&4&& -2& 28& -344& 4360&-57760&\cr
\noalign{\hrule}}\hrule} \vbox{\offinterlineskip\tabskip=0pt
\halign{\strut
\vrule#& &\hfil ~$#$\hfil &\vrule#&\hfil ~$#$ &\hfil ~$#$ &\hfil ~$#$ &\hfil ~$#$ &\hfil ~$#$ &\vrule#\cr \noalign{\hrule}
 &n=2&&    &  &  &  &  &  \cr \noalign{\hrule}
 &m  &&k=0 &1 &2 &3 &4 & \cr
\noalign{\hrule}
&-3&& 0& 0& 0& 0& 0&\cr
&-2&& 0& 0& 0& -1& 13&\cr
&-1&& 1& -1& 4& -29& 274&\cr
&1&&  1& -4   &   29&   -274&    3002&\cr
&2&&  1& -13  &  142&  -1657&   20785      &\cr
&3&&  3& -54  &758& -11058&   157347      &\cr
&4&& 10&  -253& 4608& -76068&1214324      &\cr
&5&& 40& -1274&27885&-528934&   9380474      &\cr
\noalign{\hrule}}\hrule}
$$}}}
\noindent{\ninepoint\sl \baselineskip=2pt {\bf Tab.7\ }{Disk degeneracies
  for brane $I$
in the $O(K)\rightarrow {\bf P}^2$ geometry for various choice of the ambiguity $n\in Z $. }}
\bigskip

$$\it{Integrality\ of\ the\ Bulk\ Mirror\ Map}$$

The
integrality of mirror map in the bulk, even though it has been proven
in some cases, has not been physically explained.  Here we
will connect this to the integrality of the number of domain walls $N_{k,m}$.
In the case $I_{n=-1}$ we can explicitly show that all
coefficients of the form $N_{k,1}$ are directly
related to the coefficients of the bulk mirror map and using
this relation the integrality of one follows from the other.

Define the numbers $a_i$ by
$$e^{\frac{t- {\hat t}}{2}}=1+\sum_{i=1}^\infty a_i q^i.$$
One can show the integrality of $a_i$  using the integrality
of the mirror map \inversemirrormap .  Furthermore
 one can show, by explicitly investigating the Taylor series
of $v(\hat u, q)$, that
$$N_{k,1}=P_k(a_1,\ldots,a_{k-1})-a_k,$$
where $P_k$ is a polynomial with integer coefficients in the
$a_i$. E.g. we get
$$\eqalign{
P_1&=-3\cr P_2&=30 + 12 a_1 + a_1^2\cr P_3&=-420 - 210 a_1 -
30a_1^2 - a_1^3 + 12 a_2 + 2 a_1 a_2  \cr P_4&= 6930 + 4200 a_1 +
840 a_1^2 + 60 a_1^3 + a_1^4 - 210 a_2 - 60 a_1 a_2 - 3 a_1^2 a_2
+ a_2^2 + 12 a_3 + 2 a_1 a_3\cr etc..}$$ The proof that all $P_k$
are integer polynomials is tedious and relies on some formulas for
the derivatives of $v(\hat u , q)$  for this special example.
{}From this one sees that the integrality of $N_{k,1}$ and $a_k$
are equivalent, and the integrality of $a_k$ follows from the
integrality of the mirror map in the bulk \inversemirrormap , and
vice-versa.  The integrality of mirror map in the bulk  had not
been  physically explained before.  Here, by relating it to the
integrality of numbers of domain walls $N_{k,1}$ we have found a
physical explanation for it. The integrality of $N_{k,m}$ $m\neq
1$ requires special properties of the $a_i$ and seems much more
involved.

\subsec{${\cal O}(K)\rightarrow {\bf F}_1$} Consider a CY geometry
containing the blowup of ${\bf P}^2$ at one point, which we denote
by ${\bf F}_1$. The charges describing the gauged linear
$\sigma$-model on the non-compact Calabi-Yau are
$Q_b=(-2,0,0,1,1)$ and $Q_f=(-1,1,1,-1,0)$, so the corresponding
$D$-terms are \eqn\dtermsfI{\eqalign{ |X^3|^2+|X^4|^2-
2|X^0|^2=&r_b\cr |X^1|^2+|X^2|^2- |X^0|^2-|X^3|^2=&r_f\, ,}} where
$r_b$ and $r_f$ are the areas of the base and the fiber of the
Hirzebruch surface ${\bf F}_1$. The equations are ``solved'' in
Fig.19.
 The threefold
has two classes of divisors: $H$ coming from the $\IP^2$ itself
and the exceptional divisor $E$ with $H^2=1$, $E^2=-1$ and $EH=0$.
The charge vectors (or generators of the mori-cone) correspond to
the fiber $F=H-E$ and the base, which is $E$. Note that
$Q_b+Q_f=Q_{P^2}$ with $Q_{P^2}$ is the charge for the
$O(-3)\rightarrow P^1$, and correspondingly $E+F =H$.

The A-brane charges are again such that the brane is special
Lagrangian, e.g. $q^1 = (-1,1,0,0,0)$ and $q^2 = (-1,0,0,0,1)$
which defines the moduli $c_1,c_2$:
$$|X^1|^2 - |X^0|^2 = c_1$$
$$|X^4|^2 - |X^0|^2 = c_2$$
%
%
We will consider the following 3 phases (out of a total of 8
possibilities)\foot{We have checked that all other phases also lead to
integral expansions
for disk amplitudes.}
$${\rm Phase \ \  I}\qquad
 (r_b +r_f)/2   > c_1 > 0  \quad c_2 =0$$
\eqn\fIbbranes{\rm{Phase  \ \   II} \qquad
 c_1 = 0  \quad r_b/2 > c_2 > 0,}
$${\rm Phase  \ \   III} \quad
r_f /2 > c_1 > 0 \quad c_2 = r_b$$

\bigskip
\centerline{\epsfxsize 5truein\epsfbox{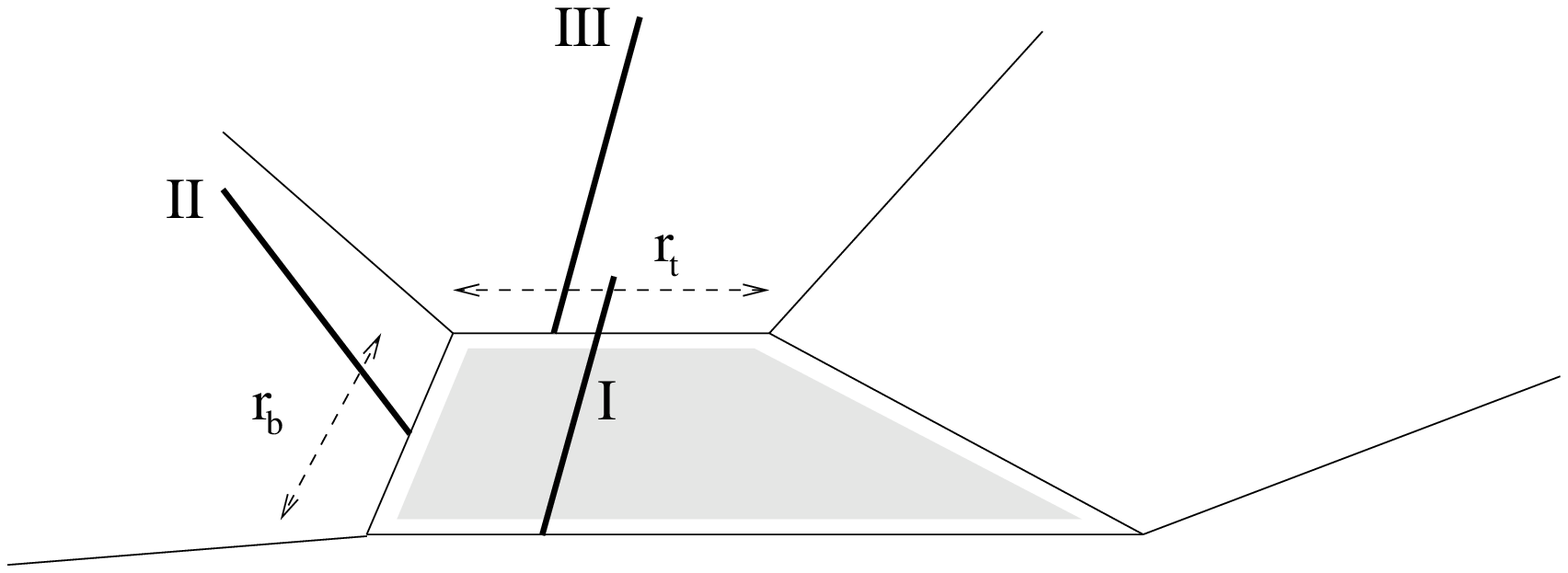}}
\rightskip 2pc
\noindent{\ninepoint\sl \baselineskip=8pt {\bf Fig.19}{ Toric base of
$O(K)\rightarrow {\bf F_1}$ and the $A-$brane in three phases.
$O(K)\rightarrow {\bf F_1}$ can be viewed as a blowup of
$O(-3)\rightarrow {\bf P^2}$ at a point by a ${\bf P^1}$ of size $r_t$.}}
\bigskip

The mirror of ${\cal O}(K)\rightarrow F_1$    is given  by
\eqn\mfI{ xz = 1+
e^u + e^{-v-u}z_f z_b + e^{-v} z_b + e^v }
where $z_b = e^{-t_b}$, $z_f = e^{-t_f}$
and we have solved the mirror relations in terms of $Y^1 = u, \; Y^4 = v$.
with $Y^0=0$.
The mirror of the brane in phase $I$ has $u \sim (t_b+t_f)/2$ as a
variable with $v \sim 0$, $II$ has $v \sim t_f/2$ as a
variable with $u \sim 0$ on the relevant leg of the toric diagram.
One finds using the same methods discussed for the
${\bf P}^1\times {\bf P}^1$ example that
$$\Delta_{u,v}=  (t_f - \hat t_f) +i \pi $$
%

The numbers of primitive disks in these two phases, listed
side by side are as follows:

\bigskip
{\vbox{\ninepoint{
$$\eqalign{
\vbox{\offinterlineskip\tabskip=0pt \halign{\strut
\vrule#& &\hfil ~$#$ \hfil &\vrule#& \hfil ~$#$ &\hfil ~$#$ &\hfil ~$#$ &\hfil ~$#$ &\hfil ~$#$ &\hfil ~$#$ \hfil & \vrule#\cr
\noalign{\hrule}
&m=-5 && II & & & & & &  \cr\noalign{\hrule}
&k_b&& k_f=0 &1 &2 &3 &4 &5& \cr \noalign{\hrule}
&5&&0& 1&-6& 14& -14& 5&  \cr
&6&&0& 2& -30& 140& -280& 252&  \cr
\noalign{\hrule}}\hrule}& \ \
\vbox{\offinterlineskip\tabskip=0pt \halign{\strut
\vrule#& & \hfil ~$#$ &\hfil ~$#$ &\hfil ~$#$ &\hfil ~$#$ &\hfil ~$#$ &\hfil ~$#$ \hfil & \vrule#\cr
\noalign{\hrule}
& I  &  &  &  &  &  &\cr\noalign{\hrule}
& k_f=0 &1 &2 &3 &4 & 5&  \cr \noalign{\hrule}
&0      & 0&0 & 0& 0& 0&  \cr
&0      & 0&0 & 0& 0&42&  \cr
\noalign{\hrule}}\hrule}}
$$}}}

{\vbox{\ninepoint{
$$\eqalign{
\vbox{\offinterlineskip\tabskip=0pt \halign{\strut
\vrule#& &\hfil ~$#$ \hfil &\vrule#& \hfil ~$#$ &\hfil ~$#$ &\hfil ~$#$ &\hfil ~$#$ &\hfil ~$#$ &\hfil ~$#$ \hfil & \vrule#\cr
\noalign{\hrule}
&m=-4 && II & & & & & &  \cr\noalign{\hrule}
&k_b&& k_f=0 &1 &2 &3 &4 &5& \cr \noalign{\hrule}
&4&&0& 1& -4& 5& -2& 0& \cr
&5&&0&2& -20& 60& -70& 28& \cr
&6&&0& 3& -68& 400& -936&  -344&\cr
\noalign{\hrule}}\hrule}& \ \
\vbox{\offinterlineskip\tabskip=0pt \halign{\strut
\vrule#& & \hfil ~$#$ &\hfil ~$#$ &\hfil ~$#$ &\hfil ~$#$ &\hfil ~$#$ &\hfil ~$#$ \hfil & \vrule#\cr
\noalign{\hrule}
& I  &  &  &  &  &  &\cr\noalign{\hrule}
& k_f=0 &1 &2 &3 &4 & 5&  \cr \noalign{\hrule}
&0& 0& 0& 0& -2& 0& \cr
&0& 0&0& 0& -14& 28&  \cr
&0& 0& 0& 0& -52& 308& \cr
\noalign{\hrule}}\hrule}}
$$}}}

{\vbox{\ninepoint{
$$\eqalign{
\vbox{\offinterlineskip\tabskip=0pt \halign{\strut
\vrule#& &\hfil ~$#$ \hfil &\vrule#& \hfil ~$#$ &\hfil ~$#$ &\hfil ~$#$ &\hfil ~$#$ &\hfil ~$#$ &\hfil ~$#$ \hfil & \vrule#\cr
\noalign{\hrule}
&m=-3 && II & & & & & &  \cr\noalign{\hrule}
&k_b&& k_f=0 &1 &2 &3 &4 &5& \cr \noalign{\hrule}
&3&&0& 1&-2& 1& 0& 0&  \cr
&4&&0& 2& -12& 20& -10& 0& \cr
&5&&0& 3& -45&170& -230& 102& \cr
&6&&0& 4& -130& 958& -2612& 2940& \cr
\noalign{\hrule}}\hrule}& \ \
\vbox{\offinterlineskip\tabskip=0pt \halign{\strut
\vrule#& & \hfil ~$#$ &\hfil ~$#$ &\hfil ~$#$ &\hfil ~$#$ &\hfil ~$#$ &\hfil ~$#$ \hfil & \vrule#\cr
\noalign{\hrule}
& I  &  &  &  &  &  &\cr\noalign{\hrule}
& k_f=0 &1 &2 &3 &4 & 5&  \cr \noalign{\hrule}
&0& 0& 0& 1& 0& 0&  \cr
&0& 0&0& 5& -10& 0&  \cr
&0& 0& 0& 14& -90& 102&  \cr
&0& 0& 0& 31&-450& 1428& \cr
\noalign{\hrule}}\hrule}}
$$}}}

{\vbox{\ninepoint{
$$\eqalign{
\vbox{\offinterlineskip\tabskip=0pt \halign{\strut
\vrule#& &\hfil ~$#$ \hfil &\vrule#& \hfil ~$#$ &\hfil ~$#$ &\hfil ~$#$ &\hfil ~$#$ &\hfil ~$#$ &\hfil ~$#$ \hfil & \vrule#\cr
\noalign{\hrule}
&m=-2 && II &  &  &  &  & &  \cr\noalign{\hrule}
&k_b&& k_f=0  &1 &2 &3 &4 &5& \cr \noalign{\hrule}
&2&&0& 1& -1& 0& 0& 0&  \cr
&3&&0& 2& -6&4& 0& 0&  \cr
&4&&0& 3& -28& 57& -32& 0& \cr
&5&&0& 4& -90&424& -664& 326& \cr
&6&&0& 5& -237& 2172& -6872& 8640& \cr
\noalign{\hrule}}\hrule}& \ \
\vbox{\offinterlineskip\tabskip=0pt \halign{\strut
\vrule#& & \hfil ~$#$ &\hfil ~$#$ &\hfil ~$#$ &\hfil ~$#$ &\hfil ~$#$ &\hfil ~$#$ \hfil & \vrule#\cr
\noalign{\hrule}
& I  &  &  &  &  &  &\cr\noalign{\hrule}
& k_f=0 &1 &2 &3 &4 & 5&  \cr \noalign{\hrule}
&0& 0& -1& 0& 0& 0& \cr
&0& 0&-2& 4& 0& 0&  \cr
&0& 0& -4& 28& -32& 0& \cr
&0& 0& -6& 112&-390& 326& \cr
&0& 0& -9& 336& -2500& 5638& \cr
\noalign{\hrule}}\hrule}}
$$}}}
{\vbox{\ninepoint{
$$\eqalign{
\vbox{\offinterlineskip\tabskip=0pt \halign{\strut
\vrule#& &\hfil ~$#$ \hfil &\vrule#& \hfil ~$#$ &\hfil ~$#$ &\hfil ~$#$ &\hfil ~$#$ &\hfil ~$#$ &\hfil ~$#$ \hfil & \vrule#\cr
\noalign{\hrule}
&m=-1 && II & & & & & &  \cr\noalign{\hrule}
&k_b&& k_f=0 &1 &2 &3 &4 &5& \cr \noalign{\hrule}
&1&&-1& 1& 0& 0& 0& 0&  \cr
&2&&0& 2& -2& 0& 0& 0&  \cr
&3&&0& 3& -15& 12& 0& 0&  \cr
&4&&0& 4& -60& 160& -104& 0&  \cr
&5&&0& 5& -175& 1080& -1995& 1085&  \cr
\noalign{\hrule}}\hrule}& \ \
\vbox{\offinterlineskip\tabskip=0pt \halign{\strut
\vrule#& & \hfil ~$#$ &\hfil ~$#$ &\hfil ~$#$ &\hfil ~$#$ &\hfil ~$#$ &\hfil ~$#$ \hfil & \vrule#\cr
\noalign{\hrule}
& I  &  &  &  &  &  &\cr\noalign{\hrule}
& k_f=0 &1 &2 &3 &4 & 5&  \cr \noalign{\hrule}
&0& 1& 0& 0& 0& 0&  \cr
&0& 1& -2& 0& 0& 0& \cr
&0& 1& -10& 12& 0& 0&  \cr
&0& 1& -30& 120& -104& 0&  \cr
&0& 1& -70& 648& -1596& 1085&  \cr
\noalign{\hrule}}\hrule}}
$$}}}
{\vbox{\ninepoint{
$$\eqalign{
\vbox{\offinterlineskip\tabskip=0pt \halign{\strut
\vrule#& &\hfil ~$#$ \hfil &\vrule#& \hfil ~$#$ &\hfil ~$#$ &\hfil ~$#$ &\hfil ~$#$ &\hfil ~$#$ &\hfil ~$#$ \hfil & \vrule#\cr
\noalign{\hrule}
&m=1 && II & & & & & &  \cr\noalign{\hrule}
&k_b&& k_f=0 &1 &2 &3 &4 &5& \cr \noalign{\hrule}
&0&&1& 0& 0& 0& 0& 0&  \cr
&1&&0& -1& 0& 0& 0& 0&  \cr
&2&&0& -2& 5& 0& 0& 0&  \cr
&3&&0& -3& 30& -40& 0& 0&  \cr
&4&&0& -4& 105& -432& 399& 0&  \cr
&5&&0& -5& 280& -2520& 6370& -4524& \cr
\noalign{\hrule}}\hrule}& \ \
\vbox{\offinterlineskip\tabskip=0pt \halign{\strut
\vrule#& & \hfil ~$#$ &\hfil ~$#$ &\hfil ~$#$ &\hfil ~$#$ &\hfil ~$#$ &\hfil ~$#$ \hfil & \vrule#\cr
\noalign{\hrule}
& I  &  &  &  &  &  &\cr\noalign{\hrule}
& k_f=0 &1 &2 &3 &4 & 5&  \cr \noalign{\hrule}
&1& 0& 0& 0& 0& 0&  \cr
&1& -1& 0& 0& 0& 0&  \cr
&1& -6& 5& 0& 0& 0&  \cr
&1& -20& 59& -40& 0& 0&  \cr
&1& -50& 356& -706& 399& 0& \cr
&1& -105& 1500& -6244& 9372& -4524&  \cr
\noalign{\hrule}}\hrule}}
$$}}}
{\vbox{\ninepoint{
$$\eqalign{
\vbox{\offinterlineskip\tabskip=0pt \halign{\strut
\vrule#& &\hfil ~$#$ \hfil &\vrule#& \hfil ~$#$ &\hfil ~$#$ &\hfil ~$#$ &\hfil ~$#$ &\hfil ~$#$ &\hfil ~$#$ \hfil & \vrule#\cr
\noalign{\hrule}
&m=2 && II & & & & & &  \cr\noalign{\hrule}
&k_b&& k_f=0 &1 &2 &3 &4 &5& \cr \noalign{\hrule}
&0&&0& 0& 0& 0& 0& 0&  \cr
&1&&0& -1& 0& 0& 0& 0&  \cr
&2&&0& -2& 7& 0& 0& 0&  \cr
&3&&0& -3& 38& -61& 0& 0& \cr
&4&&0& -4& 128& -616& 648& 0&  \cr
&5&&0& -5& 330& -3420& 9744& -7661& \cr
\noalign{\hrule}}\hrule}& \ \
\vbox{\offinterlineskip\tabskip=0pt \halign{\strut
\vrule#& & \hfil ~$#$ &\hfil ~$#$ &\hfil ~$#$ &\hfil ~$#$ &\hfil ~$#$ &\hfil ~$#$ \hfil & \vrule#\cr
\noalign{\hrule}
& I  &  &  &  &  &  &\cr\noalign{\hrule}
& k_f=0 &1 &2 &3 &4 & 5&  \cr \noalign{\hrule}
&0& 0& 0& 0& 0& 0& \cr
&1& -1& 0& 0& 0& 0&  \cr
&2& -9& 7& 0& 0& 0&  \cr
&4& -43& 100& -61& 0& 0&  \cr
&6& -147& 756& -1263& 648& 0& \cr
&9& -406& 3920& -13122& 17260& -7661&\cr
\noalign{\hrule}}\hrule}}
$$}}}
{\vbox{\ninepoint{
$$\eqalign{
\vbox{\offinterlineskip\tabskip=0pt \halign{\strut
\vrule#& &\hfil ~$#$ \hfil &\vrule#& \hfil ~$#$ &\hfil ~$#$ &\hfil ~$#$ &\hfil ~$#$ &\hfil ~$#$ &\hfil ~$#$ \hfil & \vrule#\cr
\noalign{\hrule}
&m=3 && II & & & & & &  \cr\noalign{\hrule}
&k_b&& k_f=0 &1 &2 &3 &4 &5& \cr \noalign{\hrule}
&0&&0& 0& 0& 0& 0& 0& \cr
&1&&0& -1& 0& 0& 0& 0&  \cr
&2&&0& -2& 9& 0& 0& 0&  \cr
&3&&0& -3& 48& -93& 0& 0&  \cr
&4&&0& -4& 155& -884& 1070& 0& \cr
&5&&0& -5& 390& -4682& 15134& -13257&  \cr
\noalign{\hrule}}\hrule}& \ \
\vbox{\offinterlineskip\tabskip=0pt \halign{\strut
\vrule#& & \hfil ~$#$ &\hfil ~$#$ &\hfil ~$#$ &\hfil ~$#$ &\hfil ~$#$ &\hfil ~$#$ \hfil & \vrule#\cr
\noalign{\hrule}
& I  &  &  &  &  &  &\cr\noalign{\hrule}
& k_f=0 &1 &2 &3 &4 & 5&  \cr \noalign{\hrule}
&0& 0& 0& 0& 0& 0& \cr
&1& -1& 0& 0& 0& 0&  \cr
&4& -13& 9& 0& 0& 0& \cr
&11& -85& 167& -93& 0& 0& \cr
&25& -382& 1555& -2268& 1070& 0& \cr
&49& -1344& 9813& -27584& 32323& -13257&  \cr
\noalign{\hrule}}\hrule}}
$$}}}
{\vbox{\ninepoint{
$$\eqalign{
\vbox{\offinterlineskip\tabskip=0pt \halign{\strut
\vrule#& &\hfil ~$#$ \hfil &\vrule#& \hfil ~$#$ &\hfil ~$#$ &\hfil ~$#$ &\hfil ~$#$ &\hfil ~$#$ &\hfil ~$#$ \hfil & \vrule#\cr
\noalign{\hrule}
&m=4 && II & & & & & &  \cr\noalign{\hrule}
&k_b&& k_f=0 &1 &2 &3 &4 &5& \cr \noalign{\hrule}
&0&&0& 0& 0& 0& 0& 0&  \cr
&1&&0& -1& 0& 0& 0& 0&  \cr
&2&&0& -2& 12& 0& 0& 0&  \cr
&3&&0& -3& 60& -140& 0& 0&  \cr
&4&&0& -4& 188& -1260& 1750& 0&  \cr
&5&&0& -5& 460& -6397& 23482& -22955& \cr
\noalign{\hrule}}\hrule}& \ \
\vbox{\offinterlineskip\tabskip=0pt \halign{\strut
\vrule#& & \hfil ~$#$ &\hfil ~$#$ &\hfil ~$#$ &\hfil ~$#$ &\hfil ~$#$ &\hfil ~$#$ \hfil & \vrule#\cr
\noalign{\hrule}
& I  &  &  &  &  &  &\cr\noalign{\hrule}
& k_f=0 &1 &2 &3 &4 & 5&  \cr \noalign{\hrule}
&0& 0& 0& 0& 0& 0&  \cr
&1& -1& 0& 0& 0& 0&  \cr
&6& -18& 12& 0& 0& 0&  \cr
&25& -155& 270& -140& 0& 0&  \cr
&76& -887& 3056& -3995& 1750& 0&  \cr
&196& -3873& 23040& -56429& 60021& -22955&  \cr
\noalign{\hrule}}\hrule}}
$$}}}
{\vbox{\ninepoint{
$$\eqalign{
\vbox{\offinterlineskip\tabskip=0pt \halign{\strut
\vrule#& &\hfil ~$#$ \hfil &\vrule#& \hfil ~$#$ &\hfil ~$#$ &\hfil ~$#$ &\hfil ~$#$ &\hfil ~$#$ &\hfil ~$#$ \hfil & \vrule#\cr
\noalign{\hrule}
&m=5 && II & & & & & &  \cr\noalign{\hrule}
&k_b&& k_f=0 &1 &2 &3 &4 &5& \cr \noalign{\hrule}
&0&&0& 0& 0& 0& 0& 0&  \cr
&1&&0& -1& 0& 0& 0& 0&  \cr
&2&&0& -2& 15& 0& 0& 0&  \cr
&3&&0& -3& 74& -206& 0& 0&  \cr
&4&&0& -4& 225& -1772& 2821& 0&  \cr
\noalign{\hrule}}\hrule}& \ \
\vbox{\offinterlineskip\tabskip=0pt \halign{\strut
\vrule#& & \hfil ~$#$ &\hfil ~$#$ &\hfil ~$#$ &\hfil ~$#$ &\hfil ~$#$ &\hfil ~$#$ \hfil & \vrule#\cr
\noalign{\hrule}
& I  &  &  &  &  &  &\cr\noalign{\hrule}
& k_f=0 &1 &2 &3 &4 & 5&  \cr \noalign{\hrule}
&0& 0& 0& 0& 0& 0& \cr
&1& -1& 0& 0& 0& 0&  \cr
&9& -24& 15& 0& 0& 0&  \cr
&49& -264& 421& -206& 0& 0& \cr
&196& -1876& 5700& -6841& 2821& 0&  \cr
\noalign{\hrule}}\hrule}}
$$}}}
\noindent{\ninepoint\sl \baselineskip=2pt {\bf Tab.8\ }{Disk 
degeneracies for brane $I$ and $II$
in the $O(K)\rightarrow F_1$ geometry. }}
\bigskip

We consider now the phase $III$, for which we
must change the parameterization of the curve by $u \rightarrow u' = -t_b -
u$, if we are to have the superpotential which is zero classically.
In this  phase $v$ is the transverse coordinate on the brane.
The correction to the flat coordinate is again found
by requiring integrality of the amplitude,
and we find that $$\hat v  = v + \Delta_v = v + t_f - \hat t_f.$$

The disk numbers follow:

\bigskip
{\vbox{\ninepoint{
$$\eqalign{
\vbox{\offinterlineskip\tabskip=0pt \halign{\strut
\vrule#& &\hfil ~$#$ \hfil &\vrule#& \hfil ~$#$ &\hfil ~$#$ &\hfil ~$#$ &\hfil ~$#$ &\hfil ~$#$ &\hfil ~$#$ \hfil & \vrule#\cr
\noalign{\hrule}
&m=1 &&  & & & & & &  \cr\noalign{\hrule}
&k_b&& k_f=0 &1  &2 &3 &4 & 5& \cr \noalign{\hrule}
&0&&        1& -1& 0& 0& 0& 0&\cr
&1&& 0& -2& 2& 0& 0& 0&\cr
&2&& 0& -3& 15& -12& 0& 0&\cr
&3&& 0& -4& 60& -160& 104& 0&\cr
&4&& 0& -5& 175& -1080& 1995& -1085&\cr
&5&& 0& -6& 420& -5040& 19110& -27144& \cr
&6&& 0& -7& 882& -18480& 124033& -337617&\cr
\noalign{\hrule}}\hrule}& \ \
\vbox{\offinterlineskip\tabskip=0pt \halign{\strut
\vrule#& & \hfil ~$#$ &\hfil ~$#$ &\hfil ~$#$ &\hfil ~$#$ &\hfil ~$#$ &\hfil ~$#$ \hfil & \vrule#\cr
\noalign{\hrule}
& m=-1  &  &  &  &  &  &\cr\noalign{\hrule}
& k_f=0 &1 &2 &3 &4 & 5&  \cr \noalign{\hrule}
&0& 0& 0& 0& 0& 0& \cr
&-1& 0& 0& 0& 0& 0&\cr
&0& 1& 0& 0& 0& 0& \cr
&0& 2& -5& 0& 0& 0& \cr
&0& 3& -30& 40& 0& 0& \cr
&0& 4& -105& 432& -399&  0&\cr
&0& 5& -280& 2520& -6370& 4524& \cr
\noalign{\hrule}}\hrule}}
$$}}}
{\vbox{\ninepoint{
$$\eqalign{
\vbox{\offinterlineskip\tabskip=0pt \halign{\strut
\vrule#& &\hfil ~$#$ \hfil &\vrule#& \hfil ~$#$ &\hfil ~$#$ &\hfil ~$#$ &\hfil ~$#$ &\hfil ~$#$ &\hfil ~$#$ \hfil & \vrule#\cr
\noalign{\hrule}
&m=2 &&  & & & & & &  \cr\noalign{\hrule}
&k_b&& k_f=0 &1  &2 &3 &4 & 5& \cr \noalign{\hrule}
&0&& 0& -1& 1& 0& 0& 0&\cr
&1&& 0& -2& 6& -4& 0& 0& \cr
&2&& 0& -3& 28& -57& 32& 0&\cr
&3&& 0& -4& 90& -424& 664& -326&\cr
&4&& 0& -5& 237& -2172& 6872& -8640& \cr
&5&& 0& -6& 532& -8640& 48208& -114774& \cr
&6&& 0& -7& 1072& -28578& 258516& -1023679& \cr
\noalign{\hrule}}\hrule}& \ \
\vbox{\offinterlineskip\tabskip=0pt \halign{\strut
\vrule#& & \hfil ~$#$ &\hfil ~$#$ &\hfil ~$#$ &\hfil ~$#$ &\hfil ~$#$ &\hfil ~$#$ \hfil & \vrule#\cr
\noalign{\hrule}
& m=-2  &  &  &  &  &  &\cr\noalign{\hrule}
& k_f=0 &1 &2 &3 &4 & 5&  \cr \noalign{\hrule}
& 0& 0& 0& 0& 0& 0& \cr
& 0& 0& 0& 0& 0& 0& \cr
& 0& 0& 0& 0& 0& 0& \cr
& 0& 1& 0& 0& 0& 0& \cr
& 0& 2& -7& 0& 0& 0& \cr
& 0& 3& -38& 61& 0& 0& \cr
& 0& 4& -128& 616& -648& 0&\cr
 \noalign{\hrule}}\hrule}}
$$}}}
{\vbox{\ninepoint{
$$\eqalign{
\vbox{\offinterlineskip\tabskip=0pt \halign{\strut
\vrule#& &\hfil ~$#$ \hfil &\vrule#& \hfil ~$#$ &\hfil ~$#$ &\hfil ~$#$ &\hfil ~$#$ &\hfil ~$#$ &\hfil ~$#$ \hfil & \vrule#\cr
\noalign{\hrule}
&m=3 &&  & & & & & &  \cr\noalign{\hrule}
&k_b&& k_f=0 &1  &2 &3 &4 & 5& \cr \noalign{\hrule}
&0&& 0& -1& 2& -1& 0& 0& \cr
&1&& 0& -2& 12& -20& 10& 0& \cr
&2&& 0& -3& 45& -170& 230& -102& \cr
&3&& 0& -4& 130& -958& 2612& -2940& \cr
&4&& 0& -5& 315& -4116& 19750& -41996& \cr
&5&& 0& -6& 672& -14520& 112970& -398970& \cr
&6&& 0& -7& 1302& -44073& 525031& -2854610&\cr
\noalign{\hrule}}\hrule}& \ \
\vbox{\offinterlineskip\tabskip=0pt \halign{\strut
\vrule#& & \hfil ~$#$ &\hfil ~$#$ &\hfil ~$#$ &\hfil ~$#$ &\hfil ~$#$ &\hfil ~$#$ \hfil & \vrule#\cr
\noalign{\hrule}
& m=-3  &  &  &  &  &  &\cr\noalign{\hrule}
& k_f=0 &1 &2 &3 &4 & 5&  \cr \noalign{\hrule}
& 0& 0& 0& 0& 0& 0&\cr
& 0& 0& 0& 0& 0& 0&\cr
& 0& 0& 0& 0& 0& 0&\cr
& 0& 0& 0& 0& 0& 0&\cr
& 0& 1& 0& 0& 0& 0&\cr
& 0& 2& -9& 0& 0& 0&\cr
& 0& 3& -48& 93& 0& 0&\cr
\noalign{\hrule}}\hrule}}$$}}}
{\vbox{\ninepoint{
$$\eqalign{
\vbox{\offinterlineskip\tabskip=0pt \halign{\strut
\vrule#& &\hfil ~$#$ \hfil &\vrule#& \hfil ~$#$ &\hfil ~$#$ &\hfil ~$#$ &\hfil ~$#$ &\hfil ~$#$ &\hfil ~$#$ \hfil & \vrule#\cr
\noalign{\hrule}
&m=4 &&  & & & & & &  \cr\noalign{\hrule}
&k_b&& k_f=0 &1  &2 &3 &4 & 5& \cr \noalign{\hrule}
&0&& 0& -1& 4& -5& 2& 0& \cr
&1&& 0& -2& 20& -60& 70& -28& \cr
&2&& 0& -3& 68& -400& 936& -945&  \cr
&3&& 0& -4& 180& -1912& 7910& -15030&  \cr
&4&& 0& -5& 412& -7290& 49096& -155035& \cr
&5&& 0& -6& 840& -23520& 243558& -1185830&  \cr
&6&& 0& -7& 1576& -66660& 1015960& -7246659& \cr
\noalign{\hrule}}\hrule}& \ \
\vbox{\offinterlineskip\tabskip=0pt \halign{\strut
\vrule#& & \hfil ~$#$ &\hfil ~$#$ &\hfil ~$#$ &\hfil ~$#$ &\hfil ~$#$ &\hfil ~$#$ \hfil & \vrule#\cr
\noalign{\hrule}
& m=-4  &  &  &  &  &  &\cr\noalign{\hrule}
& k_f=0 &1 &2 &3 &4 & 5&  \cr \noalign{\hrule}
& 0& 0& 0& 0& 0& 0& \cr
& 0& 0& 0& 0& 0& 0& \cr
& 0& 0& 0& 0& 0& 0& \cr
& 0& 0& 0& 0& 0& 0&\cr
& 0& 0& 0& 0& 0& 0&\cr
& 0& 1& 0& 0& 0& 0&\cr
& 0& 2& -12& 0& 0& 0&\cr
\noalign{\hrule}}\hrule}}
$$}}}
{\vbox{\ninepoint{
$$\eqalign{
\vbox{\offinterlineskip\tabskip=0pt \halign{\strut
\vrule#& &\hfil ~$#$ \hfil &\vrule#& \hfil ~$#$ &\hfil ~$#$ &\hfil ~$#$ &\hfil ~$#$ &\hfil ~$#$ &\hfil ~$#$ \hfil & \vrule#\cr
\noalign{\hrule}
&m=5 &&  & & & & & &  \cr\noalign{\hrule}
&k_b&& k_f=0 &1  &2 &3 &4 & 5&\cr \noalign{\hrule}
&0&& 0& -1& 6& -14& 14& -5&  \cr
&1&& 0& -2& 30& -140& 280& -252&  \cr
&2&& 0& -3& 95& -810& 2870& -4858&  \cr
&3&& 0& -4& 240& -3472& 20150& -56728&  \cr
&4&& 0& -5& 525& -12156& 109167& -475047&  \cr
&5&& 0& -6& 1036& -36648& 487382& -3116370&  \cr
&6&& 0& -7& 1890& -98340& 1869595& -16909871& \cr
\noalign{\hrule}}\hrule}& \ \
\vbox{\offinterlineskip\tabskip=0pt \halign{\strut
\vrule#& & \hfil ~$#$ &\hfil ~$#$ &\hfil ~$#$ &\hfil ~$#$ &\hfil ~$#$ &\hfil ~$#$ \hfil & \vrule#\cr
\noalign{\hrule}
& m=-5  &  &  &  &  &  &\cr\noalign{\hrule}
& k_f=0 &1 &2 &3 &4 & 5&\cr \noalign{\hrule}
& 0& 0& 0& 0& 0& 0&\cr
& 0& 0& 0& 0& 0& 0&\cr
& 0& 0& 0& 0& 0& 0&\cr
& 0& 0& 0& 0& 0& 0& \cr
& 0& 0& 0& 0& 0& 0& \cr
& 0& 0& 0& 0& 0& 0& \cr
& 0& 1& 0& 0& 0& 0& \cr
 \noalign{\hrule}}\hrule}}
$$}}}
\noindent{\ninepoint\sl \baselineskip=2pt {\bf Tab.9\ }{The
disk degeneracies for brane $III$ in the $O(K)\rightarrow F_1$. }}
\bigskip

\centerline{\bf Acknowledgements}

We are grateful to Volker Braun, Stefan Fredenhagen,
Kentaro Hori, Amer Iqbal, Dominic Joyce, Sheldon Katz, Chiu-Chu Liu,
Marcos Marino, Cliff Taubes, Shing-Tung Yau and Eric Zaslow
for very valuable
discussions.

The research of M.A. and C.V. is supported in part by NSF grants
PHY-9802709 and DMS 9709694. The research of A.K. is supported
in part by the GIF grant I-645-130.14/1999.

\appendix{A}{Large $N$ Limit of Chern-Simons and
Framing of the Knot}
As discussed in section 5, Wilson loop observables
of Chern-Simons theory at large $N$ are naturally encoded
in terms of the expectation values
$$\langle {\rm exp} \sum_n {1\over n} \tr U^n \tr V^n \rangle
={\rm exp}(-F(V,t,g_s))$$
where $V$ is viewed as a source and $U$ is the holonomy
of the connection along the loop.  If we are interested in
extracting the amplitudes with a single hole, we consider
terms on the right hand side of the form $\tr V^n$ in the exponent.
This will correspond to contributions where the large $N$ worldsheet
wraps $n$ times around the loop.  From the left-hand side this
is equivalent to computation of
$$\langle {1\over n} \tr U^n \rangle.$$
For a special choice of framing of the unknot this was
computed in \ov\ and it was shown that
$$\langle {1\over n} \tr U^n \rangle ={1\over n} \tr U_0^n $$
where $U_0$ is a particular element of $SU(N)$ given by
a diagonal matrix with  entries ${\rm exp}({i\pi (N+1-2r)
\over N+k})$, as $r$ ranges from $1$ to $N$. This leads,
in the leading order in $g_s$, corresponding
to disk amplitude ($g=0$ and one hole), to
$$\langle {1\over n} \tr U^n \rangle ={1\over n^2g_s }[
{\rm exp}(nt/2)-{\rm exp}(-nt/2)]$$
where in the large $t$ limit and by redefining
the $V$ (and absorbing a factor of ${\rm exp}(t/2)$
in it ) and in the large $t$ limit, we obtain the
${\bf C}^3$ answer for the disk amplitude
$${1\over g_s}\sum_n {1\over n^2} \tr V^n$$
(where we have identified in this paper $V=e^{u}$).

Now we ask what happens if we choose a different framing.
For the expectation value of the Wilson loops with holonomies
in a given representation $R$ of $SU(N)$ the answer is relatively
simple:
$$\langle \Tr_R U \rangle_p=\langle \Tr_R U \rangle_0 exp(2\pi i p {h_R
\over k+N})$$
where $p$ denotes the change in the framing from a given one
denoted by $0$ (characterized by an element of topological class
of winding of an $S^1$ over an $S^1$), and $h_R$ denotes the
Casimir of the representation.  In other words, the result of the
change in framing is a multiplication by $exp(g_s p h_R)$.

However we need to compute the correlation function for $\langle
\tr U^n \rangle$ which is the trace in the fundamental
representation of $U^n$. To find the change in this correlation
function due to change in framing, we will first have to write it
in terms of trace of $U$ in different representations, compute
each one, and multiply each one with $exp(g_s p h_R)$.

There is an identity which is useful for this purpose:
$$\tr U^n =\sum_{s=0}^{n-1}(-1)^s \Tr_{R_{n,s}}U$$
where $R_{n,s}$ denotes representations of $SU(N)$ with $n$ boxes
for the Young Tableau which look like `$\Gamma$' and which
consists of only one non-trivial row and one non-trivial column.
$s+1$ denotes the number of elements in the first column.

Combining all this we find that the coefficient of $tr V^n$
for the unknot and with framing number $p$ is given
by
$$\langle {1\over n} \tr U^n\rangle =
{1\over n}\sum_{s=0}^{n-1} (-1)^s {\rm exp}(pg_s h_{R_{n,s}})
\Tr_{R_{n,s}} U_0$$
We are interested in extracting the disk amplitude which is the
leading term for this expansion as $g_s\rightarrow 0$. Moreover we
will be interested in the limit where $t\rightarrow \infty$, but
where we rescale $V$ by a factor of $e^{t/2}$ so that we would be
computing the brane in the ${\bf C}^3$ geometry. The most
complicated aspect of this computation is finding $\Tr_{R_{n,s}}
U_0$.  We will reduce this computation to a group theory
computation as follows: Let
$$\Tr_{R_{n,s}}U_0=\sum_{\sum i n_i=n}\alpha^{R_{n,s}}_{n_i}\prod_{i=1}^{n}[tr
U_0^i]^{n_i}$$
for some group theoretic factor $\alpha^{R_{n,s}}_{n_i}$.
Using this and the leading
computation at large $N$ for the unknot
with the standard framing we can reduce the above
computation to
$$\langle {1\over n} \tr U^n\rangle =
{1\over n}\sum_{s=0}^{n-1} (-1)^s {\rm exp}(pg_s h_{R_{n,s}})
\sum_{\sum i n_i=n}\alpha^{R_{n,s}}_{n_i}\prod_{i=1}^{n}({1\over i
g_s})^{n_i}$$
Using this and the fact that up to a further redefinition of $V$
and dropping a subleading term in $g_s$, $h_{R_{n,s}} ={1\over
2}n(n-1-2s)$ we have computed the above expression for $n=1,2,3$
and found perfect match with the result of the computation done
for ${\bf C}^3$ when we shift $u\rightarrow u+pv$.  Moreover we
have verified that we obtain for all $n$ a polynomial of degree
$n-1$ in $p$ in agreement with the results of ${\bf C}^3$.
Furthermore we have verified\foot{This involves using the identity
${n^{n-2}\over n! 2^{n-1}}\sum_{r=1}^{n}
(-1)^{r+1}{(n-2r+1)^{n-1}\over (n-r)!(r-1)!}={n^{n-2}\over n!}$.}\
 that the leading power in $p$ agrees
between both computations for all $n$.

\listrefs
\end